\documentclass[fleqn,usenatbib]{mnras}

\usepackage{newtxtext,newtxmath}
\usepackage[T1]{fontenc}

\DeclareRobustCommand{\VAN}[3]{#2}
\let\VANthebibliography\thebibliography
\def\thebibliography{\DeclareRobustCommand{\VAN}[3]{##3}\VANthebibliography}

\usepackage{graphicx}	% Including figure files
\usepackage[flushleft]{threeparttable}
\usepackage{booktabs}
\usepackage{amsmath}
\usepackage{subfigure}
\usepackage{pdflscape}
\usepackage{array}
\usepackage{tabularx}
\usepackage{verbatim}
\usepackage{colortbl}
\usepackage{hyperref}
\usepackage{float}
\usepackage{adjustbox}
\usepackage{multirow}
\usepackage{rotating}

\title[TOI-2374\,b and TOI-3071\,b]{TOI-2374\,b and TOI-3071\,b: two metal-rich sub-Saturns well within the Neptunian desert\thanks{This paper includes data gathered with the 6.5 meter Magellan Telescopes located at Las Campanas Observatory, Chile.}}

\author[A. Hacker et al.]{
Alejandro~Hacker$^{1}$,\thanks{E-mail: ahacker@unsam.edu.ar}
Rodrigo~F.~Díaz$^{1}$,
David~J.~Armstrong$^{2,3}$,
Jorge~Fern\'andez~Fern\'andez$^{2,3}$,
\newauthor
Simon~Müller$^{4}$,
Elisa~Delgado-Mena$^{5}$,
S\'ergio~G.~Sousa$^{5}$,
Vardan~Adibekyan$^{5,6}$,
Keivan~G.~Stassun$^{7}$,
\newauthor
Karen~A.~Collins$^{8}$,
Samuel~W.~Yee$^{9,8}$,
Daniel Bayliss$^{2}$,
Allyson~Bieryla$^{8}$,
Fran\c{c}ois Bouchy$^{10}$,
\newauthor
R.~Paul~Butler$^{11}$,
Jeffrey~D.~Crane$^{12}$,
Xavier Dumusque$^{10}$,
Joel~D.~Hartman$^{9}$,
Ravit~Helled$^{4}$,
Jon~Jenkins$^{13}$,
\newauthor
Marcelo~Aron~F.~Keniger$^{2,3}$,
Hannah~Lewis$^{14}$,
Jorge~Lillo-Box$^{15}$,
Michael~B.~Lund$^{16}$,
\newauthor
Louise,\ D.\ Nielsen$^{17,18}$,
Ares~Osborn$^{19,20}$,
David~Osip$^{21}$,
Martin~Paegert$^{8}$,
Don~J.~Radford$^{22}$,
\newauthor
Nuno~C.~Santos$^{5,6}$,
Sara Seager$^{23,24,25}$,
Stephen~A.~Shectman$^{12}$,
Gregor~Srdoc$^{26}$,
Paul A. Str{\o}m$^{2,3}$,
\newauthor
Thiam-Guan~Tan$^{27}$,
Johanna~K.~Teske$^{11}$,
Michael~Vezie$^{23,24}$,
David~Watanabe$^{28}$,
Cristilyn~N.~Watkins$^{8}$,
\newauthor
Peter~J.~Wheatley$^{2,3}$
Joshua~N.~Winn$^{9}$,
Bill~Wohler$^{29,13}$
and Carl~Ziegler$^{30}$
\\
The authors' affiliations are shown in Appendix \ref{sec:affiliations}.
}

\date{Accepted  2024 June 01. Received 2024 May 29; in original form 2024 May 08}

\pubyear{2023}

\begin{document}
\label{firstpage}
\pagerange{\pageref{firstpage}--\pageref{lastpage}}
\maketitle

\begin{abstract}
We report the discovery of two transiting planets detected by the Transiting Exoplanet Survey Satellite (TESS), TOI-2374 b and TOI-3071 b, orbiting a K5V and an F8V star, respectively, with periods of 4.31 and 1.27 days, respectively. We confirm and characterize these two planets with a variety of ground-based and follow-up observations, including photometry, precise radial velocity monitoring, and high-resolution imaging. The planetary and orbital parameters were derived from a joint analysis of the radial velocities and photometric data. 
%Inference was performed using samples from the posterior distribution obtained using a Hamiltonian Monte Carlo algorithm.
We found that the two planets have masses of $(57 \pm 4)$ $M_\oplus$ or $(0.18 \pm 0.01)$ $M_J$, and $(68 \pm 4)$ $M_\oplus$ or $(0.21 \pm 0.01)$ $M_J$, respectively, and they have radii of $(6.8 \pm 0.3)$ $R_\oplus$ or $(0.61 \pm 0.03)$ $R_J$ and $(7.2 \pm 0.5)$ $R_\oplus$ or $(0.64 \pm 0.05)$ $R_J$, respectively. These parameters correspond to sub-Saturns within the Neptunian desert, both planets being hot and highly irradiated, with $T_{\rm eq} \approx 745$ $K$ and $T_{\rm eq} \approx 1812$ $K$, respectively, assuming a Bond albedo of 0.5. TOI-3071\,b has the hottest equilibrium temperature of all known planets with masses between $10$ and $300$ $M_\oplus$ and radii less than $1.5$ $R_J$.
%has a considerably smaller radius than that of all previously detected exoplanets with masses between $10$ and $300$ $M_\oplus$ and similar equilibrium temperature (1800 $\pm$ 100 K).
By applying gas giant evolution models we found that both planets, especially TOI-3071\,b, are very metal-rich. This challenges standard formation models which generally predict lower heavy-element masses for planets with similar characteristics. We studied the evolution of the planets' atmospheres under photoevaporation and concluded that both are stable against evaporation due to their large masses and likely high metallicities in their gaseous envelopes.

%TOI-2374\,b is stable against evaporation but TOI-3071\,b, which is twice as close to its star and receives a higher X-ray flux, is rapidly losing its atmosphere and may lose it completely before its star leaves the main sequence.
\end{abstract}


%%%%%%%%%%%%%%%%%%%%%%%%%%%%%%%%%%%%%%%%%%%%%%%%%%

%%%%%%%%%%%%%%%%% BODY OF PAPER %%%%%%%%%%%%%%%%%%

\section{Introduction}

%EXOPLANETS & TESS
Since the discovery of 51 Pegasi b \citep{mayorqueloz}, more than 5500 exoplanets have been confirmed, thanks largely to detections from ground based RV surveys such as HARPS \citep[High Accuracy Radial velocity Planet Searcher;][]{harps} , transiting surveys such as WASP \citep[Wide Angle Search for Planets;][]{Pollacco2006} and space telescopes such as CoRot \citep[Convection, Rotation, and planetary Transits;][]{baglin2008}, Kepler \citep{Borucki2010} and, most recently, the Transiting Exoplanet Survey Satellite \citep[{TESS},][]{Ricker2015}. TESS was launched in 2018 with the objective of detecting exoplanets around bright nearby stars by monitoring flux variations of hundreds of thousands of stars in the solar neighborhood, covering $\sim 95\%$ of the sky. Furthermore, it has been optimized to find planets around stars that are bright enough to be able to extract meaningful information from radial velocity (RV) observations from the ground (particularly, planetary mass determination) and conduct atmospheric studies with the James Webb Space Telescope \citep[{JWST},][]{Gardner2006} and other facilities. Since its launch, TESS has confirmed 398 exoplanets with thousands of project candidates still awaiting their nature to be established. \footnote{https://exoplanetarchive.ipac.caltech.edu/}

%NEPTUNE DESERT

During the lifetime of the Kepler and TESS missions, a notable absence of exoplanets was noticed within a defined region of period-radius and period-mass space, the so-called "Neptunian Desert" (also known as the “hot Neptune desert”, “subJovian desert”, and “evaporation desert”), which roughly encompasses Neptunian-sized planets (approximately $2\,\textrm{R}_{\oplus} < R_p < 9\,\textrm{R}_{\oplus}$ and $0.03\,\textrm{M}_{\rm J} < M_p < 0.8\,\textrm{M}_{\rm J}$) with periods up to $\sim5$ days \citep{SzaboKiss2011, Mazeh2016, OwenWu2017, OwenLai2018, Deeg2023}. This should not be due to an observational bias, as planets with short period and intermediate size are easily discovered by transit surveys like Kepler. Several theories have been put forward to explain the existence of the desert and its boundaries in the parameter space \citep[e.g.,][]{OwenLai2018,Vissapragada2022}: the lower boundary could be caused by photoevaporation stripping away their gaseous H/He envelopes, therefore reducing their radii/masses and leaving behind a dense core; while the upper boundary seems to be stable against photoevaporation, and may instead be understood as a “tidal disruption barrier”: in order to form planets below and left of the boundary through high-eccentricity migration, they would have to come so close to their host star and no longer be able to succesfully circularize and stabilize.

Many of these planets are readily observable with missions such as TESS and JWST, and also with ground-based spectroscopic instruments such as HARPS \citep{mayor2003} and ESPRESSO \citep{Pepe2021} due to their close orbits and short periods. The HARPS-NOMADS programme (PI Armstrong, 1108.C-0697) aims to significantly increase the number of planet confirmations in this parameter space with precise masses and radii. With these parameters we can constrain the densities and thus the internal structures of these planets. The few planets found in this desert are likely to have undergone unusual formation and/or evolutionary processes compared to those in more populated parameter spaces so this analysis could lead towards an improved understanding of the formation and evolution of the Neptunian desert.

%HOT SATURNS
Residing in the Neptunian desert with masses between $0.1$ and $0.4$ $M_J$, hot Saturns are a group whose study can help us further understand the divergent formation pathways of small planets and gas giants: on the one hand, they could be the smallest planets formed via runaway gas accretion. They can therefore provide information on the limits of the core accretion mechanism %on the basis that they are intrinsically less common than hot Jupiters and hot Neptunes
\citep{Petigura_2018} . On the other hand, in accordance with the trend that associates the presence of short-period planets and of large planets with higher host star metallicity \citep{Mulders_2016,dong_2018}, \citet{Petigura_2018} found that stars hosting hot sub-Saturns (planets with radii between 4 and 8 $R_\oplus$ and periods between 1 and 10 days) have the highest mean stellar metallicity among all planet hosts. This evidence suggests some kind of mechanism connected to high stellar metallicity that leads to these short-period large planets. Lastly, because their lower surface gravities lead to larger atmospheric scale heights than those of typical hot Jupiters, hot Saturns are some of the best targets for transmission spectroscopy observations \citep{Wakeford_2018}.

%THIS WORK
We present here the detection and characterisation of TOI-2374\,b and TOI-3071\,b, two hot sub-Saturns transiting a K5V and an F8V star, respectively. The observations leading to the detection and confirmation of the two planets include photometry from TESS and ground-based telescopes, spectroscopy from HARPS and PFS and high-spatial resolution imaging from SOAR. The details of these observations are outlined in Section \ref{observations}. In Section \ref{analysis} we present the analysis of these observations, including the determination of stellar parameters and the joint modelling used to constrain the planetary parameters. In Section \ref{resultsdiscussion} we present the results and discuss the nature of the two targets, including the position of the planets in the Neptune desert and in mass-radius parameter space. This section also includes an analysis of the internal structure and evaporation history of the planet. Finally, in Section \ref{conclusion} we report the conclusions of our work.

\section{Observations}\label{observations}

\subsection{Photometry}

\subsubsection{\it TESS photometry}\label{tess}

\par TESS observed TOI-2374, with TESS Input Catalog (TIC) ID 439366538, in sector 1 (UT 2018 July 25–August 22) with a 30 minute cadence and in sector 28 (UT 2020 July 30–August 26) with a 10 minute cadence, both on camera 1 CCD 4. TOI-2374\,b was detected in the Full-frame images (FFIs) by the MIT's Quick-Look Pipeline \citep[QLP,][]{Huang2020} and alerted as a TESS object of interest (TOI) on 28 October 2020\footnote{\url{https://exofop.ipac.caltech.edu/tess/target.php?id=439366538}} \citep{guerrero2021}. The detection gave a period of $4.31362 \pm 0.00001$\,d, an epoch of $2458326.555387$\, days in BJD, and a depth of $9.490 \pm 0.008$\,parts-per-thousand (ppt).

TOI-3071 (TIC ID 452006073) was monitored by TESS in sectors 10 (UT 2019 March 26–April 22) and 37 (UT 2021 April 2–28) with camera 2 CCD 2 with a 30 and 10 minute cadence FFI, respectively. The QLP detected the candidate and it was alerted as a TOI on 4 June 2021\footnote{\url{https://exofop.ipac.caltech.edu/tess/target.php?id=452006073}}. The detection gave a period of $1.2671 \pm 0.0003$\,d, an epoch of $2459331.9673 \pm 0.0017$\, days in BJD, and a depth of $2.580 \pm 0.002$\,ppt. It was subsequently observed in sector 64 (UT 2023 April 6-May 4) on camera 2 CCD 1 as a 2-minute cadence target. The Science Processing Operations Center (SPOC) at NASA Ames Research Center conducted a transit with a noise-compensating matched filter \citep{Jenkins2002,jenkins2010,jenkins_2020} and detected the transit signature of TOI-3071\,b. An initial limb-darkened model fit was performed \citep{Li2019} and a suite of diagnostic tests were conducted help make or break the planetary nature of the signal \citep{Twicken_2018}. The candidate transit signature passed all these diagnostic tests including the difference image centroiding test, which constrained the location of the host star to within 2.1 +/- 2.6 arc sec of the transit source.

For observations of TOI-2374 from TESS sectors 1 and 28 and observations of TOI-3071 from sectors 10 and 37, we downloaded the light curves computed by the QLP and used the Kepler Spline Simple Aperture Photometry (KSPSAP) flux, which is detrended by applying a high-pass filter to remove low-frequency variability from stellar activity or instrumental noise \citep{Huang2020}. We then used the \texttt{lightkurve} python package \citep{lightkurve} to remove all data points flagged as being affected by excess noise (corresponding to stray light from Earth or the Moon in the camera field of view or scattered light). For TOI-3071 observations from sector 64, we downloaded the photometry provided by the SPOC pipeline, and used the Presearch Data Conditioning Simple Aperture Photometry (PDCSAP), from which common trends and artefacts, as well as an estimated contamination from blended sources, have been removed by the SPOC Presearch Data Conditioning (PDC) algorithm \citep{Twicken2010,Smith2012,Stumpe2012,Stumpe2014}. No further detrending of the light curves was deemed necessary as they are relatively flat across the whole time series. The median-normalised light curves for both targets were used in the joint model outlined in section \ref{jointfit}. %and are shown in figures \ref{fig:LC_TOI-2374_TESS} and \ref{fig:LC_TOI-3071}.

\par
\begin{figure}
    \centering
    \includegraphics[width=\columnwidth]{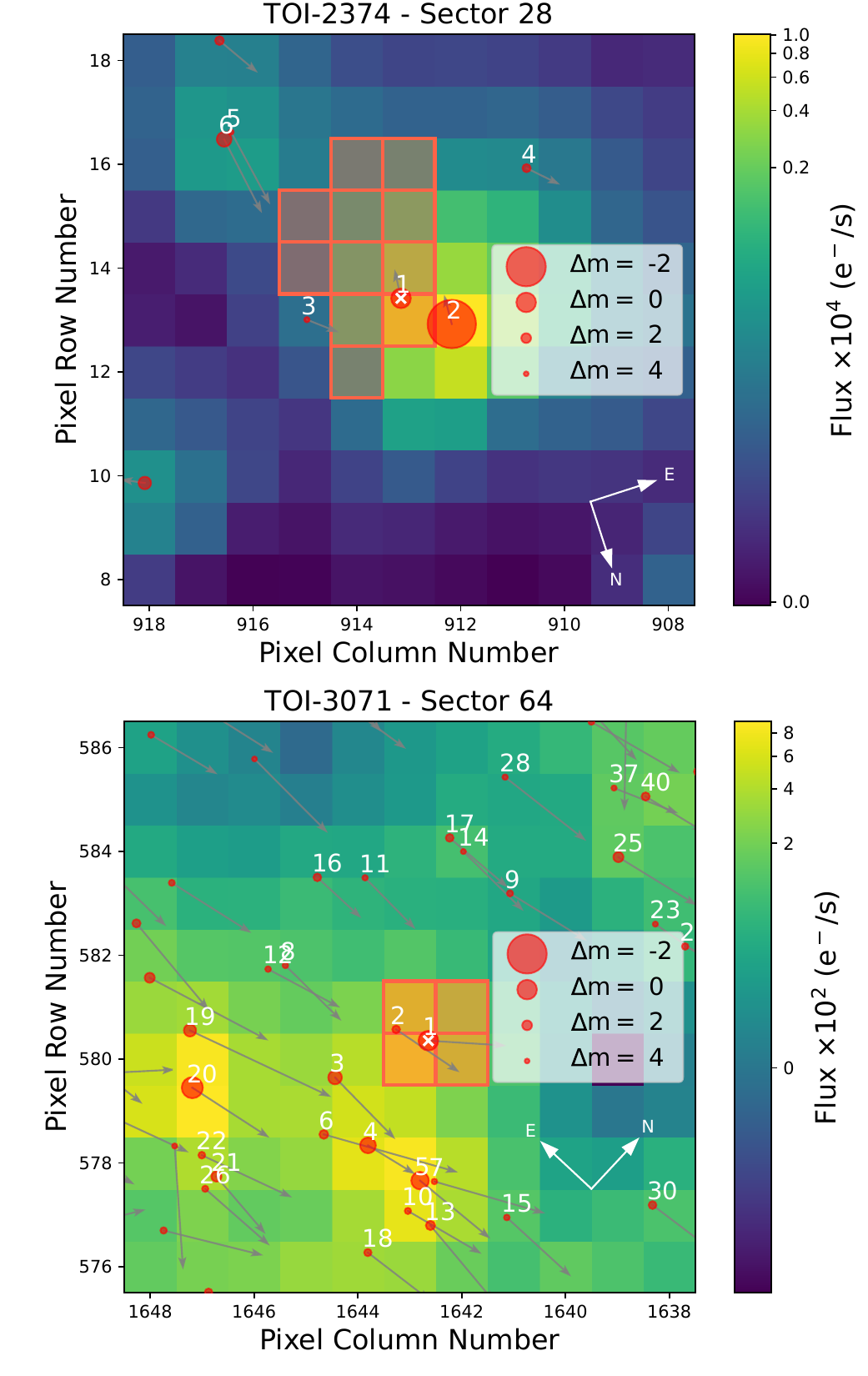}
    \caption{Target Pixel Files (TPF) for TOI-2374 from {\it TESS} S28 (top) and TOI-3071 from S64 (bottom). Targets are marked as a white cross. Other {\it Gaia} DR3 sources within a limit of 4 {\it Gaia} magnitudes difference are marked as red circles, and are numbered in distance order from the targets. The aperture mask is outlined and shaded in red. This figure was created with {\tt tpfplotter} \citep{Aller2020}.}
    \label{fig:tpf}
\end{figure}

  TIC contamination ratios of 6.65 and 0.62 are reported for TOI-2374 and TOI-3071 respectively. These values are computed as the nominal flux from the contaminants divided by the flux from the source. The large values reported for both targets are most likely due to close sources as can be seen in the TESS Target Pixel Files images shown in Figure \ref{fig:tpf} (created with {\tt tpfplotter}\footnote{Publicly available at \url{https://github.com/jlillo/tpfplotter}}, \citealt{Aller2020}). The TPFs show several sources of potential contamination around the target stars (particularly TOI-2374, which has a very close source two magnitudes brighter). Lightcurves from QLP and SPOC data processing pipelines are approximately corrected for blending from nearby stars. The SPOC pipeline computes a crowding metric using a series of assumptions described in Section 2.3.11 of \citet{Stumpe2012}. The QLP uses a different set of assumptions described in step 7 in \citet{Huang2020}. To further measure the extent of this blending and contribute to the correction, follow up ground-based photometry was necessary; the details of these observations are described below. Additionally, the TPF for TOI-2374 shows that both the target and the bright companion have the same proper motion direction so they might be bounded companions.

\subsubsection{Light Curve Follow-up Observations} \label{gb-photometry}

The TESS pixel scale is $\sim 21\arcsec$ pixel$^{-1}$ and photometric apertures typically extend out to roughly 1 arcminute, generally causing multiple stars to blend in the TESS photometric aperture. This is precisely the case for TOI-2374 and TOI-3071 (Fig.\ref{fig:tpf}). To determine the true source of the TESS detection, we acquired ground-based time-series follow-up photometry of the field around both TOI-2374 and TOI-3071 as part of the TESS Follow-up Observing Program \citep[TFOP;][]{collins:2018}\footnote{https://tess.mit.edu/followup}. The follow-up light curves are also used to confirm the transit depth and thus the TESS photometric deblending factor and refine the TESS ephemeris. We used the {\tt TESS Transit Finder}, which is a customized version of the {\tt Tapir} software package \citep{Jensen:2013}, to schedule our transit observations. The observations are summarised below and confirm the transit-like events detected by TESS to be occurring within the small TOI-2374 and TOI-3071 follow-up photometric apertures.

%TOI-2374.01
\paragraph{LCOGT}
We observed two full transit windows of TOI-2374\,b on UTC 2021 May 10 and 2021 May 18 in Sloan $i'$ and Sloan $g'$ bands, respectively, using the Las Cumbres Observatory Global Telescope \citep[LCOGT;][]{Brown:2013} 1.0\,m network nodes at South Africa Astronomical Observatory (SAAO) and Siding Spring Observatory (SSO). The images were calibrated by the standard LCOGT {\tt BANZAI} pipeline \citep{McCully:2018} and differential photometric data were extracted using {\tt AstroImageJ} \citep{Collins:2017}. We used circular photometric apertures with radius $5.1\arcsec$ for the $i'$ band observations and $5.4\arcsec$ for the $g'$ band observations. The target star apertures excluded all of the flux from the nearest known neighbor in the Gaia DR3 catalog (Gaia DR3 6828814283414902784), which is $\sim22\arcsec$ northeast of TOI-2374. Each light curve is included in the global modelling described in section \ref{jointfit}.

\paragraph{Brierfield Private Observatory} \label{sssec:brierfield_obs}
We observed one full transit window of TOI-2374\,b on UTC 2021 May 18 in B band using the Brierfield Observatory, located near Bowral, New S. Wales, Australia. The 0.36\,m telescope is equipped with a $4096\times4096$ Moravian 16803 camera. The image scale after binning 2$\times$2 is $1.47\arcsec$ pixel$^{-1}$, resulting in a $50\arcmin\times50\arcmin$ field of view. The photometric data were extracted using the {\tt AstroImageJ} ({\tt AIJ}) software package \citep{Collins:2017}. Circular photometric apertures with radius $7.4\arcsec$ were used as they excluded all of the flux from the nearest known neighbor.

%TOI-3071.01
\paragraph{PEST}
We observed a egress window of TOI-3071\,b in Rc band on UTC 2021 June 04 using the Perth Exoplanet Survey Telescope (PEST), located near Perth, Australia. The 0.3 m telescope is equipped with a $5544\times3694$ QHY183M camera.  Images are binned $2\times2$ in software giving an image scale of $0.7\arcsec$ pixel$^{-1}$ resulting in a $32\arcmin\times21\arcmin$ field of view. A custom pipeline based on {\tt C-Munipack}\footnote{http://c-munipack.sourceforge.net} was used to calibrate the images and extract the differential photometry. We used circular photometric apertures with radius $6.4\arcsec$. The target star aperture included the flux from the nearest known neighbor in the Gaia DR3 catalog (Gaia DR3 5342880462297608192), which is $\sim3.8\arcsec$ east of TOI-3071. The PEST observation was not included in the modeling due to the lack of ingress coverage and low signal to noise detection.

\subsection{Spectroscopy}

In this section we describe in detail the high-precision RV measurements used in this paper. These observations are presented in figures \ref{fig:rv_2374} and \ref{fig:rv_3071}. In section \ref{sec:RVanalisis} we perform a further exploration of this data.

\subsubsection{HARPS RVs}\label{harps}

We obtained radial velocity (RV) measurements of TOI-2374 and TOI-3071 with the High Accuracy Radial velocity Planet Searcher (HARPS) spectrograph mounted on the ESO 3.6\,m telescope at the La Silla Observatory in Chile \citep{harps}.
Under the HARPS-NOMADS large programme (ID 1108.C-0697, PI: Armstrong), a total of 21 spectra were obtained between 9 April 2022 and 7 July 2022 for TOI-2374 and 14 spectra between 19 March and 2 April 2022 for TOI-3071. The instrument (with resolving power $R = 115\,000$) was used in high-accuracy mode (HAM). The exposure time for the TOI-2374 observations was 1800 s. This resulted in a signal to noise ratio (SNR) of between 22 and 35 per pixel. The exposure time for the TOI-3071 observations was 2400 s, resulting in a SNR of between 10 and 25 per pixel.
The data were reduced using the standard offline HARPS data reduction pipeline, and a K5 template (TOI-2374) or G2 template (TOI-3071) was used to form a weighted cross-correlation function (CCF) to determine the RV values \citep{Baranne1996,Pepe2002}. The line bisector (BIS) and full-width at half-maximum (FWHM) were measured using  methods by \citet{Boisse2011}.

Three bad data points were removed from the TOI-2374 dataset. One of them (BJD = 2459711.8) was reduced incorrectly and the other two (BJD = 2459712.8 and BJD = 2459716.8) contained large errors due to bad weather during the observations. The achieved instrumental precision for the rest of the RV measurements is $\approx 4$~m~s$^{-1}$. The observations are presented in tables \ref{tab:harps_3071} and \ref{tab:harps}.

\subsubsection{PFS RVs}\label{pfs}

We observed TOI-2374 with the Planet Finder Spectrograph (PFS; \citealt{PFS_Crane2006,PFS_Crane2008,PFS_Crane2010}) on the Magellan/Clay telescope at Las Campanas Observatory in Chile. We obtained seven observations of this target between 2021 Sep 14 and 2021 Nov 17 through an iodine cell, with relatively short exposure times of 5 minutes. These observations were made in 3x3 binning mode to reduce readout noise. We also obtained an iodine-free template spectrum at higher signal to noise, which we used to extract relative RVs. The RVs were extracted using the pipeline described in \citet{PFS_Butler1996}, achieving a typical instrumental RV precision of $\approx 3$~m~s$^{-1}$. We also quantified potential variations in the spectral line profiles using bisector inverse slopes (BIS) measurements. These were derived using the procedures described in \citet{Hartman2019}, applied to the iodine-free orders of each spectrum. The BIS measurements are shifted such that the median bisector span for each order is zero. All measurements are presented in Table \ref{tab:pfs}.

\subsubsection{TRES}\label{TRES}

We obtained two reconnaissance spectra on UT 2021 July 21 and UT 2021 August 5 of TOI-2374 using the Tillinghast Reflector Echelle Spectrograph \citep[TRES,][]{gaborthesis} located at the Fred Lawrence Whipple Observatory (FLWO) in Arizona, USA. TRES has a resolving power of $\sim$44,000 and operates in the wavelength range 390-910 nm. The spectra were extracted as described in \citet{buchhave2010}.

\subsection{High resolution imaging}

\begin{figure}
    \centering
    \includegraphics[width=\columnwidth]{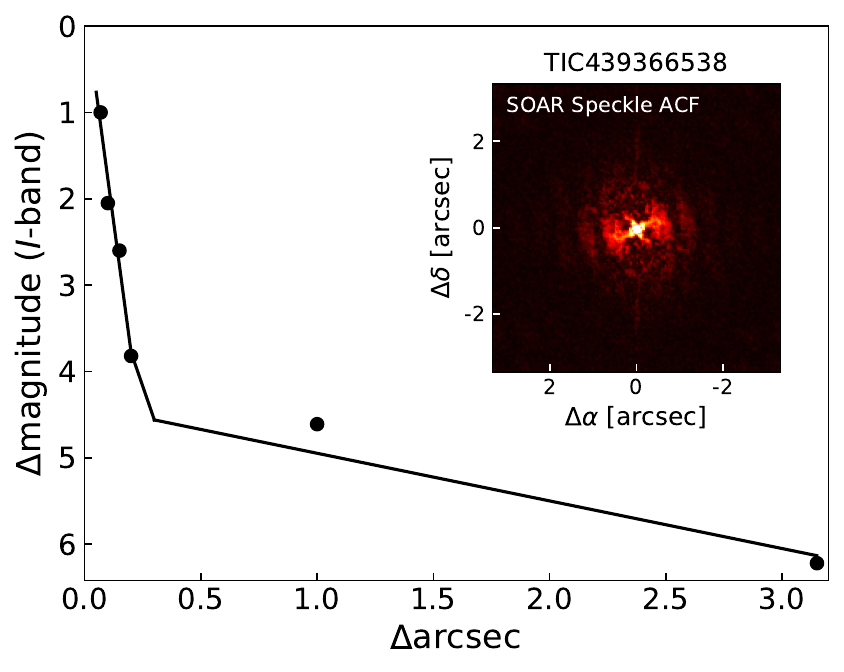}
    \caption{SOAR observation of TOI-2374. Black circles correspond to measured data points and the black lines show the fit in two different separation regimes. The $5\sigma$ detection sensitivity is plotted with the speckle imaging auto-correlation functions inset.}
    \label{fig:soar_2374}
\end{figure}

\begin{figure}
    \centering
    \includegraphics[width=\columnwidth]{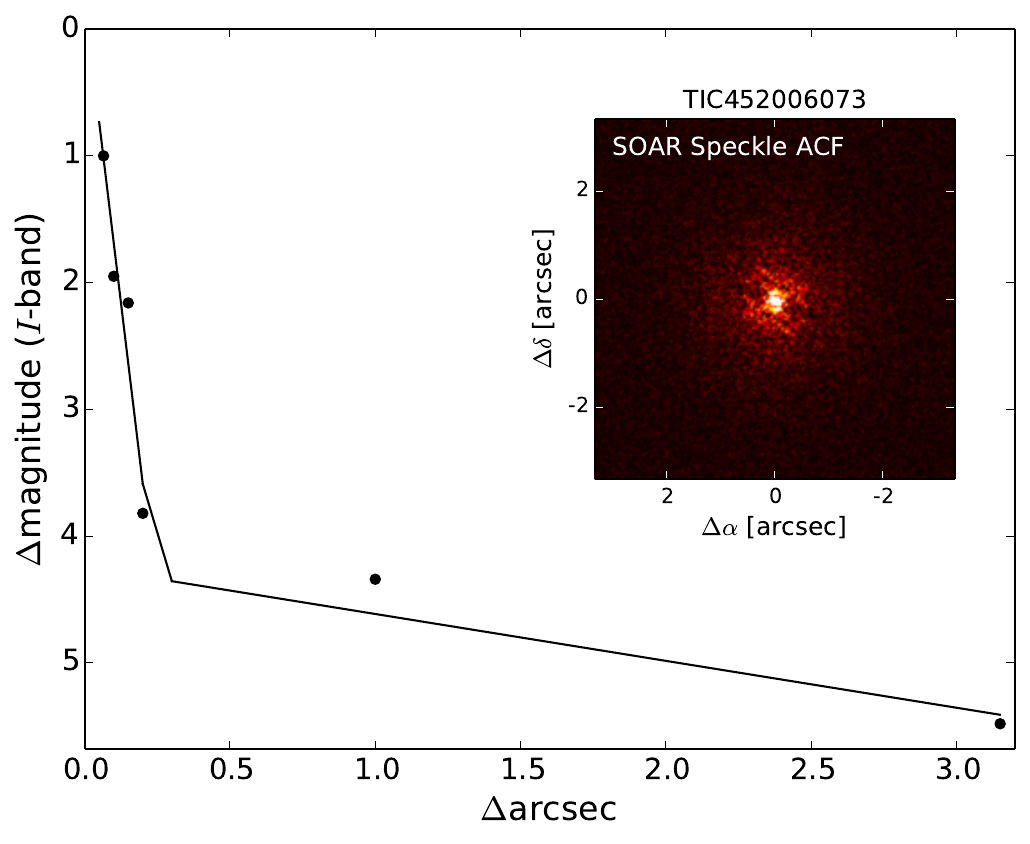}
    \caption{SOAR observation of TOI-3071. Black circles correspond to measured data points and the black lines show the fit in two different separation regimes. The $5\sigma$ detection sensitivity is plotted with the speckle imaging auto-correlation functions inset.}
    \label{fig:soar_3071}
\end{figure}

High-angular resolution imaging is needed to search for nearby sources that can contaminate the TESS photometry, resulting in an underestimated planetary radius, or be the source of astrophysical false positives, such as background eclipsing binaries \citep[][for example]{daemgen2009,LilloBox2012}. We searched for stellar companions to TOI-2374 with speckle imaging on the \mbox{4.1-m} Southern Astrophysical Research (SOAR) telescope \citep{tokovinin} on 3 December 2020 UT, observing in Cousins I-band, a similar visible bandpass as TESS. This observation was sensitive to a 4.7-magnitude fainter star at an angular distance of 1 arcsec from the target. More details of the observations within the SOAR TESS survey are available in \citet{ziegler}. The $5\sigma$ detection sensitivity and speckle auto-correlation functions from the observation are shown in Figure \ref{fig:soar_2374}. No nearby stars were detected within 3\arcsec of TOI-2374 in the SOAR observations.

TOI-3071 was similarly observed by SOAR on 20 March 2022, sensitive to a 5.5-magnitude fainter star at a 1 arcsec separation from the target. The $5\sigma$ detection sensitivity and speckle auto-correlation functions from this observation are shown in Figure \ref{fig:soar_3071}. We detected no nearby stars within 3\arcsec of TOI-3071 in the SOAR observations.

\section{Analysis}
\label{analysis}

\subsection{Stellar parameters}            
\label{stellar-parameters}

Because the final physical parameters of the planets depend directly on the values of the stellar parameters, here we perform several independent methods to measure and derive a range of stellar parameters for TOI-3071 and TOI-2374. 

\subsubsection{Spectroscopic parameters}
\label{stellar-spectroscopic-parameters}

The stellar spectroscopic parameters ($T_{\mathrm{eff}}$, $\log g$, microturbulence, [Fe/H]) were estimated using the ARES+MOOG methodology from the respective combined HARPS spectrum of each star. The methodology is described in detail in \citet[][]{Sousa-21, Sousa-14, Santos-13}. We used the latest version of ARES \footnote{The last version, ARES v2, can be downloaded at https://github.com/sousasag/ARES} \citep{Sousa-07, Sousa-15} to consistently measure the equivalent widths (EW) on the list of iron lines. For TOI-3071 we used the list of lines presented in \citet[][]{Sousa-08}, while for TOI-2374 we used the list of lines presented in \citet[][]{Tsantaki-2013} which is more appropriate for stars cooler than 5200 K. In the analysis we use a minimization process to find the ionization and excitation equilibrium to converge for the best set of spectroscopic parameters. This process makes use of a grid of Kurucz model atmospheres \citep{Kurucz-93} and the radiative transfer code MOOG \citep{Sneden-73}. We also derived a more accurate trigonometric surface gravity using recent GAIA data following the same procedure as described in \citet[][]{Sousa-21}. We estimated rotational velocity $v\sin i$ values by performing spectral synthesis (with the same code and model atmospheres as for stellar parameters) of two iron lines in the 6705\AA\ region. We fix the macroturbulence velocity with the empirical formula by \citet{Doyle2014}, which provides v$_{mac}$=4.6 km~s$^{-1}$ for TOI-3071. Cool stars are outside of such calibration, hence we adopted a value v$_{mac}$=2 km~s$^{-1}$ for TOI-2374. Both stars seem to be slow rotators with $v \sin i$ = 2.0 km~s$^{-1}$ for TOI-3071 and $v \sin i$ $<$ 0.5 km~s$^{-1}$ for TOI-2374. %We used a method similar to the one presented in \citet{santos2002} to derive rotational velocity $v\sin i$ values from the FWHM measurements and the [Fe/H] estimates. For TOI-2374, the value obtained is outside the range for which it was calibrated, and must therefore be treated with caution. In particular, the slow rotational velocity implies that the line is not resolved in HARPS spectra and the measured $v\sin i$ is therefore only a rough lower limit. 
The stellar parameters are presented in tables \ref{stepartable2374} and \ref{stepartable3071}.

\par Independently, we derived the stellar atmospheric parameters for TOI-2374 from the TRES spectra using the Stellar Parameter Classification \citep[SPC,][]{buchhave2012} tool. SPC cross correlates an observed spectrum against a grid of synthetic spectra based on Kurucz atmospheric models \citep{kurucz1992}. With this method we obtained $T_{\mathrm{eff}} = 4937 \pm 50$ K, $\log g = 4.61 \pm 0.10$ and $[\mathrm{Fe/H}] = 0.24 \pm 0.08$, which are in agreement (within errors) with the results derived from the HARPS spectra.

\subsubsection{Abundances}
\label{stellar-abundances}

Stellar abundances of the elements were derived using the same tools and models as for stellar parameter determination under the assumption of local thermodynamic equilibrium. For the derivation of chemical abundances of refractory elements we closely followed the methods described in e.g. \citet{Adibekyan-12, Adibekyan-15, Delgado-14, Delgado-17}. Abundances of the volatile elements, C and O, were derived following the method of \cite{Delgado-21} and \cite{Bertrandelis-15}. Nine out of the 14 individual spectra of TOI-3071 were contaminated around the 6300.3 \AA{} oxygen line and had to be discarded before being coadded, leading to a lower S/N spectrum and thus a higher error in oxygen abundance. These two elements could not be derived for TOI-2374 because their lines are very weak in cool stars spectra. All the [X/H] ratios are obtained by doing a differential analysis with respect to a high S/N solar (Vesta) spectrum from HARPS. The abundances of the elements are shown in table \ref{tab:abundances} in the appendix.

\begin{table}
\centering
\caption{An overview of stellar properties for TOI-2374.}
\label{stepartable2374}
\begin{tabular}{lcl}
\toprule
\textbf{Property [unit]} & \textbf{Value} & \textbf{Source}\\
\midrule
\multicolumn{3}{l}{\textbf{Identifiers}} \\
Name                                  & TOI-2374                      & --    \\
TIC ID                                  & 439366538                       & TICv8.2   \\
Gaia ID                                  & 6828814283414902912             & Gaia DR3 \\
\midrule
\multicolumn{3}{l}{\textbf{Astrometric properties}} \\
R.A (J2000.0)         & 21:17:59.59       & Gaia DR3 \\
Dec (J2000.0)         & -22:02:59.36      & Gaia DR3 \\
Parallax [mas]             & 7.3592 $\pm$ 0.018   & Gaia DR3 \\
Distance [pc]              & 134.655 $\pm$ 0.8695 & TICv8.2 \\
\midrule
\multicolumn{3}{l}{\textbf{Photometric properties}} \\
T mag                  & $11.2057 \pm 0.0061$            &  TICv8.2 \\
V mag                  & $12.095 \pm 0.080$                &  TICv8.2\\
G mag                  & $11.818 \pm 0.003$                &  Gaia DR3\\
J mag                  & $10.387 \pm 0.022$                &  2MASS\\
K mag                  & $9.793 \pm 0.023$                &  2MASS \\

\midrule
\multicolumn{3}{l}{\textbf{Atmosfpheric properties}} \\
$T_\mathrm{eff}$ [K]                  & $4802  \pm 97$                & This work\\
$\log g$ [cgs]                        & $4.52  \pm 0.05$              & Gaia DR3 \\
$\log g$ [cgs]                        & $4.14  \pm 0.24$              & This work \\
$v_\mathrm{mic}$ [km~s$^{-1}$]        & $0.62  \pm 0.18$              & This work\\
$[\mathrm{Fe/H}]$ [dex]               & $0.15  \pm 0.04$              & This work\\
$v \sin i$ [km~s$^{-1}$]                & $< 0.5$              & This work\\
$\log R'_{\rm HK}$                & $-4.98 \pm 0.06$              & This work\\
\midrule
%\multicolumn{3}{l}{\textbf{Abundances}} \\

%$[\mathrm{Mg/H}]$ [dex]               & $ 0.07 \pm 0.09$              & This work\\
%$[\mathrm{Al/H}]$ [dex]               & $ 0.17 \pm 0.07$              & This work\\
%$[\mathrm{Si/H}]$ [dex]               & $ 0.21 \pm 0.09$              & This work\\
%$[\mathrm{Ti/H}]$ [dex]               & $ 0.20 \pm 0.12$              & This work\\
%\midrule
\multicolumn{3}{l}{\textbf{Physical parameters}} \\
$R_\mathrm{s}$ [$R_\odot$]            & $0.69  \pm 0.02$              & This work\\
$M_\mathrm{s}$ [$M_\odot$]            & $0.75  \pm 0.01$              & This work\\  $P_{\rm rot}$ [days]            & $50.2  \pm 2.8$              & This work\\
$\tau$ [Gyr]            & $4.10  \pm 0.63$              & This work\\

\bottomrule
\end{tabular}
\begin{tablenotes}
    %\item $^{*}$ This $v \sin i$ value is outside the range for which it was calibrated and constitutes only a rough lower limit for the rotational velocity.
	\item \textbf{Sources:} TICv8.2 \citep{Stassun2019}, 2MASS \citep{Skrutskie2006}, Gaia Data Release 3 \citep{GaiaDR3}.
\end{tablenotes}
\end{table}

\begin{table}
\centering
\caption{An overview of stellar properties for TOI-3071.}
\label{stepartable3071}
\begin{tabular}{lcl}
\toprule
\textbf{Property [unit]} & \textbf{Value} & \textbf{Source}\\
\midrule
\multicolumn{3}{l}{\textbf{Identifiers}} \\
Name                                  & TOI-3071                      & --    \\
TIC ID                                  & 452006073                      & TIC v8.2   \\
Gaia ID                                  & 5342880462319699200            & Gaia DR3 \\
\midrule
\multicolumn{3}{l}{\textbf{Astrometric properties}} \\
R.A (J2000.0)         & 11:33:06.98       & Gaia DR3 \\
Dec (J2000.0)         & -56:30:12.27     & Gaia DR3 \\
Parallax [mas]             & 2.0101 $\pm$ 0.0097  & Gaia DR3 \\
Distance [pc]              & 485.219 $\pm$ 8.8845 & TICv8.2 \\
\midrule
\multicolumn{3}{l}{\textbf{Photometric properties}} \\
T mag                 & $11.8706 \pm 0.0061$          &  TICv8.2 \\
V mag                  & $12.383 \pm 0.057$                &  TICv8.2 \\
G mag                  & $12.267 \pm 0.002$                &  Gaia DR3 \\
J mag                  & $11.314 \pm 0.024$                &  TICv8.2 \\
K mag                  & $11.005 \pm 0.021$                &  TICv8.2 \\
\midrule
\multicolumn{3}{l}{\textbf{Atmosfpheric properties}} \\
$T_\mathrm{eff}$ [K]                  & $6177  \pm 62$                & This work\\
$\log g$ [cgs]                        & $4.37  \pm 0.10$              & This work \\
$\log g$ [cgs]                        & $4.32  \pm 0.02$              & Gaia DR3 \\
$v_\mathrm{mic}$ [km~s$^{-1}$]        & $1.29  \pm 0.02$              & This work\\
$[\mathrm{Fe/H}]$ [dex]               & $0.35  \pm 0.04$              & This work\\
$v \sin i$ [km~s$^{-1}$]                & $2.0 \pm 0.5$              & This work\\
$\log R'_{\rm HK}$                & $-5.15 \pm 0.26$              & This work\\
\midrule
%\multicolumn{3}{l}{\textbf{Abundances}} \\
%$[\mathrm{Mg/H}]$ [dex]               & $ 0.30 \pm 0.06$              & This work\\
%$[\mathrm{Al/H}]$ [dex]               & $ 0.30 \pm 0.08$             & This work\\
%$[\mathrm{Si/H}]$ [dex]               & $ 0.33 \pm 0.04$              & This work\\
%$[\mathrm{Ti/H}]$ [dex]               & $ 0.33 \pm 0.04$             & This work\\
%$[\mathrm{N/H}]$ [dex]               & $ 0.36 \pm 0.02$             & This work\\

%\midrule
\multicolumn{3}{l}{\textbf{Phyisical parameters}} \\
$R_\mathrm{s}$ [$R_\odot$]            & $1.31  \pm 0.04$              & This work\\
$M_\mathrm{s}$ [$M_\odot$]            & $1.29  \pm 0.02$              & This work\\
$P_{\rm rot}$ [days]            & $26.6  \pm 6.4$              & This work\\
$\tau$ [Gyr]            & $4.91  \pm 0.83$              & This work (gyro) \\
$\tau$ [Gyr]            & $0.95  \pm 0.14$              & This work (clocks $\&$ PARAM)\\
\bottomrule
\end{tabular}
\begin{tablenotes}
	\item \textbf{Sources:} TICv8.2 \citep{Stassun2019}, Gaia Data Release 3 \citep{GaiaDR3}.
\end{tablenotes}
\end{table}

\subsubsection{Physical parameters}
\label{stellar-physical-parameters}

%Stellar radius Stellar mass

\begin{figure}
    \centering
    \includegraphics[width=0.9\columnwidth]{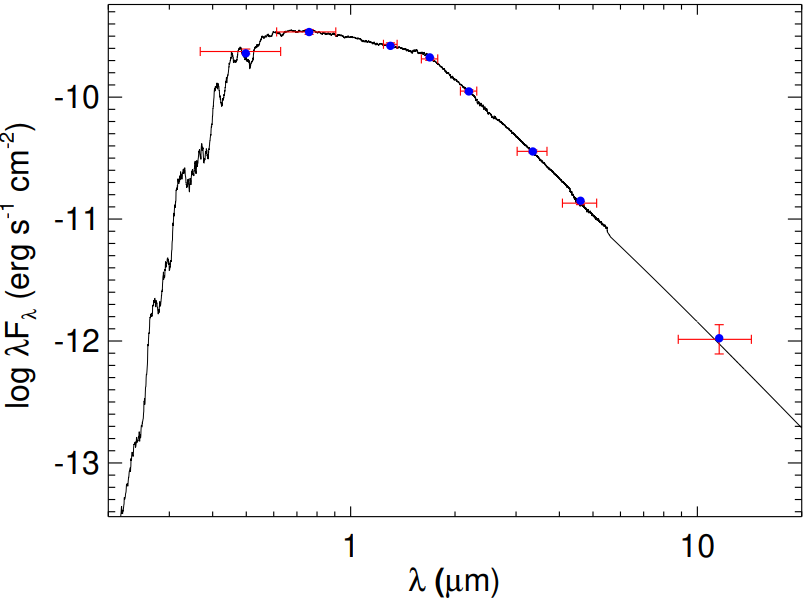}
    \caption{Spectral energy distribution of TOI-2374. Red symbols represent the observed photometric measurements, where the horizontal bars represent the effective width of the passband. Blue symbols are the model fluxes from the best-fit PHOENIX atmosphere model (black).}
    \label{fig:sed_2374}
\end{figure}

\begin{figure}
    \centering
    \includegraphics[width=0.9\columnwidth]{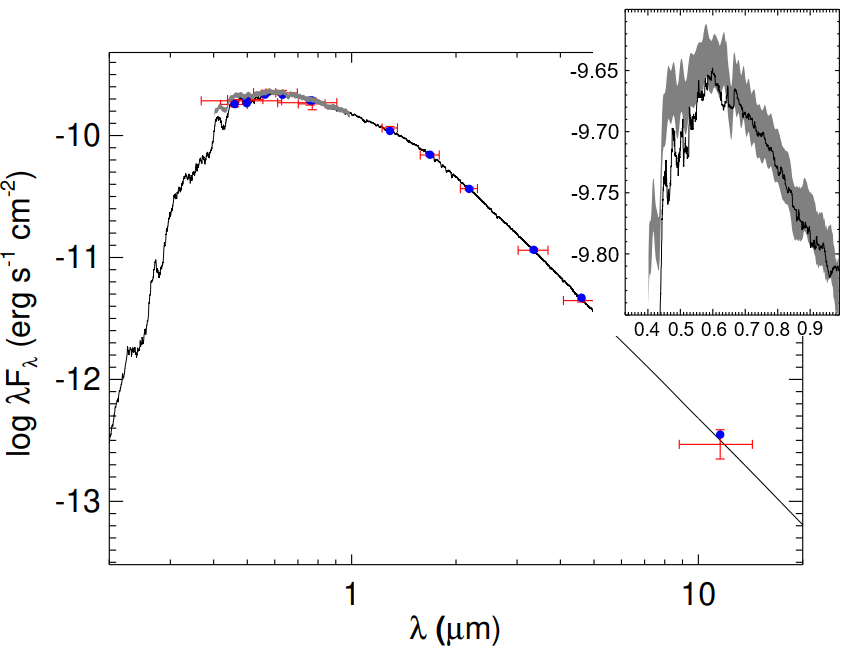}
    \caption{Spectral energy distribution of TOI-3071. Red symbols represent the observed photometric measurements, where the horizontal bars represent the effective width of the passband. Blue symbols are the model fluxes from the best-fit PHOENIX atmosphere model (black). The absolute flux-calibrated {\it Gaia\/} spectrum is shown as a grey swathe in the inset figure.}
    \label{fig:sed_3071}
\end{figure}

From the parameters obtained in the spectroscopic analyisis we estimated the stellar radius and mass using the calibrations presented in \citet[][]{Torres2010} (see tables \ref{stepartable2374} and \ref{stepartable3071}).
In addition, we performed an analysis of the broadband spectral energy distribution (SED) of the star together with the {\it Gaia\/} DR3 parallax, in order to determine an empirical measurement of the stellar radius \citep{StassunTorres2016,Stassun2017,Stassun2018}. We pulled the $JHK_S$ magnitudes from {\it 2MASS}, the W1--W3 magnitudes from {\it WISE}, and the $G_{\rm BP} G_{\rm RP}$ magnitudes from {\it Gaia}. We also utilized the absolute flux-calibrated {\it Gaia\/} spectrum where available. Together, the available photometry spans the full stellar SED over the wavelength range 0.4--10~$\mu$m (see figures~\ref{fig:sed_2374} and \ref{fig:sed_3071}).  We performed a fit using PHOENIX stellar atmosphere models \citep{husser2013}, adopting from the spectroscopic analysis the effective temperature ($T_{\rm eff}$), metallicity ([Fe/H]), and surface gravity ($\log g$). We fitted for the extinction $A_V$, limited to the maximum line-of-sight value from the Galactic dust maps of \citet{Schlegel1998}. The resulting fits (figures~\ref{fig:sed_2374} and \ref{fig:sed_3071}) have $A_V = 0.02 \pm 0.02$ and $0.20 \pm 0.03$ for TOI-2374 and TOI-3071, respectively, with a reduced $\chi^2$ of 1.3 and 1.4, respectively. Integrating the (unreddened) model SED gives the bolometric flux at Earth, $F_{\rm bol} = 4.981 \pm 0.058 \times 10^{-10}$ erg~s$^{-1}$~cm$^{-2}$ and $3.291 \pm 0.038 \times 10^{-10}$ erg~s$^{-1}$~cm$^{-2}$, respectively. Taking the $F_{\rm bol}$ together with the {\it Gaia\/} parallax directly gives the bolometric luminosity, $L_{\rm bol} = 0.2867 \pm 0.0034$~L$_\odot$ and $2.539 \pm 0.032$~L$_\odot$, respectively. The Stefan-Boltzmann relation then gives the stellar radius, $R_\star = 0.774 \pm 0.025$~R$_\odot$ and $1.393 \pm 0.030$~R$_\odot$, respectively. These estimates differ by $2.6 \sigma$ and $1.7 \sigma$ with the ones obtained from the spectroscopic analysis, respectively. Given that the two sets disagree by less than $3 \sigma$ and that the former analysis works with the full spectrum, we decided to use the ones estimated from the calibrations in \citet[][]{Torres2010} as input for the joint model described in section \ref{jointfit}.

%logR'HK, Prot, ages.

We used the relations stated in \citet{Noyes1984} to derive chromospheric activity index $\log R'_{\rm HK}$ values from the HARPS measurements of the Mount-Wilson S-index. With these values we obtained estimates of the rotational period of the stars $P_{\rm rot}$. For both stars, $P_{\rm rot}$ is consistent with the maximum rotation period calculated with the $v \sin i$ estimates, $P_{\rm rot}/\sin i > 70$ d for TOI-2374 and $33 \pm 8$ d for TOI-3071. We used Eq. 3 and Eq. 16 in \citet{Barnes2007} to calculate the gyrochronological stellar ages from the rotation periods, giving 4.34\,$\pm$\,0.67 Gyr and 4.9\,$\pm$\,0.8 Gyr for TOI-2374 and TOI-3071, respectively.
\par Moreover, we obtained two other independent estimates of the stellar ages $\tau$: first, we used the web interface PARAM 1.3 \footnote{\url{http://stev.oapd.inaf.it/cgi-bin/param_1.3}}, which estimates the basic intrinsic parameters of stars given their photometric and spectroscopic data following the method described in \citet{daSilva-2006}. This resulted in stellar age estimates of 4.8\,$\pm$\,4.2 Gyr and 0.8\,$\pm$\,0.6 Gyr for TOI-2374 and TOI-3071, respectively. As expected, the age determination based on evolutionary models is extremely uncertain for main-sequence stars. The age of TOI-2374 formally agrees with the one from gyrochronology. On the other hand, PARAM provides a much younger age for TOI-3071. We then used the chemical abundances of some elements to derive ages through the so-called chemical clocks (i.e. certain chemical abundance ratios which have a strong correlation for age). We applied the 3D formulas described in Table 10 of \citet{Delgado-19}, which also consider the variation in age produced by the effective temperature and iron abundance. The chemical clocks [Y/Mg], [Y/Zn], [Y/Ti], [Y/Si], [Y/Al], [Sr/Ti], [Sr/Mg] and [Sr/Si] were used from which we obtain a weighted average age of 1.1\,$\pm$\,0.2 Gyr for TOI-3071. This age is again much younger than the age from gyrochornology. The ages from chemical clocks for cool stars such as TOI-2374 must be taken with caution since the calibrations are obtained using hotter stars. Therefore, we choose to apply the 2D formulas considering the abundance ratios and the metallicity \citep[see Table 8 in][]{Delgado-19} which provide a weighted average age estimate of 3.3\,$\pm$\,0.7 Gyr. Ages from each individual clock are shown in Table \ref{tab:clocks}. %\textcolor{red}{NOTE: using the 1D formulas, just considering the abundance ratios, we get an age of 5.0$\pm$1.2 Gyr which might be more in agreement with other methods you have used.}
We include in Table \ref{stepartable2374} the geometric average of the three stellar age estimates for TOI-2374. The error is given by the deviation of the three values. In Table \ref{stepartable3071} we report two values for the age of TOI-3071, corresponding to the two disagreen results: the gyrochronological age and the weighted average of the ages derived with PARAM and with the chemical clocks.

%\par \textcolor{red}{Concerning the rotational period of the stars, we performed a periodogram analysis on the detrended light curves of both targets and found no significant periodic signals.}

\begin{figure}
    \centering
    \includegraphics[width=0.9\columnwidth]{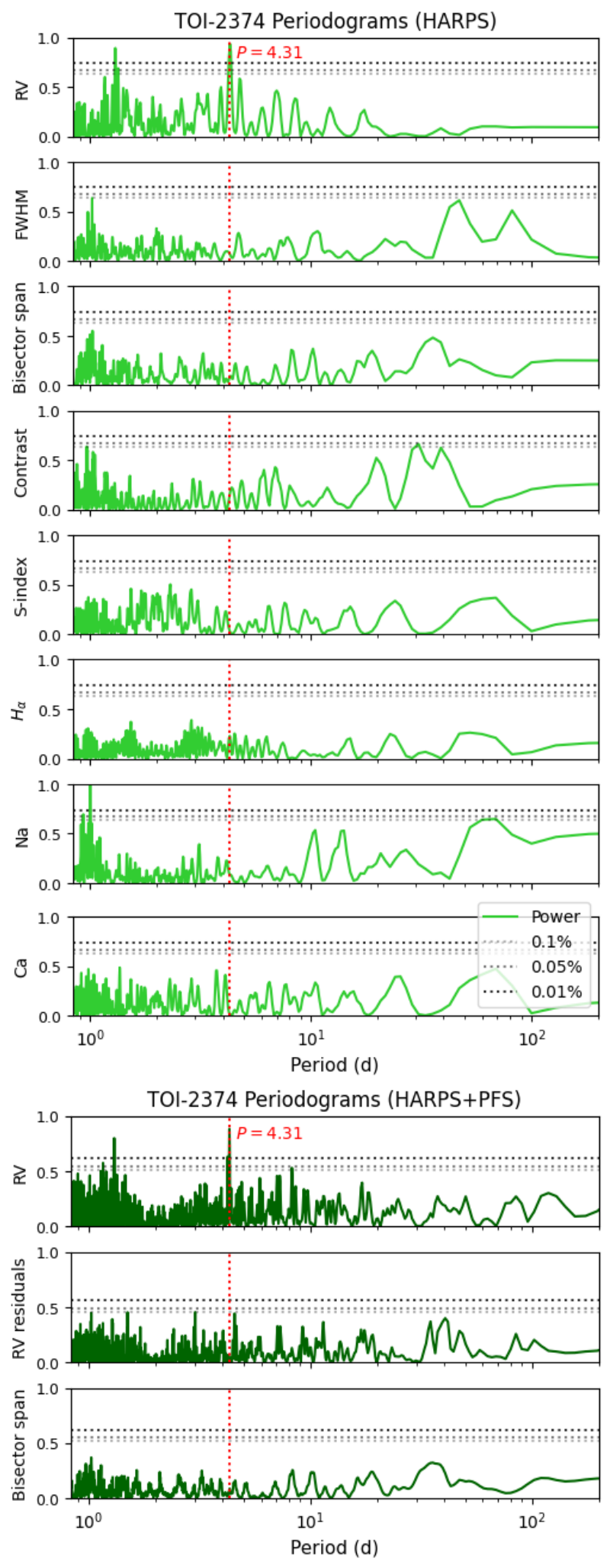}
    \caption{GLS Periodograms \citep{zechmeister_kurster}  for the TOI-2374 HARPS data (top) and HARPS + PFS data (bottom). The highest peak in the RV periodogram, corresponding to the orbital period of TOI-2374\,b is denoted by a dashed vertical red line. The 0.1, 0.05, and 0.01 per cent False Alarm Probabilities (FAP) are calculated using the approach of \citet{baluev} and are shown as dashed horizontal lines. In the HARPS + PFS figures, the second periodogram corresponds to the the radial velocity residuals after the best fit model for a Keplerian orbit has been removed.}
    \label{fig:periodogram_2374}
\end{figure}

\begin{figure}
    \centering
    \includegraphics[width=0.9\columnwidth]{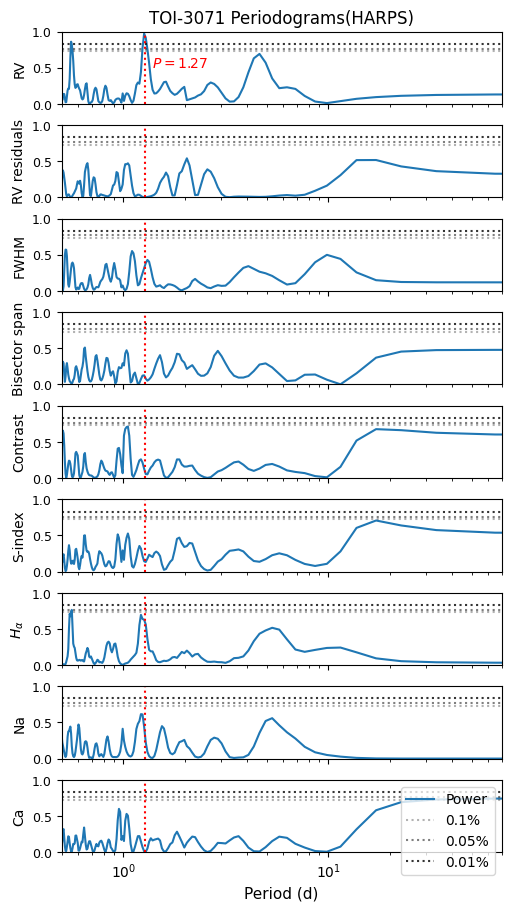}
    \caption{GLS Periodograms \citep{zechmeister_kurster} for the TOI-3071 HARPS data. The highest peak in the RV periodogram, corresponding to the orbital period of TOI-3071\,b is denoted by a dashed vertical red line. The 0.1, 0.05, and 0.01 per cent False Alarm Probabilities (FAP) are calculated using the approach of \citet{baluev} and are shown as dashed horizontal lines. The second periodogram corresponds to the the radial velocity residuals after the best fit model for a Keplerian orbit has been removed.}
    \label{fig:periodogram_3071}
\end{figure}

\begin{figure*}
    \centering
    \includegraphics[width=0.9\textwidth]{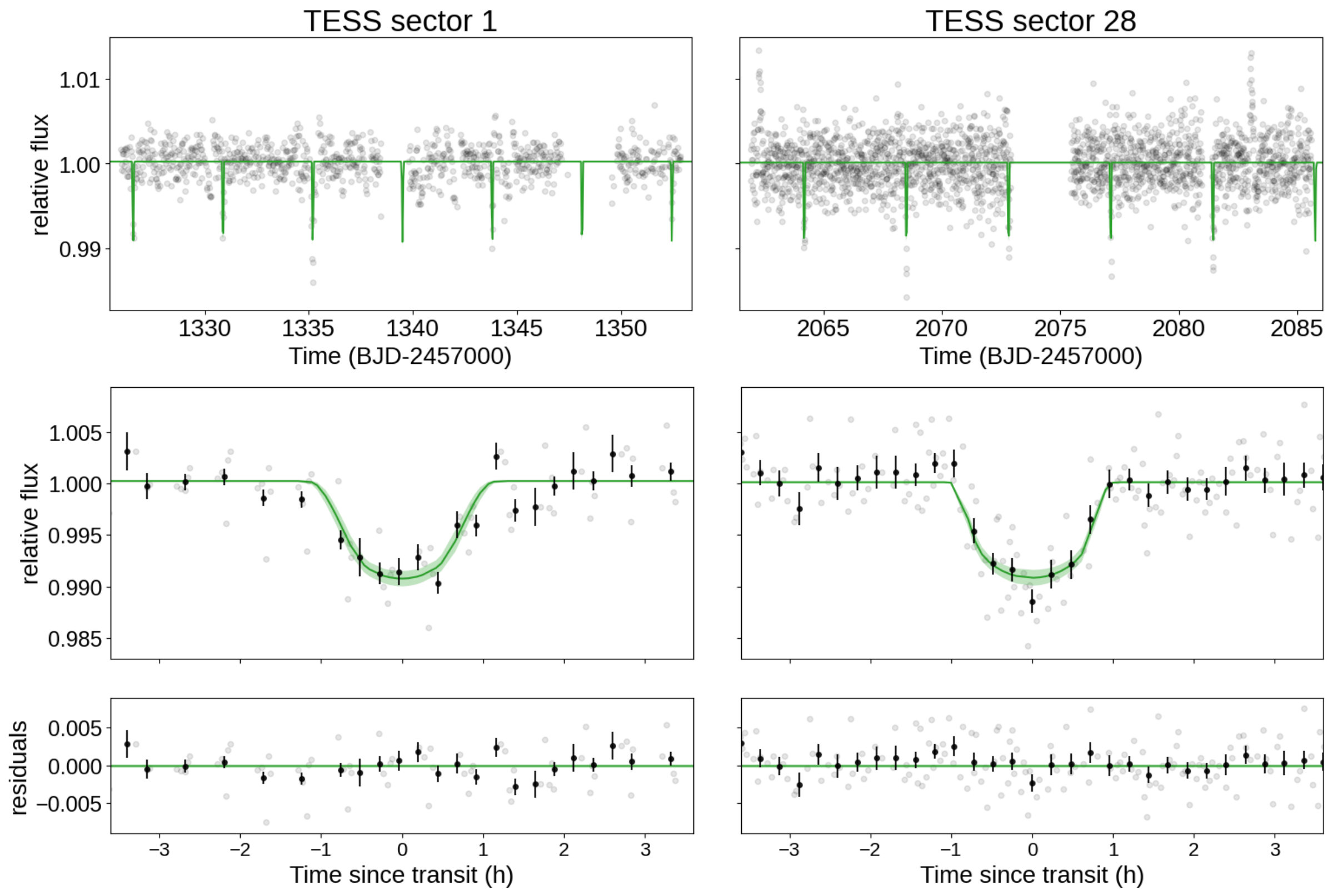}
    \caption{Light curves of TOI-2374.  
    {\it TESS} light curve for Sectors 1 and 28 (grey dots), with time given as Barycentric Julian Date (BJD). The top plots show the TESS QLP KSPSAP light curves from sector 1 (30 minute cadence) and sector 28 (10 min cadence). Overplotted in green is the best-fit solution to the global model resulting from the analysis described in section \ref{jointfit}. The middle plots show this same data phase folded. The binned points indicate the mean of each bin, with error bars representing the standard error of the mean with the binned data shown in black. Residuals of this fit can be found in the bottom plots.}
    \label{fig:LC_TOI-2374_TESS}
\end{figure*}

\begin{figure*}
    \centering
    \includegraphics[width=\textwidth]{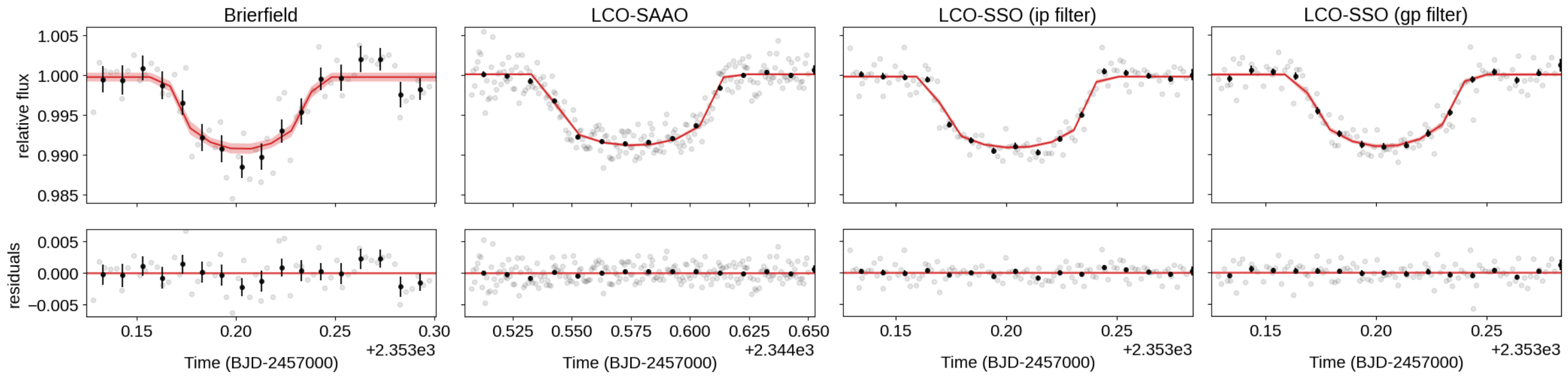}
    \caption{Light curves of TOI-2374 from ground-based telescopes. From left to right, the light curves correspond to the Brierfield Observatory, the LCO-SAAO, the LCO-SSO with ip filter, and the LCO-SSO with gp filter. The data are shown as grey dots, with binned values in black. Overplotted in red is the best fit model from the joint fit described in section \ref{jointfit}. The bottom plots show the residuals of this fit for each lightcurve.}
    \label{fig:LC_TOI-2374_ground}
\end{figure*}

\begin{figure*}
    \centering
    \includegraphics[width=0.9\textwidth]{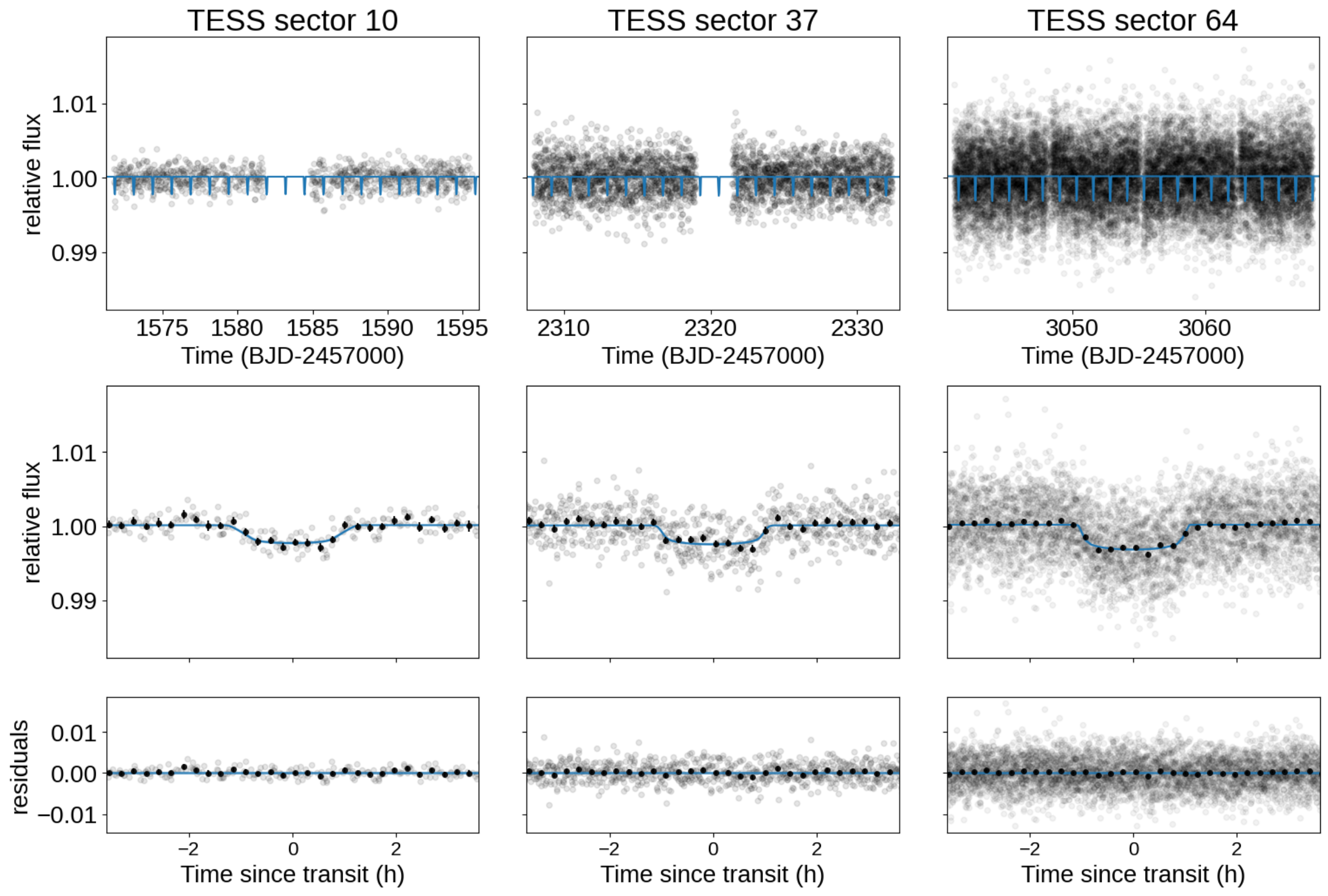}
    \caption{Light curves of TOI-3071.  
    {\it TESS} light curve for Sectors 10, 37 and 64 (grey dots), with time given as Barycentric Julian Date (BJD). The top plots show the TESS QLP KSPSAP light curves from sector 10 (30 minute cadence) and sector 37 (10 minute cadence) and SPOC PDCSAP light curves from sector 64 (2 minute cadence). Overplotted in blue is the best-fit solution to the global model resulting from the analysis described in section \ref{jointfit}. The middle plots show this same data phase folded. The binned points indicate the mean of each bin, with error bars representing the standard error of the mean with the binned data shown in black. Residuals of this fit can be found in the bottom plots.}
    \label{fig:LC_TOI-3071}
\end{figure*}

\begin{figure}
    \centering
    \includegraphics[width=\columnwidth]{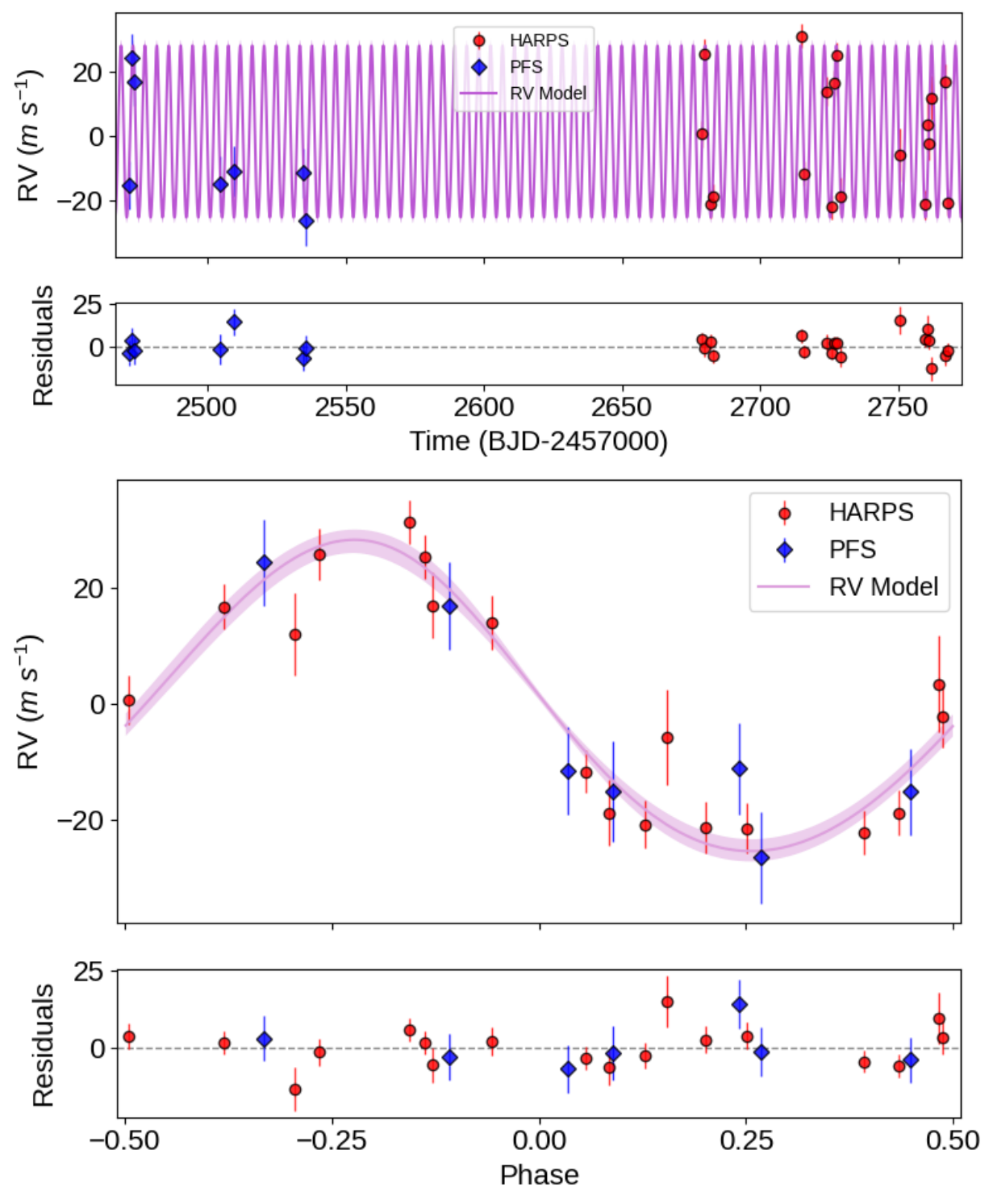}
    \caption{Precise RV measurements of TOI-2374. HARPS measurements are shown as red circles and PFS measurements as blue diamonds. The model plotted is the MCMC median of the global model, corresponding to the orbital period of TOI-2374b. Empirically derived linear and quadratic trends and per-instrument offset $\gamma$  have been subtracted from the raw RV measurements. The error bars represent the reported uncertainty and the empirically derived per-instrument jitter, added in quadrature. These RV measurements are also listed in tables \ref{tab:harps} (HARPS) and \ref{tab:pfs} (PFS). Top: RV measurements plotted against time. Bottom: phase-folded RV measurements.}
    \label{fig:rv_2374}
\end{figure}

\begin{figure}
    \centering
    \includegraphics[width=\columnwidth]{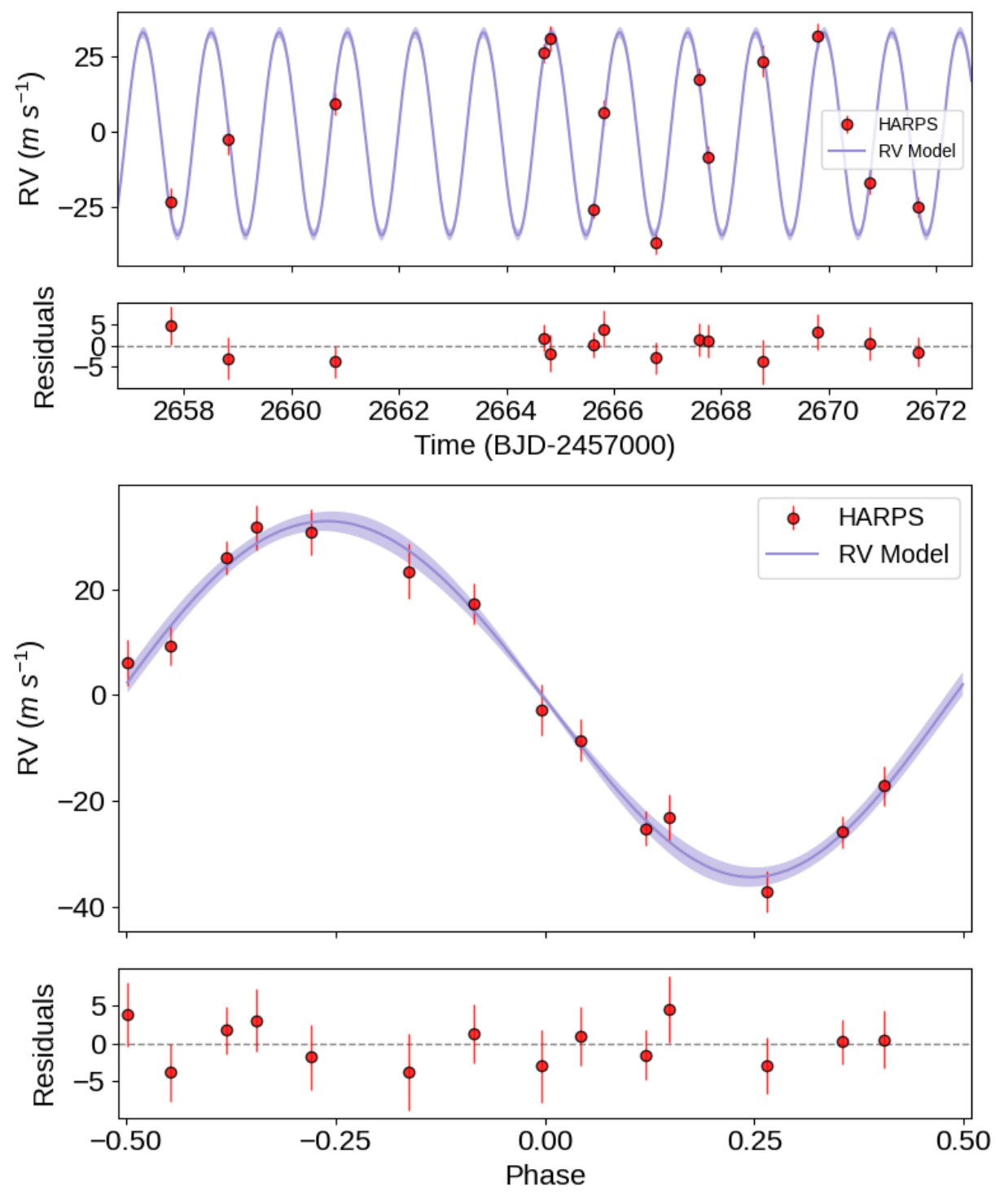}
    \caption{Precise RV measurements of TOI-3071. The model plotted is the MCMC median of the global model, corresponding to the orbital period of TOI-3071b. Empirically derived linear and quadratic trends and per-instrument offset $\gamma$  have been subtracted from the raw RV measurements. The error bars represent the reported uncertainty and the empirically derived per-instrument jitter, added in quadrature. These RV measurements are also listed in Table \ref{tab:harps_3071}. Top: RV measurements plotted against time. Bottom: phase-folded RV measurements.}
    \label{fig:rv_3071}
\end{figure}

\subsection{RV analysis}\label{sec:RVanalisis}

We carried out a preliminary stage of RV data exploration in two parts: periodogram analysis through the DACE platform\footnote{https://dace.unige.ch} and testing of linear RV models with different stellar activity indicators as model covariates. These examinations showed no indication of stellar activity in either of the systems. However, the time span is short for TOI-3071.

Despite having a relatively small number of points, we were able to find the planets of both systems independently of photometry. Generalised Lomb-Scargle Periodograms \citep[GLS,][]{zechmeister_kurster} of the RV data and stellar activity indicators are shown in Figures \ref{fig:periodogram_2374} (TOI-2374) and \ref{fig:periodogram_3071} (TOI-3071). The RV periodograms present peaks above the 0.01 per cent False Alarm Probability at the expected planetary periods. After removing the best fit model for a Keplerian orbit, the periodograms of the RV residuals show no further significant peaks, therefore ruling out the detection of a second planet in each system from these time series.

When looking at periodograms of TOI-3071 stellar activity indicators, we found that the bisector span, contrast and Ca-index show power at low frequency. However, as the time span is relatively short (13.9 days), we cannot constrain the period. We also found one peak in the $H_\alpha$ periodogram (and, to a lesser extent, in the Na periodogram) that is similar to the planetary period but is not large enough to be deemed significant. No other stellar activity indicator shows signs of significant frequencies in this system. As to TOI-2374, there is a significant peak around 1 day in the Na periodogram, which corresponds to a 1 d alias of a low frequency signal, which disappears when we substract a long-term drift from the timeseries. Other than that, this star does not show signs of any other activity.

We computed the Pearson’s R statistic for the RV data and the stellar activity indicators and found no significant correlation between those variables for any system. As for the correlations between the stellar activity indicators and the RV residuals after removing the best fit for a Keplerian orbit, the highest value obtained was R$\sim 0.61$ between TOI-2374 RV residuals and its FWHM, but it strongly depends on a single point (BJD = 2459750.72), without which the correlation disappeared and the Keplerian model stays essentially unchanged. The same situation is seen between the TOI-3071 RV residuals and its contrast, which shared a R of $\sim 0.61$, but strongly depends on four points. No other indicator shows significant correlation with the RV residuals for none of the systems.

To further examine the effects of stellar activity on the RV data, we tested if the RV curves of our joint model described in section \ref{jointfit} should include linear terms corresponding to any of the stellar activity indicators. For this we built simple linear models where the dependent variable was the radial velocity and the covariates were each of the available stellar activity indicators, placed individually in each linear model. We performed an F-test for each model compared to a constant RV model and none of them obtained a p-value smaller than 0.01. We then incorporated into each model a circular orbit with a period fixed to the planetary period found and we repeated the F-test comparing them to the model that only included the circular orbit. Again, none of the stellar activity indicators obtained a p-value smaller than 0.01. This shows that it is not necessary to incorporate linear terms with these covariates to explain the RV data. 

\subsection{Joint modelling}\label{jointfit}

For each of the TOI-2374 and the TOI-3071 planetary systems, we used the {\tt exoplanet} package \citep{exoplanet:exoplanet} to model the RV data and light curves simultaneously. {\tt exoplanet} utilises the probabilistic programming package {\tt PyMC3} \citep{exoplanet:pymc3} and the light curve modelling package {\tt Starry} \citep{exoplanet:luger18}. The radial velocity mean for each spectroscopic instrument was subtracted from the data. QLP and ground-based light curves were already normalized from the pipelines to have an out-of-transit flux of one. The TESS-SPOC light curves were normalized by dividing them by the median of the flux. All timestamps were converted to the time system used by {\it TESS}, i.e. BJD\,-\,2457000 (BJD-TDB). 
\par To model the planetary transits, we used a limb-darkened transit model utilising the quadratic limb-darkening parameterisation in \citet{exoplanet:kipping13} and a Keplerian orbit model, which was parameterised for the planet in terms of the orbital period $P$, the epoch $t_0$, the eccentricity $e$, the argument of periastron $\omega$, the impact parameter $b$, the stellar radius $R_\star$ and the stellar mass $M_\star$. These parameters were then input into light curve models created with {\tt Starry}, alongside parameter $R_p$ (planetary radius), hyperparameter $t_{\mathrm{exp}}$ (the exposure time of the instrument) and the time series of the data $t$. The RV model was computed from the Keplerian orbit model by adding the semi-amplitude $K$ of the RV signal as a parameter. Based on the analysis described in the previous section, we modeled only one planet per system and found no need to include a Gaussian process as a model for the noise given that periodogram analysis of RV and LC data showed no indication of stellar activity or signals from stellar rotation at sufficiently significant frequencies. Instead, we assumed the errors are normal and included an additional white noise term for every instrument, which was added as a free nonnegative jitter term in quadrature with the reported errors.
%This term encapsulates any uncharacterised signal or noise that is perceived as white noise in the RV data, for example instrumental effects and short-scale stellar activity. 

%PRIORS 
All prior distributions set on the parameters fit in this model are given in tables \ref{tab:jointfit2374} and \ref{tab:jointfit3071}. We put Gaussian priors on the stellar radius and mass informed by the values reported in section \ref{stellar-physical-parameters}. We used values given by the QLP detection for the epoch and period to inform our Gaussian priors for log-$P$ and $t_0$ in both planets, and depth values from the same pipeline to inform the Gaussian prior on log-$k$, where $k$ is the planet-to-star radius ratio $k = \frac{R_p}{R_\star}$, with a mean of $0.5 \log$-depth. We inflated the widths of these priors to $\sigma=1$. Because it is well known that sampling directly the eccentricity and the argument of periastron can be problematic for most MCMC samplers \citep{Parmentier2018}, we sampled for $\sqrt{e}\sin\omega$ and $\sqrt{e}\cos\omega$ with a uniform prior within a unit disk. This leads to a uniform prior on $e$ as noted by \citet{Anderson2011}. For the impact parameter we chose a uniform prior between 0 and $1+k$. For each lightcurve, we put a Gaussian prior on the transit normalisation factor $f_0$ (the light curve flux level out of transit) with a mean of 1 and standard deviation of 0.1. For the jitter parameter of each instrument we used a wide Gaussian prior, the mean of which was the log of the error
median on each light curve. We used independent limb-darkening parameters for each filter, parameterised following \citet{exoplanet:kipping13}. For each spectroscopic instrument, we introduced a constant offset term to the reported RVs. We also sampled for a linear and quadratic trend in the RV data. We used uniform priors for log-$K$ and for the log-jitter of each RV instrument. 

%DILUTION
We added one dilution factor for each TESS sector as in \citet{dilution}.\footnote{For the light curve model, we used $M = f_0(f_t T + f_c)/(f_t + f_c)$, with $M$ being the model to be compared with the observations, $f_0$ being the transit normalisation factor, $f_t$ being the flux of the target star in the aperture, $T$ being the transit model, and $f_c$ being the flux of the contaminant star in the aperture. The dilution factor is defined as $d = f_c/(f_t + f_c)$.} Here, we use this parameter to complement the contamination correction carried out by the pipelines, so we refer to it dilution correction factor $d$. We used a Gaussian prior for $d$ with a mean of 0 and standard deviation of 0.2. In this way, positive dilution correction values would correspond to a pipeline undercorrection, while negative values indicate an overcorrection. For the ground-based light curves we fixed the dilution correction factor to zero, thus assuming that these light curves are not contaminated at all. 
% we expect the data to be able to determine their dilution factor due to the availability of ground-based light curves. 

%SAMPLE

We used {\tt exoplanet} to maximise the log probability of the model. Two data points from TESS sector 28 were sigma-clipped during the optimization. The fit values obtained were then used as the starting point of the {\tt PyMC3} sampler, which draws samples from the posterior using a variant of Hamiltonian Monte Carlo, the No-U-Turn Sampler (NUTS). We allowed for 10000 burn-in samples which were discarded, and then 30000 steps with 10 chains. The MCMC chains for all model parameteres have Gelman-Rubin statistics (rank normalized split-$\hat{R}$) \citep{rhat} very close to unity ($<1.01$), effective sample sizes (bulk-ESS) larger than 9100 and tail-ESS larger than 4600, so we have a good level of confidence that the chains mixed well and we have a stable estimate of uncertainty. The median model and the interval between the 16th and 84th percentiles are shown in figures \ref{fig:LC_TOI-2374_TESS}–\ref{fig:rv_3071}. We present the parameters' marginal posterior sample median for the TOI-2374 and TOI-3071 systems from the joint fits in table \ref{tab:params}.

\section{Results \& discussion}\label{resultsdiscussion}

\begin{table*}
    \small
    \renewcommand{\arraystretch}{1.4}
    \caption{Planetary and system parameters of TOI-2374\,b and TOI-3071\,b. The values given are the medians, 16th and 84th percentiles of the MCMC-derived marginalized posterior distribution. Further parameters from the joint fit model can be found in tables \ref{tab:jointfit2374} and \ref{tab:jointfit3071}.}
	\label{tab:params}
	\begin{threeparttable}
	\begin{tabular}{lccl}
	\toprule
	\textbf{Parameter}                      & \textbf{(unit)}           & \textbf{Value}        & \textbf{Source} \\
	\midrule
	\multicolumn{4}{l}{\textbf{TOI-2374\,b}} \\
    Period $P$                              & (days)                    & $4.31361\pm0.00001$ & Joint fit\\
	Full transit duration $T_{dur}$         & (hours)                   & $1.38\pm0.07$           & Joint fit (derived) \\
    Epoch $t_0$     & (BJD-2457000)             & $1326.564 \pm 0.002$    & Joint fit \\
	Radius $R_p$                            & (R$_{\oplus}$)            & $6.81\pm0.30$    & Joint fit \\
    Planet-to-star radius ratio $R_p/R_\star$ & -                         & $0.091\pm0.002$       & Joint fit \\
	Impact parameter $b$                    & -                         & $0.651^{+0.047}_{-0.063}$    & Joint fit\\  
	Inclination $i$                         & ($^{\circ}$)              & $87.5^{+0.3}_{-0.2}$      & Joint fit\\
	Eccentricity $e$                        & -                         & $\leq 0.13^{*}$                 & Joint fit\\
	Radial velocity semi-amplitude $K$      & (ms$^{-1}$)               & $27 \pm 2$                & Joint fit\\
 	$\sqrt{e}\sin{\omega}$      & -               & $0.20^{+0.06}_{-0.09}$                & Joint fit\\
   	$\sqrt{e}\cos{\omega}$      & -               & $0.10^{+0.13}_{-0.17}$                & Joint fit\\
 	Minimum mass $M_p \sin{i}$                              & (M$_{\oplus}$)            & $56.59^{+3.49}_{-4.26}$            & Joint fit (derived) \\
	Mass $M_p$                              & (M$_{\oplus}$)            & $56.64^{+3.49}_{-4.27}$            & Joint fit (derived) \\
	Bulk density $\rho$                     & (g\,cm$^{-3}$)            & $0.98^{+0.15}_{-0.13}$       & Joint fit (derived) \\
	Semi-major axis $a$                     & (AU)                      & $ 0.0471 \pm 0.0002$      & Joint fit (derived) \\
    System scale $a/R_{\star}$                  & -                         & $14.7\pm0.4$    & Joint fit (derived) \\
	Equilibrium temperature $T_{\rm eq}$ (albedo $\zeta = 0.0$)& (K)                       & $885\pm22$        & Joint fit (derived) \\
 	Equilibrium temperature $T_{\rm eq}$ ($\zeta = 0.3$) & (K)                       & $810\pm20$        & Joint fit (derived) \\
  	Equilibrium temperature $T_{\rm eq}$ ($\zeta = 0.5$) & (K)                       & $744\pm18$        & Joint fit (derived) \\
	\midrule
	\multicolumn{4}{l}{\textbf{TOI-3071\,b}} \\
	    Period $P$                              & (days)                    & $1.266938\pm0.000002$ & Joint fit\\
	Full transit duration $T_{dur}$         & (hours)                   & $1.85^{+0.08}_{-0.07}$           & Joint fit (derived) \\
    Epoch $t_0$     & (BJD-2457000)             & $1594.608 \pm 0.002$    & Joint fit \\
	Radius $R_p$                            & (R$_{\oplus}$)            & $7.16^{+0.57}_{-0.51}$    & Joint fit \\
    Planet-to-star radius ratio $R_p/R_\star$ & -                         & $0.050^{+0.003}_{-0.003}$       & Joint fit \\
	Impact parameter $b$                    & -                         & $0.56^{+0.07}_{-0.08}$    & Joint fit\\  
	Inclination $i$                         & ($^{\circ}$)              & $82.1^{+1.3}_{-1.1}$      & Joint fit\\
	Eccentricity $e$                        & -                         & $\leq 0.09^*$                 & Joint fit\\
	%The argument of periastron $\omega$     & ($^{\circ}$)              &$1.5^{+1.1*}_{-4.2}$                 & Joint fit\\
	Radial velocity semi-amplitude $K$      & (ms$^{-1}$)               & $33.7\pm1.7$                & Joint fit\\
 	$\sqrt{e}\sin{\omega}$      & -               & $-0.11^{+0.10}_{-0.07}$                & Joint fit\\
   	$\sqrt{e}\cos{\omega}$      & -               & $0.07^{+0.16}_{-0.14}$                & Joint fit\\
 	Minimum mass $M_p \sin{i}$                              & (M$_{\oplus}$)            & $67.6\pm3.5$            & Joint fit (derived) \\
	Mass $M_p$                              & (M$_{\oplus}$)            & $68.2\pm3.5$            & Joint fit (derived) \\
	Bulk density $\rho$                     & (g\,cm$^{-3}$)            & $1.02^{+0.26}_{-0.22}$       & Joint fit (derived) \\
	Semi-major axis $a$                     & (AU)                      & $ 0.0249 \pm 0.0001$      & Joint fit (derived) \\
    System scale $a/R_{\star}$                  & -                         & $4.11\pm0.13$    & Joint fit (derived) \\
	Equilibrium temperature $T_{\rm eq}$ (albedo $\zeta = 0.0$)& (K)                       & $2155\pm40$        & Joint fit (derived) \\
 	Equilibrium temperature $T_{\rm eq}$ ($\zeta = 0.3$) & (K)                       & $1971\pm37$        & Joint fit (derived) \\
  	Equilibrium temperature $T_{\rm eq}$ ($\zeta = 0.5$) & (K)                       & $1812\pm34$        & Joint fit (derived) \\
	\bottomrule
 	\end{tabular}
	\begin{tablenotes}
	\item $T_{dur}$ is approximated by $\frac{P}{\pi}\sin^{-1}\left[\frac{R_\star}{a}\frac{\sqrt{(1+k)^2-b^2}}{\sin{i}} \right]\frac{\sqrt{1-e^2}}{1+e\sin{\omega}}$ where $k = R_p/R_\star$ \citep{winn2014transits}.
    \item $^*$ Computed as the $95\%$ $HDI$ upper bound for the posterior of $e$. Because the orbit is essentially circular, $\omega$ is not well-determined.
	\end{tablenotes}
	\end{threeparttable}
\end{table*}

\begin{figure}
    \centering
    \includegraphics[width=\columnwidth]{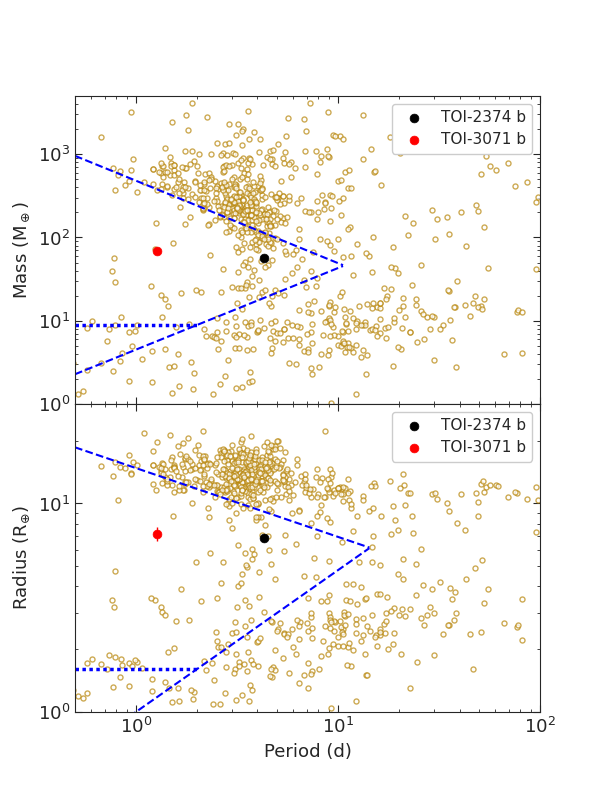}
    \caption{TOI-2374 b (black marker) and TOI-3071 b (red marker) in the context of the Neptunian Desert regions in Mass vs Period (top panel) and Radius vs Period (bottom panel). The dashed blue lines show the delineation of the Neptunian desert from \citet{Mazeh2016}, whereas the horizontal dotted blue lines show the updated lower limits for periods $\leq2$ days from \citet{Deeg2023}. Known planets were sourced from the NASA exoplanet archive (\url{https://exoplanetarchive.ipac.caltech.edu/}) on 28 May 2024.}
    \label{fig:neptunian_desert}
\end{figure}

The combined analysis of high resolution spectroscopy and space- and ground-based photometry allowed the determination of the masses and radii of planetary companions for the two target systems. 
%In the TPFs wee see that TOI-2374 has the same proper motion direction as its bright companion so they might be bounded.
We determined that both planets are sub-Saturns with mass and radius precision of $4-7 \%$ (see Table \ref{tab:params}). From this we inferred bulk densities of $0.98^{+0.15}_{-0.13}$\,g\,cm$^{-3}$ for TOI-2374\,b and $1.02^{+0.26}_{-0.22}$\,g\,cm$^{-3}$ for TOI-3071\,b. These parameters place both planets within in the Neptunian desert, with TOI-3071\,b being much deeper in than TOI-2374\,b (Fig.\ \ref{fig:neptunian_desert}). We found the eccentricity of both targets to be consistent with 0. The $95\%$ highest density interval (HDI) for the eccentricity ranges from 0 to 0.13 and 0.09 for TOI-2374\,b and TOI-3071\,b, respectively. We did not find any statistically significant long-term trends in the RV measurements of either star.

We note that the MCMC posterior distributions for the TESS light curves dilution correction factors are narrower than the priors (see figures \ref{fig:dilution_hist_2374} and \ref{fig:dilution_hist_3071} and tables \ref{tab:jointfit2374} and \ref{tab:jointfit3071} in the Appendix). For TOI-2374, the posterior modes of the QLP dilution correction factors are systematically negative, hinting to an overestimation of the dilution factor estimated by this pipeline. On the other hand, in TOI-3071, the posterior mean of the QLP dilution correction factor is positive, suggesting that for this star, the correction procedure underestimates the dilution. Finally, our model also hints to an overestimation of the dilution factor in the SPOC lightcurve of TOI-3071, as the posterior mean of $d$ is negative. Exploring the reasons of these behaviours is outside the scope of this article. Additionaly, in all cases the 83$\%$ HDI contains the null correction factor value. In any case, our model accounts for the uncertainty in the dilution factors estimated by the pipelines on the determination of the planetary radii.

%\par Additionally, we ran a model with the same features and priors as the base model described in Sect. \ref{jointfit} but with the stellar parameters informed by the SED analysis (Sect. \ref{sec:sed}). This model gives different estimates for the planetary radii: $R_p = 8.19^{+0.40}_{-0.38}$\,R$_{\oplus}$ for TOI-2374\,b ($1.9$-$\sigma$ difference with the base model) and $R_p = 8.77^{+0.45}_{-0.41}$\,R$_{\oplus}$ for TOI-3071\,b ($1$-$\sigma$ difference). The rest of the parameters are statistically indistinguishable. Since the planetary radii (and all the other parameters) differ by less than $2 \sigma$ and for the reasons discussed in Sect. \ref{sec:sed}, we use the parameters of the base model for the subsequent analyses.

\subsection{Comparison to other exoplanets}
\label{sec:comparison}

\begin{figure*}
    \centering
    \includegraphics[width=0.9\textwidth]{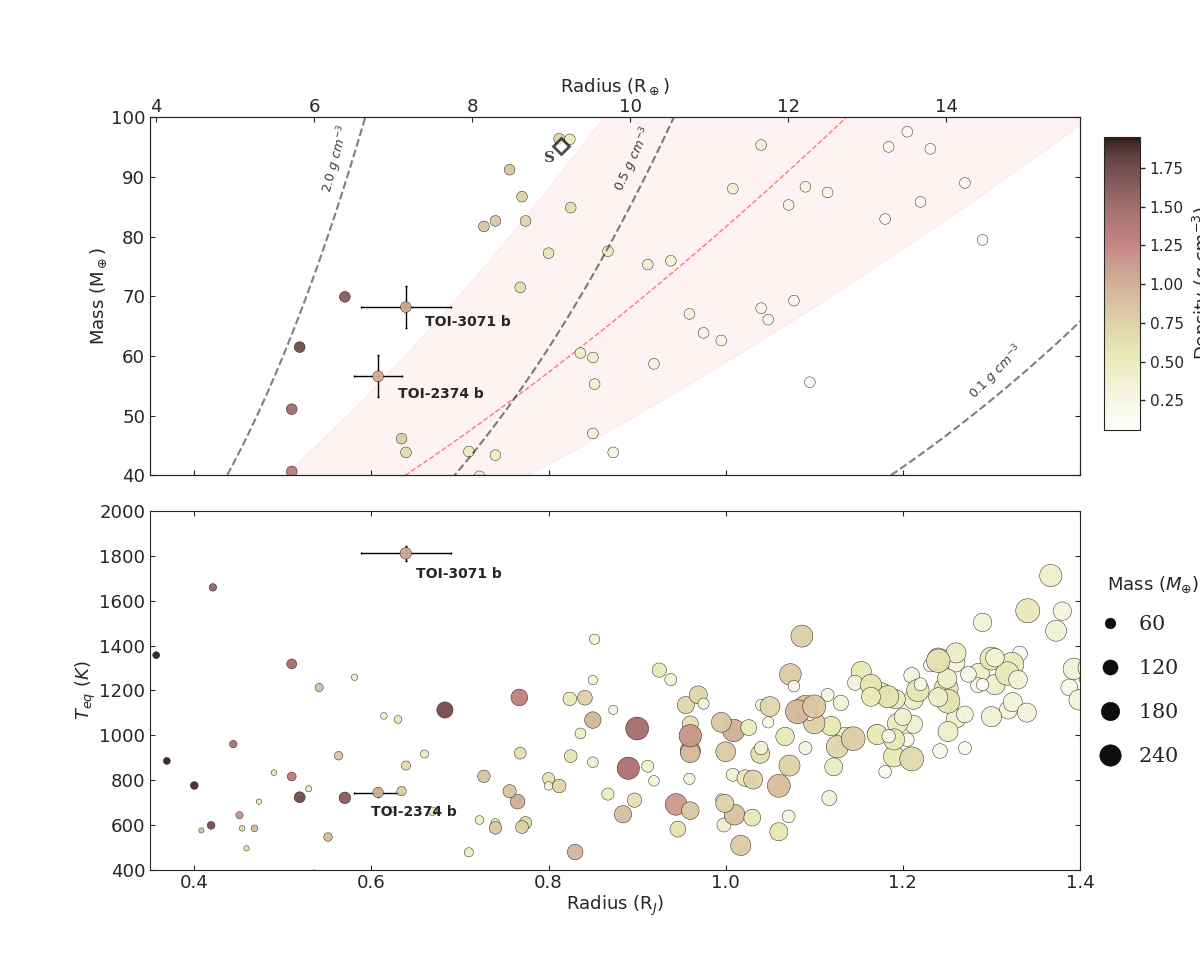}
    \caption{\textit{Top}: mass-radius diagram of the confirmed planet population with $P<20$ $d$, masses $40<M_P<100$ $M_\oplus$ and bulk densities $\rho < 2$ $g~cm^{-3}$, with radius determined to better than $20\%$ precision and masses determined to better than $30\%$ precision. The dashed lines show constant density values. For comparison, the position of Saturn in the plot is represented with a white diamond marker. The bulk densities for each planet are color coded. The R-M empirical relation found by \citet{Otegi2020} is shown by a red dashed line with the $1 \sigma$ regions colored. \textit{Bottom}: equilibrium temperatures and radii of the same set of exoplanets, extending the mass range to $10<M_P<300$ $M_\oplus$. The bulk densities for each planet are color coded and the markers are sized according to their mass.}
    \label{fig:mass_radius_teq}
\end{figure*}

In order to understand how TOI-2374 b and TOI-3071 b fit into the landscape of known planets, we compare them to planets of similar size, mass, and period ($P<20$ $d$; $40<M_P<100$ $M_{\oplus}$; $\rho~<~2$~$g~cm^{-3}$) with mass precision better than $30\%$ and radius precision better than $20\%$ (top panel of Fig. \ref{fig:mass_radius_teq}). This includes all “hot” giant planets below Saturn’s mass. We use the parameters from the planetary systems table on the NASA Exoplanet Archive \footnote{\url{https://exoplanetarchive.ipac.caltech.edu/}} accessed on 28 May 2024. The empirical R-M relation for volatile-rich planets ($\rho < 3.3$ $g~cm^{-3}$) found by \citet{Otegi2020} is also shown in the diagram. The bottom panel of Fig. \ref{fig:mass_radius_teq} shows the equilibrium temperature ($T_{\rm eq}$) vs radius plane for the same subset of exoplanets, but extending the mass range to $10<M_P<300$ $M_\oplus$. $T_{\rm eq}$ is calculated from the planetary orbital semi-major axes $a$ and the host-star effective temperature $T_\star$ and radius $R_\star$ assuming full day-night heat distribution, according to

\begin{equation}
    T_{\rm eq} = T_\star\sqrt{\frac{R_\star}{2a}}(1-\zeta)^{1/4}
\end{equation}

\noindent where $\zeta$ is the Bond albedo of the planet. For this analysis we assumed $\zeta = 0.5$, similar to that of Jupiter \citep{2018NatCo...9.3709L}, to calculate $T_{\rm eq}$ for TOI-2374 b, TOI-3071 b and every other planet in the dataset (even if they already had a reported $T_{\rm eq}$).

Both planets are hot and highly irradiated, with $T_{\rm eq} \approx 744$ K for TOI-2374 b and $T_{\rm eq} \approx 1812$ K for TOI-3071 b. TOI-2374 b is slightly below the usual inflated hot-Jupiter cutoff of $F_* \approx 2 \times 10^8$ erg/s/cm$^2$ \citep{2011ApJ...736L..29M}, and it is therefore not expected to be anomalously inflated. On the other hand, TOI-3071 b is well inside the hot-Jupiter instellation regime. However, previous studies suggest that planets with masses below about $0.4 M_J$ show a weaker response to high instellation fluxes compared to more massive planets, and are less likely to be strongly inflated \citep{2018AJ....155..214T,2018A&A...616A..76S}. Therefore, standard evolution models of giant planets may reflect the radius more accurately than for higher mass planets, for which the inflation is known to be underestimated. From Fig. \ref{fig:mass_radius_teq} it is clear that TOI-3071 b has a similar radius to previously detected giant exoplanets with similar masses, despite having a much higher equilibrium temperature. Besides hydrogen and helium, the planet is therefore expected to contain a large amount of heavier elements.

\subsection{Inferred heavy-element masses}
\label{sec:heavymasses}

Both planets have available masses, radii and ages, which make them viable targets for the estimation of their heavy-element masses \citep[e.g.,][]{2011ApJ...736L..29M,2016ApJ...831...64T,2023A&A...669A..24M}. The bulk heavy-element mass of a giant planet is a key property since it can be used to test planet formation models and provide additional constraints on the possible formation pathways \citep[e.g.,][]{2006A&A...453L..21G,2017AREPS..45..359J,2018ApJ...865...32H}. Unfortunately, it cannot be measured directly and must be inferred from a combination of measurements and giant-planet evolution models. To estimate the composition of the two planets, we modeled their evolution with a modified version of the stellar evolution code Modules for Experiments in Stellar Astrophysics (MESA; \citet{2011ApJS..192....3P,2013ApJS..208....4P,2015ApJS..220...15P,2018ApJS..234...34P,2019ApJS..243...10P,2023ApJS..265...15J}). \citet{2020ApJ...903..147M, 2020A&A...638A.121M} modified the equation of state of MESA to make it suitable to model giant planets that contain a significant amount of heavy elements. Here, we used a recently updated version \citep{2024arXiv240316273M}, which uses a new hydrogen-helium equation of state that includes non-ideal interactions in the hydrogen-helium mixture \citep{2021ApJ...917....4C}. These effects have been shown to lead to a significant change in the cooling of planets \citep{2023A&A...672L...1H}. We assumed a core-envelope structure and a 50-50 water-rock composition for the heavy elements. The radiative opacity was from \citet{Freedman2014}, and to account for the instellation fluxes we used a gray atmospheric model developed for irradiated planets \citep{2010A&A...520A..27G,2014A&A...562A.133P}.

We first calculated the planetary evolution assuming that they have the same composition as their stars. Both planets are much smaller compared to these predictions, demonstrating that they must have super-stellar metallicities (see solid lines in Fig. \ref{fig:age_radius_stellar} in Appendix).

To quantify the metal enrichment of the planets, we used the standard Monte-Carlo approach \citep[see, e.g.,][for a review]{2023FrASS..1079000M}: sample planets were created by drawing from the observation prior distributions for the mass, radius, and stellar age. For the mass and radius, we used normal distributions informed by the parameters' marginal posterior sample median and standard deviation from the joint fit (Sect. \ref{jointfit}), and for the age we used a uniform distribution based on the estimates from the stellar parameters analysis (Sect. \ref{stellar-physical-parameters}). The evolution models were then used to calculate the cooling for different heavy-element masses $M_z$ to find the one that agrees with the observations. This process is repeated until there is enough data to estimate the posterior distribution for the heavy-element mass. For TOI-3071b we used two different stellar age priors (from chemical clocks and gyrochronology) and then combined their posteriors for the final result assuming that the priors are equally likely. The MCMC approach required the calculation of a large number of evolution models. Since this would have been too slow, we followed a similar approach as outlined in \citet{2021MNRAS.507.2094M}. For each planet, we calculated a grid of evolution models (in the planet mass-metallicity space) and then used 2d piecewise monotonic cubic interpolation to generate new cooling tracks.

\begin{figure}
    \centering
    \includegraphics[width=0.75\columnwidth]{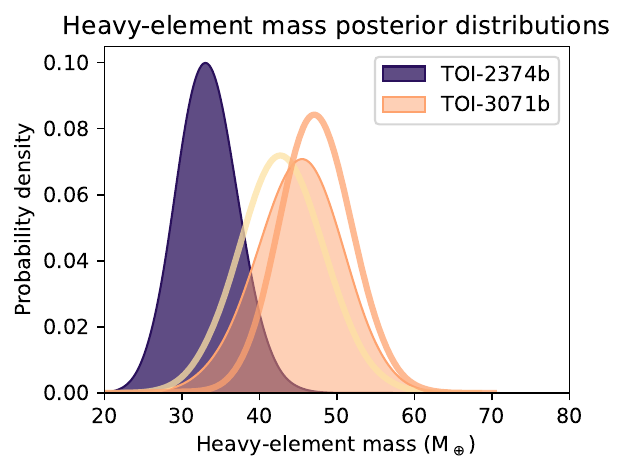}
    \caption{Inferred heavy-element mass posteriors for the two planets. For TOI-3071b, the two curves show the results for the two different stellar age estimates (chemical clocks and gyrochronology), and the filled area is their combined posterior. Both planets are heavy-element rich, with a bulk heavy-element fraction of $Z \gtrsim 0.6$.}
    \label{fig:heavy_element_mass_posteriors}
\end{figure}

The resulting posterior distributions of the inferred heavy-element masses are shown in Fig. \ref{fig:heavy_element_mass_posteriors}. We find heavy-element masses (mass fractions) of $M_z = 33.3 \pm 3.8 M_\oplus$ ($Z = 0.59 \pm 0.05$) for TOI-2374 b and $M_z = 45.1 \pm 5.5 M_\oplus$ ($Z = 0.66 \pm 0.07)$ for TOI-3071 b. For comparison, Saturn, the most similar planet in the solar system, has an estimated bulk heavy-element mass (mass fraction) of $M_z \simeq 30 M_\oplus$ ($Z \simeq 0.3$). The exact value is unknown and depends on several assumptions when modeling Saturn's interior. Both targets are metal-rich, with TOI-3071b being more enriched both in absolute and relative terms (see Figure \ref{fig:heavy_element_mass_fraction_posteriors} for posterior distributions of the heavy-element masses in mass fractions). TOI-3071b in particular has an extremely high bulk metallicity and is about 60-70\% heavy elements (water and rocks).

The fact that these planets have such high metallicities challenges standard formation models which generally tend to predict lower heavy-element masses \citep{1996Icar..124...62P,2014prpl.conf..643H,2017AREPS..45..359J,2018A&A...612A..30B}. Interestingly, both planets correspond to masses below Saturn's mass, which was recently suggested to be the transition mass to become giant planets \citep{2023A&A...675L...8H}. In that view, the formation of a giant planet can last for a few million years where gas accretion is being delayed due to continuous accretion of planetesimals \citep{2018NatAs...2..873A,2022Icar..37814937H}. Such a formation scenario also suggests that planets below about Saturn's mass are expected to be metal-rich in composition, consistent with our estimates for TOI-2374\,b and TOI-3071\,b.
An alternative explanation is a post-formation enrichment due to planetesimal accretion or collisions between planetary embryos \citep{2009ApJ...696.1348M,2020A&A...633A..33S,2020MNRAS.498..680G}. However, such collisions have difficulty enriching planets with tens of Earth masses of heavy elements.

Based on the models from \citet{Santos2015,Santos2017,2021Sci...374..330A}, we used the stellar abundances to calculate the summed mass fraction of all heavy elements available as planetary building blocks. The values are $Z_{TOI-2374} = 0.018 \pm 0.03$ and $Z_{TOI-3071b} = 0.024 \pm 0.03$, respectively. For comparison, the same models yield $Z_\odot = 0.013 \pm 0.01$ for solar abundances. Both mass fractions are above the solar value, which hints at a possible connection between the stellar abundances and the high inferred bulk metallicities. 

Finally, we note that measurements of the atmospheric chemical composition could be used to further constrain the planetary formation, evolution, and interior  \citep[e.g.,][]{2014PNAS..11112601B,2019ApJ...874L..31T,2022ExA....53..323H}.

\subsection{Evaporation history}\label{sec:evap}

\subsubsection{Internal structure}
\label{sec:jorge-evap-structure}

To model the evaporation histories of the two planets, we assumed a simpler two-layer internal structure with a rocky core surrounded by a pure H/He envelope.
This model can be described with four quantities: the rocky core mass M$_\text{core}$ and core radius R$_\text{core}$, and the gaseous envelope thickness R$_\text{env}$ and envelope mass fraction $\text{f}_\text{env}$, defined as the ratio of envelope mass to planet mass $\text{M}_\text{env}/\text{M}_\text{p}$.
In order to determine these parameters, we adopted the empirical mass-radius relation by \citet{Otegi2020} for the rocky core, and the envelope structure model by \citet{ChenRogers16:envelope-model} for the gaseous atmosphere, which is based on MESA simulations.
Uisng this model, we thus determined envelope mass fractions of $38\pm5$~\% for TOI-2374\,b and $26\pm6$~\% for TOI-3071\,b.
These correspond to core (heavy-element) masses of M$_\text{core}=34.7\pm3.7$ for TOI-2374\,b, and M$_\text{core}=50.5\pm4.9$ for TOI-3071\,b, both consistent with the heavy-element masses determined in Sect.~\ref{sec:heavymasses} within $1\sigma$.

\begin{figure*}
    \centering
    \includegraphics[width=0.8\textwidth]{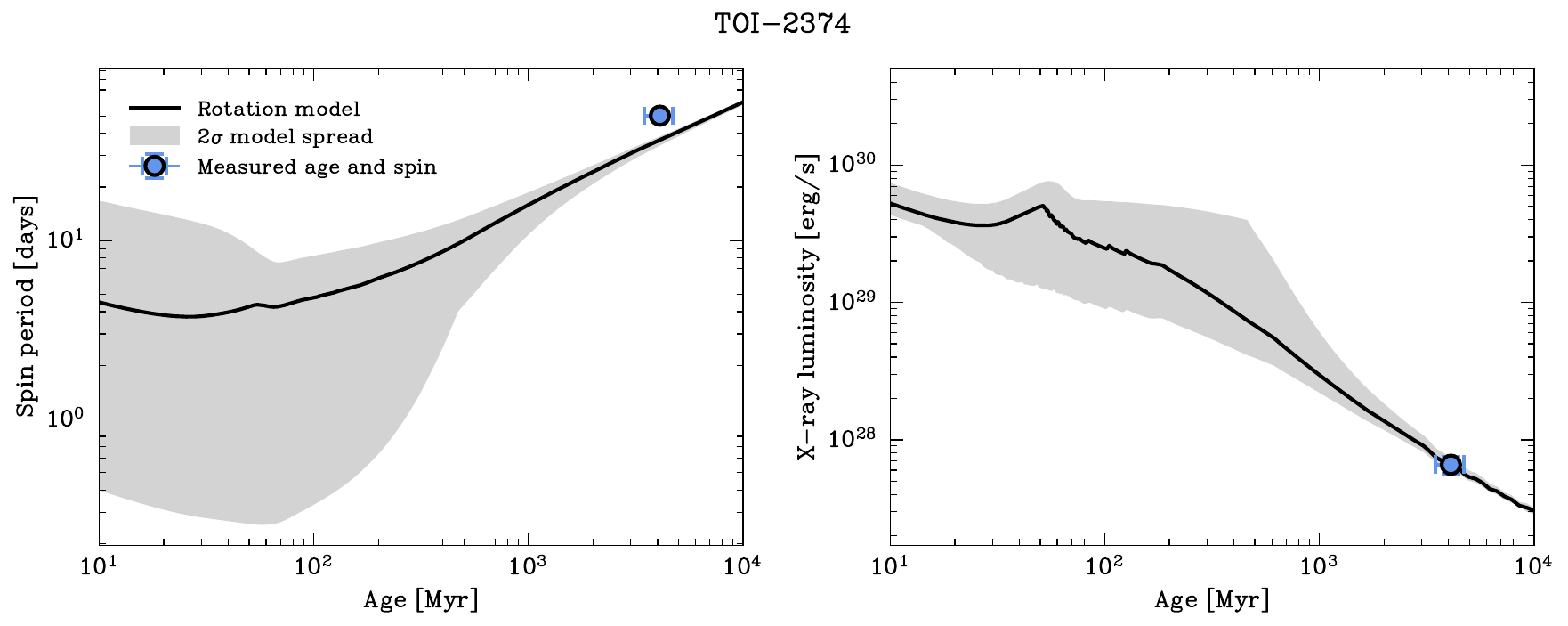}

    \caption{
        \textbf{Left panel:} spin period evolution of TOI-2374 using the models by \citet{Johnstone21:rotation-model}. The model mean is shown as a solid black line, and the $2\sigma$ spread as a grey shaded region. TOI-2374 is shown as a blue circle.
        \textbf{Right panel:} evolution models of the X-ray luminosity following the left panel. The current X-ray luminosity of the star predicted by the model is shown as a blue circle.
    }
    \label{fig:toi2374starevo}
\end{figure*}

\begin{figure*}
    \centering
    \includegraphics[width=0.8\textwidth]{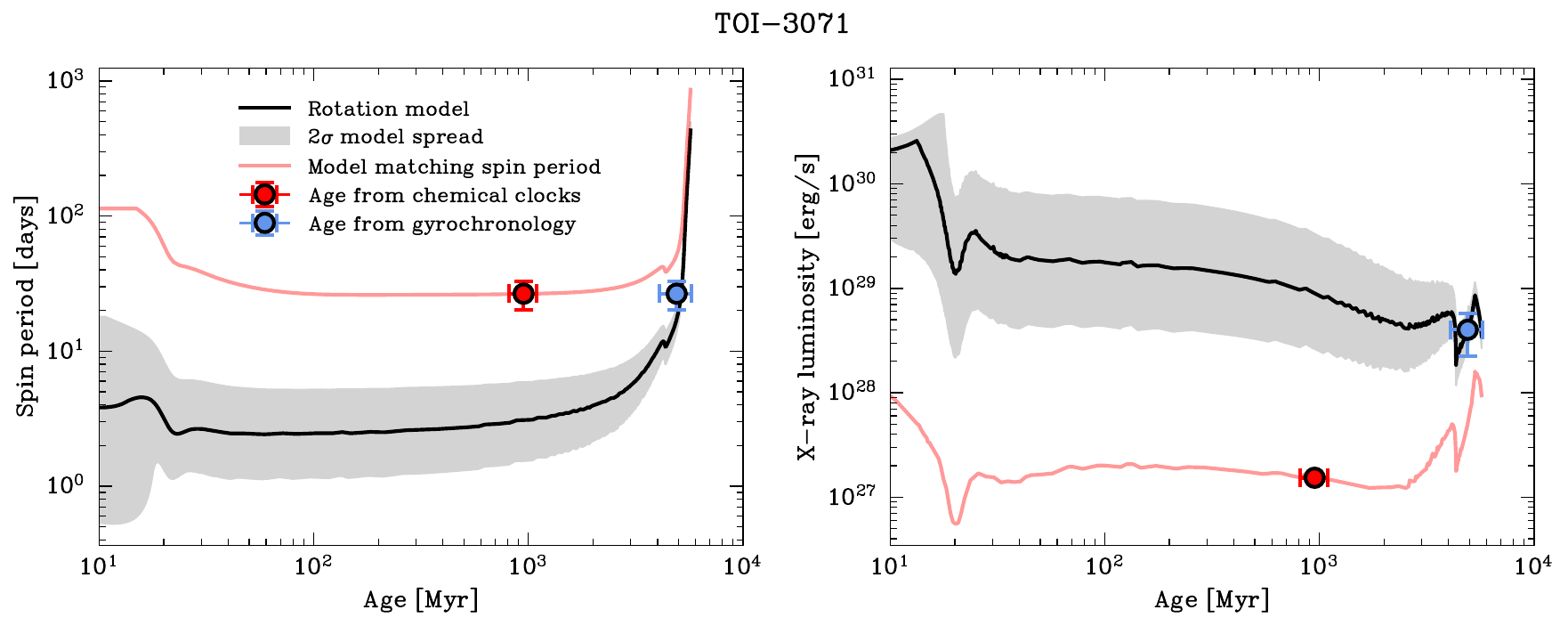}
    \caption{
        \textbf{Left panel:} spin period evolution of TOI-3071 using the models by \citet{Johnstone21:rotation-model}, following Figure~\ref{fig:toi2374starevo}.
        Due to the disagreement between the age estimated from gyrochronology (blue circle) and the age from the chemical clocks and \texttt{PARAM} isochrones (red circle), we plotted an additional spin evolution model fitted to the age from chemical clocks (red line).
        \textbf{Right panel:} X-ray evolution of TOI-3071 using the models by \citet{Johnstone21:rotation-model}, following Figure~\ref{fig:toi2374starevo}.
        The X-ray evolution model fitted to the age from the chemical clocks is shown as a red line.
    }
    \label{fig:toi3071starevo}
\end{figure*}

\subsubsection{Evolution modelling}
\label{sec:jorge-evap-evolution}

Simulating the evolution of a planet's atmosphere under photoevaporation requires knowledge of its X-ray irradiation history.
We can estimate the X-ray and EUV emission (together, XUV) of a star from its spin period, as the two are related via rotation--activity relations \citep[e.g.][]{Wright11:rotation-xrays, Pizzolato03:xray-rotation}.
We adopted the stellar rotation evolution models by \citet{Johnstone21:rotation-model}, which simulate the evolution of the angular momentum of stars constrained by observations of young star clusters.
Moreover, to estimate the EUV component, we adopted the scaling relation by \citet{King21:euv-gyr}, which are based on Solar TIMED/SEE data.
We plot the evolution of the spin period and corresponding XUV emission of TOI-2374 and TOI-3071 on Figures \ref{fig:toi2374starevo} and \ref{fig:toi3071starevo}, respectively.

Regarding TOI-2374, we found the rotational models by \citet{Johnstone21:rotation-model} are consistent with the measured age and spin period of the star.

For TOI-3071, we considered the two age estimates separately: an age of 1~Gyr from the chemical clocks and \texttt{PARAM} isochrones, and an age of 5~Gyr from gyrochronology (see Sect.~\ref{stellar-physical-parameters}).

We found the choice of age had a strong impact on the X-ray emission history of the star.
The younger age is more consistent with a short spin period of\,2-4\,d, whereas the older age from gyrochronology is more consistent with a slower spin period of\,20-30\,d, as the star has had more time to spin down.
The spin period of the star, however, is constrained to 26\,d, and so adopting the younger age would imply that the star must be unusually slowly rotating and X-ray faint (Fig.~\ref{fig:toi3071starevo}, red line).

For that reason, we adopted two scenarios for the X-ray emission history of TOI-3071: (1) the \textit{high activity} scenario where we adopted the predicted evolution from the models by \citet{Johnstone21:rotation-model}, consistent with the older gyrochronological age (see Fig.~\ref{fig:toi3071starevo}, black line); and (2) the \textit{low activity} scenario, where we adopted a fainter emission history model consistent with the younger age from the 3D chemical clocks (Fig.~\ref{fig:toi3071starevo}, red line).

\begin{figure*}
    \centering
    \includegraphics[width=0.8\textwidth]{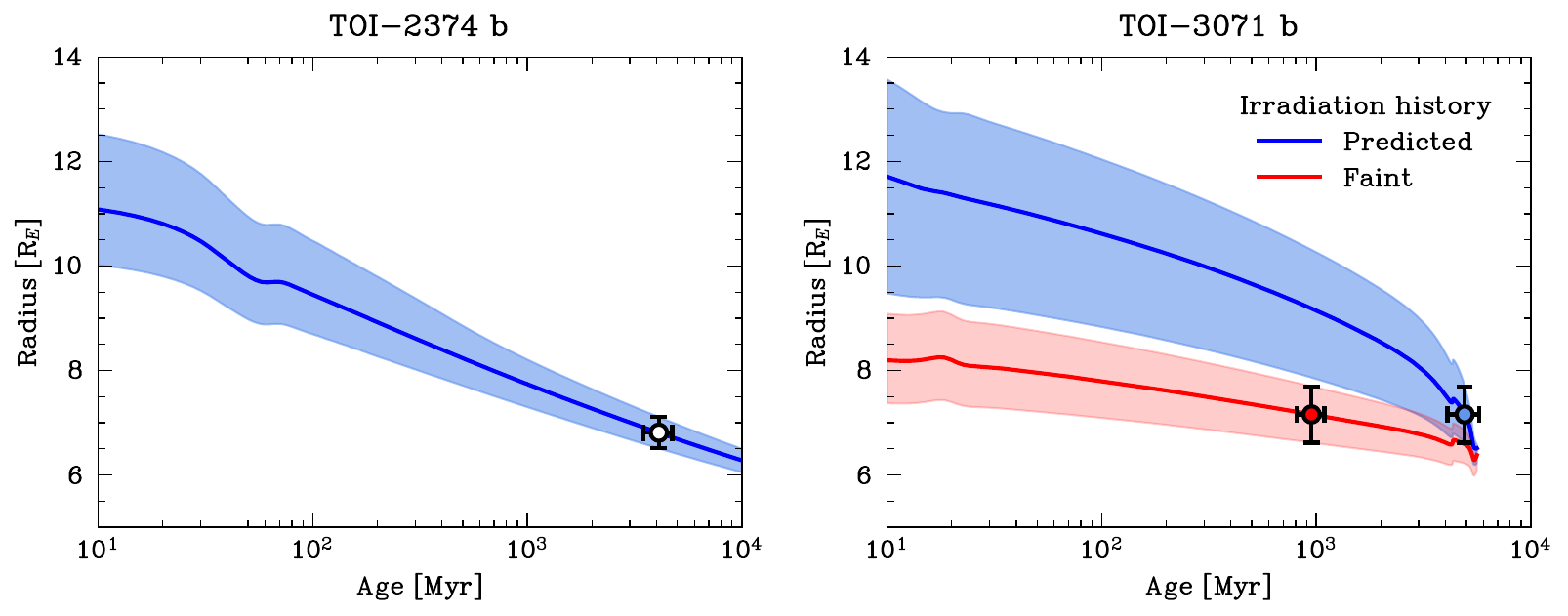}
    \caption{
        \textbf{Left panel:} Past and future evolution of the atmosphere of TOI-2374\,b under photoevaporation. The solid blue line shows the expected evolution of the planet, whereas the shaded region represents possible evaporation histories taking into account the uncertainties in the planet's mass, radius, and XUV irradiation. The planet is plotted as a white circle.
        \textbf{Right panel:} Plot of the past and future evolution of the atmosphere of TOI-3071\,b under photoevaporation, following the left panel. The area shaded in blue shows evolution under an X-ray irradiation history consistent with the stellar age derived from gyrochronology (5~Gyr), whereas the area shaded in red shows evolution under a fainter irradiation history consistent with the stellar age derived from chemical clocks (1~Gyr).
    }
    \label{fig:jorge-planets-evo}
\end{figure*}

We then simulated the past and future evaporation histories of the planets with the \texttt{photoevolver} code \citep{Fernandez23:evap}.
The simulation is built upon three ingredients: (1) the XUV emission history of the star, which determines the X-ray irradiation on the planet, (2) a mass loss formulation, which determines the amount of gas lost due to XUV irradiation, and (3) an envelope structure model, which relates the mass of the gaseous atmosphere to its thickness.
Therefore, on each time step, the X-ray luminosity of the star at that age is used to calculate the mass loss rate, which is then used to recalculate the mass and size of the planet.

We adopted the models by \citet{Johnstone21:rotation-model} for the stellar XUV emission history (described above), and the model by \citet{ChenRogers16:envelope-model} for the structure of the gaseous atmosphere, described in Section~\ref{sec:jorge-evap-structure}.
Regarding the mass loss formulation, we adopted the model by \citet{Kubyshkina18:mass-loss-model}.
We evolved the planets' atmospheres using the 4th order Runge-Kutta algorithm \citep{Dormand80:rk45} as the integration method with a variable time step no larger than 1 Myr.
We ran the simulation backwards to the age of 10 Myr and forwards to 10 Gyr. In the case of the F-type TOI-3071, we stop the simulation at 6 Gyr, the approximate age at which the star leaves the main sequence \citep{Choi16:mist-models}.

Moreover, for each of the two planets, we also explored a range of possible evaporation histories taking into account the uncertainties on the planet's mass, radius, and XUV irradiation.
% We evolved the planets under two additional scenarios in which they would experience the maximum and minimum evaporation rates allowed by the uncertainties in the simulation parameters.
To maximise the evaporation rate, we picked a radius on the higher end of its $1\sigma$ uncertainty and a mass on the lower end of its $1\sigma$ uncertainty, which minimises the density and maximises the escape rates. We also picked the highest XUV emission history allowed by the model's $2\sigma$ spread, as shown on Figures \ref{fig:toi2374starevo} and \ref{fig:toi3071starevo}.
Likewise, to minimise the evaporation rate, we picked the opposite parameters: values for the mass and radius within their $1\sigma$ errors that maximise density under the faintest XUV irradiation history allowed by the model's $2\sigma$ spread.

\subsubsection{Simulation results}
\label{sec:jorge-evap-results}

The evaporation histories of TOI-2374\,b and TOI-3071\,b are plotted on Figure~\ref{fig:jorge-planets-evo} (left and right panels, respectively).

We found that TOI-2374\,b is stable against evaporation, as its envelope mass fraction changes only by a few percent across its lifetime. The evolution of its radius is thus largely driven by thermal contraction. We also constrained its current mass loss rates to $0.4-0.8\times10^{8}$\,g s$^{-1}$.

On the other hand, TOI-3071\,b is more susceptible to escape despite its higher mass. This is due to being twice as close to its star compared to TOI-2374\,b, as well as the higher X-ray luminosity of its F-type host star.
Under the higher X-ray irradiation history consistent with the gyrochronological age (5~Gyr), our simulations show that the planet is reaching the point of losing its atmosphere completely when its star starts evolving off the main sequence (see Fig.~\ref{fig:jorge-planets-evo}), with an envelope mass fraction of 10-20\% at 6\,Gyr down from 40-60\% initially.
Under this X-ray irradiation history, the planet currently experiences mass loss rates of $7.5-40.7\times10^{8}$\,g s$^{-1}$.

However, under the fainter irradiation history consistent with the much younger age derived from chemical clocks (1~Gyr), TOI-3071\,b experiences lower escape rates throughout its life and is stable against evaporation. We constrained its current mass loss rates to $0.9-2.6\times10^{8}$\,g s$^{-1}$, one order of magnitude lower compared to the high activity scenario.
% especially since F and G dwarfs stars experience a decline in their X-ray emission after the main sequence stage of their lives \citep[e.g.][]{Pizzolato00:xrays-evolved-stars}.

In Sect.~\ref{sec:heavymasses}, we presented more detailed internal structure models to estimate the bulk metallicity of the two planets. These models cannot predict the atmospheric composition, which can only be determined by observations. However, since both are metal-rich, they are also likely to have high-metallicity atmospheres. Such high-metallicity atmospheres experience much lower escape rates as heavier species require more energy to remove \citep{Wilson22:toi-1064, OwenJackson12:xray-evap}.
Since we only considered a pure H/He composition in our atmospheric evolution models, our results in Fig.~\ref{fig:jorge-planets-evo} constitute an upper bound to the evaporation histories of the planets.

For TOI-2374,b, the atmosphere is stable against evaporation throughout its life, even when assuming a pure hydrogen envelope. TOI-3071,b experiences moderate escape rates and is significantly affected by evaporation only as its host star transitions out of the main sequence, risking total atmospheric loss. However, heavy-mass planets like TOI-3071\,b could present a metallicity gradient in their gaseous envelopes, with a relatively pure H/He upper atmosphere that progresses towards metal-rich layers deeper in the atmosphere.
If the planet experienced stronger escape rates earlier in its life, either from photoevaporation or Roche-lobe overflow, the pure H/He upper layers of the envelope would be stripped off first, exposing the deeper metal-rich layers which would then curtail atmospheric escape \citep{OwenMurrayClay18, Hoyer23:LTT-9779-metal-rich}.

Therefore, the presence of heavy species in the hydrogen-rich envelope of TOI-3071\,b at the present time would lead to diminished escape rates, making its atmosphere more resistant to evaporation and thus likely stable. As a result, we find the two planets are not significantly affected by photoevaporation and are thus able to hold onto their gaseous envelopes throughout their lifetimes.
% We thus predict the two planets are stable against evaporation due to their large masses and high metallicities in their gaseous envelopes.

\subsection{Prospects for atmospheric characterization}

The transmission spectroscopy metric (TSM) is defined as

\begin{equation}
    TSM = \text{(scale factor)}\times \frac{R_p^3 T_{\rm eq}}{M_p R_\star^2} \times 10^{-m_J/5}
\end{equation}

\noindent where $m_J$ is the apparent magnitude of the host star in
the J band and the scale factor is $1.15$ for $4<R_p<10$ $R_\oplus$ \citep{Kempton_2018}.

We calculated the targets' TSM using the equilibrium temperatures for albedos of $\zeta = 0.5$ and obtained a value of $84\pm13$ for TOI-2374 b and $36\pm9$ for TOI-3071 b. The former is slightly below the threshold of 90 set by \citet{Kempton_2018} for sub-jovians to be considered high quality atmospheric characterization targets. However, if we assume lower albedos of $0.3$ and $0.0$, we obtain TSM values of $91\pm15$ and $100 \pm 16$ for TOI-2374\,b, respectively. For TOI-3071\,b the TSM is always lower than 50. As mentioned above, atmospheric measurements of these targets could provide key information to further constrain the planetary formation, evolution and interior structure.

\section{Conclusion}\label{conclusion}

We have presented the discovery and characterisation of two new transiting hot Saturns orbiting the K-type and F-type stars TOI-2374 and TOI-3071, respectively. Our analysis includes photometry from TESS, follow up ground-based photometry and spectroscopy from HARPS and PFS. The combination of this data in a joint light curve and radial velocity model allowed the confirmation of the planets and the determination of their orbital and physical properties. 

\par Both planets are hot sub-Saturns in close, virtually circular orbits. They have similar masses and radii and are located within the Neptunian desert. Due to its closer orbit, TOI-3071\,b is much deeper in the desert than TOI-2374\,b. This, and the fact that TOI-3071 is a much hotter star and emits higher XUV flux, leads to TOI-3071\,b having a much higher equilibrium temperature, higher even than that of all previously detected giant exoplanets with similar masses or radii. Both planets are very metal rich, with TOI-3071\,b being more enriched both in absolute and relative terms. By studying the evolution of the planets’ atmospheres under photoevaporation we find that both are stable against evaporation. 

Both planets are amenable to further follow-up with JWST, as despite the relatively low TSM of TOI-3071\,b, it is a fairly bright (11.87 T mag) target undergoing atmospheric evaporation. Meanwhile, TOI-2374\,b has a higher TSM and lies close to the Continuous Viewing Zone of both TESS and JWST for ease of observation scheduling. Measuring the atmospheric chemical compositions would prove very useful to further constrain the formation, evolution and interior of the planets.

\section*{Acknowledgements}

% people

VA is supported by FCT - Fundação para a Ciência e a Tecnologia through national funds by the following grants: UIDP/04434/2020 (\url{https://doi.org/10.54499/UIDP/04434/2020}); UIDB/04434/2020 (\url{https://doi.org/10.54499/UIDB/04434/2020}); 2022.06962.PTDC (\url{http://doi.org/10.54499/2022.06962.PTDC}).

KAC and CNW acknowledge support from the TESS mission via subaward s3449 from MIT.

FB carried out this work within the framework of the National Centre of Competence in Research PlanetS supported by the Swiss National Science Foundation. 

XD acknowledges the support from the European Research Council (ERC) under the European Union’s Horizon 2020 research and innovation programme (grant agreement SCORE No 851555) and from the Swiss National Science Foundation under the grant SPECTRE (No 200021\_215200)

JL-B was partly funded by grants LCF/BQ/PI20/11760023, Ram\'on y Cajal fellowship with code RYC2021-031640-I, and the Spanish MCIN/AEI/10.13039/501100011033 grant PID2019-107061GB-C61.

NCS is co-funded by the European Union (ERC, FIERCE, 101052347). Views and opinions expressed are however those of the author(s) only and do not necessarily reflect those of the European Union or the European Research Council. Neither the European Union nor the granting authority can be held responsible for them. This work was supported by FCT - Fundação para a Ciência e a Tecnologia through national funds and by FEDER through COMPETE2020 - Programa Operacional Competitividade e Internacionalização by these grants: UIDB/04434/2020; UIDP/04434/2020.

% LCOGT

This work makes use of observations from the LCOGT network. Part of the LCOGT telescope time was granted by NOIRLab through the Mid-Scale Innovations Program (MSIP). MSIP is funded by NSF.

% ExoFOP

This research has made use of the Exoplanet Follow-up Observation Program (ExoFOP; DOI: 10.26134/ExoFOP5) website, which is operated by the California Institute of Technology, under contract with the National Aeronautics and Space Administration under the Exoplanet Exploration Program.

This research has made use of the NASA Exoplanet Archive, which is operated by the California Institute of Technology, under contract with the National Aeronautics and Space Administration under the Exoplanet Exploration Program.

% TESS

This paper made use of data collected by the TESS mission and are publicly available from the Mikulski Archive for Space Telescopes (MAST) operated by the Space Telescope Science Institute (STScI). 

We acknowledge the use of public TESS data from pipelines at the TESS Science Office and at the TESS Science Processing Operations Center. 

Resources supporting this work were provided by the NASA High-End Computing (HEC) Program through the NASA Advanced Supercomputing (NAS) Division at Ames Research Center for the production of the SPOC data products.

Funding for the TESS mission is provided by NASA's Science Mission Directorate. KAC and CNW acknowledge support from the TESS mission via subaward s3449 from MIT.

This research was funded in part by the UKRI, (Grants ST/X001121/1, EP/X027562/1). For the purpose of open access, the author has applied a Creative Commons Attribution (CC BY) licence to any Author Accepted Manuscript version arising from this submission.

\section*{Data Availability}

The {\it TESS} data are available from the Mikulski Archive for Space Telescopes (MAST), at \url{https://heasarc.gsfc.nasa.gov/docs/tess/data-access.html}. The other photometry from LCOGT, Briefrield Private Observatory and PEST, and the high-resolution imaging data, are available for public download from the ExoFOP-TESS archive at \url{https://exofop.ipac.caltech.edu/tess/target.php?id=439366538} for TOI-2374 and at \url{https://exofop.ipac.caltech.edu/tess/target.php?id=452006073} for TOI-3071. The PFS RV data is also available here. The HARPS RV data is shown in tables \ref{tab:harps_3071} and \ref{tab:harps} and the full HARPS data products can be found on the ESO archive.  The model code underlying this article will be shared on reasonable request to the corresponding author.


\bibliographystyle{mnras}
\bibliography{paper}

\begin{thebibliography}{}
\makeatletter
\relax
\def\mn@urlcharsother{\let\do\@makeother \do\$\do\&\do\#\do\^\do\_\do\%\do\~}
\def\mn@doi{\begingroup\mn@urlcharsother \@ifnextchar [ {\mn@doi@}
  {\mn@doi@[]}}
\def\mn@doi@[#1]#2{\def\@tempa{#1}\ifx\@tempa\@empty \href
  {http://dx.doi.org/#2} {doi:#2}\else \href {http://dx.doi.org/#2} {#1}\fi
  \endgroup}
\def\mn@eprint#1#2{\mn@eprint@#1:#2::\@nil}
\def\mn@eprint@arXiv#1{\href {http://arxiv.org/abs/#1} {{\tt arXiv:#1}}}
\def\mn@eprint@dblp#1{\href {http://dblp.uni-trier.de/rec/bibtex/#1.xml}
  {dblp:#1}}
\def\mn@eprint@#1:#2:#3:#4\@nil{\def\@tempa {#1}\def\@tempb {#2}\def\@tempc
  {#3}\ifx \@tempc \@empty \let \@tempc \@tempb \let \@tempb \@tempa \fi \ifx
  \@tempb \@empty \def\@tempb {arXiv}\fi \@ifundefined
  {mn@eprint@\@tempb}{\@tempb:\@tempc}{\expandafter \expandafter \csname
  mn@eprint@\@tempb\endcsname \expandafter{\@tempc}}}

\bibitem[\protect\citeauthoryear{{Adibekyan}, {Sousa}, {Santos}, {Delgado
  Mena}, {Gonz{\'a}lez Hern{\'a}ndez}, {Israelian}, {Mayor}  \&
  {Khachatryan}}{{Adibekyan} et~al.}{2012}]{Adibekyan-12}
{Adibekyan} V.~Z.,  {Sousa} S.~G.,  {Santos} N.~C.,  {Delgado Mena} E.,
  {Gonz{\'a}lez Hern{\'a}ndez} J.~I.,  {Israelian} G.,  {Mayor} M.,
  {Khachatryan} G.,  2012, \mn@doi [\aap] {10.1051/0004-6361/201219401}, \href
  {http://adsabs.harvard.edu/abs/2012A%26A...545A..32A} {545, A32}

\bibitem[\protect\citeauthoryear{{Adibekyan} et~al.,}{{Adibekyan}
  et~al.}{2015}]{Adibekyan-15}
{Adibekyan} V.,  et~al., 2015, \mn@doi [\aap] {10.1051/0004-6361/201527120},
  \href {http://adsabs.harvard.edu/abs/2015A%26A...583A..94A} {583, A94}

\bibitem[\protect\citeauthoryear{{Adibekyan} et~al.,}{{Adibekyan}
  et~al.}{2021}]{2021Sci...374..330A}
{Adibekyan} V.,  et~al., 2021, \mn@doi [Science] {10.1126/science.abg8794},
  \href {https://ui.adsabs.harvard.edu/abs/2021Sci...374..330A} {374, 330}

\bibitem[\protect\citeauthoryear{{Alibert} et~al.,}{{Alibert}
  et~al.}{2018}]{2018NatAs...2..873A}
{Alibert} Y.,  et~al., 2018, \mn@doi [Nature Astronomy]
  {10.1038/s41550-018-0557-2}, \href
  {https://ui.adsabs.harvard.edu/abs/2018NatAs...2..873A} {2, 873}

\bibitem[\protect\citeauthoryear{{Aller}, {Lillo-Box}, {Jones}, {Miranda}  \&
  {Barcel{\'o} Forteza}}{{Aller} et~al.}{2020}]{Aller2020}
{Aller} A.,  {Lillo-Box} J.,  {Jones} D.,  {Miranda} L.~F.,   {Barcel{\'o}
  Forteza} S.,  2020, \mn@doi [\aap] {10.1051/0004-6361/201937118}, \href
  {https://ui.adsabs.harvard.edu/abs/2020A&A...635A.128A} {635, A128}

\bibitem[\protect\citeauthoryear{{Almenara} et~al.,}{{Almenara}
  et~al.}{2022}]{dilution}
{Almenara} J.~M.,  et~al., 2022, Photodynamical analysis of the nearly resonant
  planetary system WASP-148 - Accurate transit-timing variations and mutual
  orbital inclination, \mn@doi{10.1051/0004-6361/202142964}, \url
  {https://doi.org/10.1051/0004-6361/202142964}

\bibitem[\protect\citeauthoryear{{Anderson} et~al.,}{{Anderson}
  et~al.}{2011}]{Anderson2011}
{Anderson} D.~R.,  et~al., 2011, \mn@doi [\apjl] {10.1088/2041-8205/726/2/L19},
  \href {https://ui.adsabs.harvard.edu/abs/2011ApJ...726L..19A} {726, L19}

\bibitem[\protect\citeauthoryear{Baglin, Auvergne, Barge, Deleuil  \&
  Michel}{Baglin et~al.}{2008}]{baglin2008}
Baglin A.,  Auvergne M.,  Barge P.,  Deleuil M.,   Michel E.,  2008, \mn@doi
  [Proceedings of the International Astronomical Union]
  {10.1017/S1743921308026252}, 4, 71–81

\bibitem[\protect\citeauthoryear{{Baluev}}{{Baluev}}{2008}]{baluev}
{Baluev} R.~V.,  2008, \mn@doi [\mnras] {10.1111/j.1365-2966.2008.12689.x},
  \href {https://ui.adsabs.harvard.edu/abs/2008MNRAS.385.1279B} {385, 1279}

\bibitem[\protect\citeauthoryear{{Baranne} et~al.,}{{Baranne}
  et~al.}{1996}]{Baranne1996}
{Baranne} A.,  et~al., 1996, \aaps, \href
  {https://ui.adsabs.harvard.edu/abs/1996A&AS..119..373B} {119, 373}

\bibitem[\protect\citeauthoryear{{Barnes}}{{Barnes}}{2007}]{Barnes2007}
{Barnes} S.~A.,  2007, \mn@doi [\apj] {10.1086/519295}, \href
  {https://ui.adsabs.harvard.edu/abs/2007ApJ...669.1167B} {669, 1167}

\bibitem[\protect\citeauthoryear{{Bertran de Lis}, {Delgado Mena}, {Adibekyan},
  {Santos}  \& {Sousa}}{{Bertran de Lis} et~al.}{2015}]{Bertrandelis-15}
{Bertran de Lis} S.,  {Delgado Mena} E.,  {Adibekyan} V.~Z.,  {Santos} N.~C.,
  {Sousa} S.~G.,  2015, \mn@doi [\aap] {10.1051/0004-6361/201424633}, \href
  {http://adsabs.harvard.edu/abs/2015A%26A...576A..89B} {576, A89}

\bibitem[\protect\citeauthoryear{{Bitsch}, {Morbidelli}, {Johansen}, {Lega},
  {Lambrechts}  \& {Crida}}{{Bitsch} et~al.}{2018}]{2018A&A...612A..30B}
{Bitsch} B.,  {Morbidelli} A.,  {Johansen} A.,  {Lega} E.,  {Lambrechts} M.,
  {Crida} A.,  2018, \mn@doi [\aap] {10.1051/0004-6361/201731931}, \href
  {https://ui.adsabs.harvard.edu/abs/2018A&A...612A..30B} {612, A30}

\bibitem[\protect\citeauthoryear{{Boisse}, {Bouchy}, {H{\'e}brard}, {Bonfils},
  {Santos}  \& {Vauclair}}{{Boisse} et~al.}{2011}]{Boisse2011}
{Boisse} I.,  {Bouchy} F.,  {H{\'e}brard} G.,  {Bonfils} X.,  {Santos} N.,
  {Vauclair} S.,  2011, \mn@doi [\aap] {10.1051/0004-6361/201014354}, \href
  {https://ui.adsabs.harvard.edu/abs/2011A&A...528A...4B} {528, A4}

\bibitem[\protect\citeauthoryear{{Borucki} et~al.,}{{Borucki}
  et~al.}{2010}]{Borucki2010}
{Borucki} W.~J.,  et~al., 2010, \mn@doi [Science] {10.1126/science.1185402},
  \href {https://ui.adsabs.harvard.edu/abs/2010Sci...327..977B} {327, 977}

\bibitem[\protect\citeauthoryear{{Brown} et~al.,}{{Brown}
  et~al.}{2013}]{Brown:2013}
{Brown} T.~M.,  et~al., 2013, \mn@doi [\pasp] {10.1086/673168}, \href
  {https://ui.adsabs.harvard.edu/abs/2013PASP..125.1031B} {125, 1031}

\bibitem[\protect\citeauthoryear{{Buchhave} et~al.,}{{Buchhave}
  et~al.}{2010}]{buchhave2010}
{Buchhave} L.~A.,  et~al., 2010, \mn@doi [\apj] {10.1088/0004-637X/720/2/1118},
  \href {https://ui.adsabs.harvard.edu/abs/2010ApJ...720.1118B} {720, 1118}

\bibitem[\protect\citeauthoryear{Buchhave et~al.,}{Buchhave
  et~al.}{2012}]{buchhave2012}
Buchhave L.~A.,  et~al., 2012, Nature, 486, 375

\bibitem[\protect\citeauthoryear{{Burrows}}{{Burrows}}{2014}]{2014PNAS..11112601B}
{Burrows} A.~S.,  2014, \mn@doi [Proceedings of the National Academy of
  Science] {10.1073/pnas.1304208111}, \href
  {https://ui.adsabs.harvard.edu/abs/2014PNAS..11112601B} {111, 12601}

\bibitem[\protect\citeauthoryear{{Butler}, {Marcy}, {Williams}, {McCarthy},
  {Dosanjh}  \& {Vogt}}{{Butler} et~al.}{1996}]{PFS_Butler1996}
{Butler} R.~P.,  {Marcy} G.~W.,  {Williams} E.,  {McCarthy} C.,  {Dosanjh} P.,
   {Vogt} S.~S.,  1996, \mn@doi [\pasp] {10.1086/133755}, \href
  {https://ui.adsabs.harvard.edu/abs/1996PASP..108..500B} {108, 500}

\bibitem[\protect\citeauthoryear{{Chabrier} \& {Debras}}{{Chabrier} \&
  {Debras}}{2021}]{2021ApJ...917....4C}
{Chabrier} G.,  {Debras} F.,  2021, \mn@doi [\apj] {10.3847/1538-4357/abfc48},
  \href {https://ui.adsabs.harvard.edu/abs/2021ApJ...917....4C} {917, 4}

\bibitem[\protect\citeauthoryear{{Chen} \& {Rogers}}{{Chen} \&
  {Rogers}}{2016}]{ChenRogers16:envelope-model}
{Chen} H.,  {Rogers} L.~A.,  2016, \mn@doi [\apj]
  {10.3847/0004-637X/831/2/180}, \href
  {https://ui.adsabs.harvard.edu/abs/2016ApJ...831..180C} {831, 180}

\bibitem[\protect\citeauthoryear{{Choi}, {Dotter}, {Conroy}, {Cantiello},
  {Paxton}  \& {Johnson}}{{Choi} et~al.}{2016}]{Choi16:mist-models}
{Choi} J.,  {Dotter} A.,  {Conroy} C.,  {Cantiello} M.,  {Paxton} B.,
  {Johnson} B.~D.,  2016, \mn@doi [\apj] {10.3847/0004-637X/823/2/102}, \href
  {https://ui.adsabs.harvard.edu/abs/2016ApJ...823..102C} {823, 102}

\bibitem[\protect\citeauthoryear{{Collins}, {Kielkopf}, {Stassun}  \&
  {Hessman}}{{Collins} et~al.}{2017}]{Collins:2017}
{Collins} K.~A.,  {Kielkopf} J.~F.,  {Stassun} K.~G.,   {Hessman} F.~V.,  2017,
  \mn@doi [\aj] {10.3847/1538-3881/153/2/77}, \href
  {http://adsabs.harvard.edu/abs/2017AJ....153...77C} {153, 77}

\bibitem[\protect\citeauthoryear{{Collins}, {Quinn}, {Latham}, {Christiansen},
  {Ciardi}, {Dragomir}, {Crossfield}  \& {Seager}}{{Collins}
  et~al.}{2018}]{collins:2018}
{Collins} K.,  {Quinn} S.~N.,  {Latham} D.~W.,  {Christiansen} J.,  {Ciardi}
  D.,  {Dragomir} D.,  {Crossfield} I.,   {Seager} S.,  2018, in American
  Astronomical Society Meeting Abstracts \#231. p. 439.08

\bibitem[\protect\citeauthoryear{{Crane}, {Shectman}  \& {Butler}}{{Crane}
  et~al.}{2006}]{PFS_Crane2006}
{Crane} J.~D.,  {Shectman} S.~A.,   {Butler} R.~P.,  2006, in {McLean} I.~S.,
  {Iye} M.,  eds,  Society of Photo-Optical Instrumentation Engineers (SPIE)
  Conference Series Vol. 6269, Society of Photo-Optical Instrumentation
  Engineers (SPIE) Conference Series. p. 626931, \mn@doi{10.1117/12.672339}

\bibitem[\protect\citeauthoryear{{Crane}, {Shectman}, {Butler}, {Thompson}  \&
  {Burley}}{{Crane} et~al.}{2008}]{PFS_Crane2008}
{Crane} J.~D.,  {Shectman} S.~A.,  {Butler} R.~P.,  {Thompson} I.~B.,
  {Burley} G.~S.,  2008, in {McLean} I.~S.,  {Casali} M.~M.,  eds,  Society of
  Photo-Optical Instrumentation Engineers (SPIE) Conference Series Vol. 7014,
  Ground-based and Airborne Instrumentation for Astronomy II. p. 701479,
  \mn@doi{10.1117/12.789637}

\bibitem[\protect\citeauthoryear{{Crane}, {Shectman}, {Butler}, {Thompson},
  {Birk}, {Jones}  \& {Burley}}{{Crane} et~al.}{2010}]{PFS_Crane2010}
{Crane} J.~D.,  {Shectman} S.~A.,  {Butler} R.~P.,  {Thompson} I.~B.,  {Birk}
  C.,  {Jones} P.,   {Burley} G.~S.,  2010, in {McLean} I.~S.,  {Ramsay} S.~K.,
    {Takami} H.,  eds,  Society of Photo-Optical Instrumentation Engineers
  (SPIE) Conference Series Vol. 7735, Ground-based and Airborne Instrumentation
  for Astronomy III. p. 773553, \mn@doi{10.1117/12.857792}

\bibitem[\protect\citeauthoryear{Daemgen, Hormuth, Brandner, Bergfors, Janson,
  Hippler  \& Henning}{Daemgen et~al.}{2009}]{daemgen2009}
Daemgen S.,  Hormuth F.,  Brandner W.,  Bergfors C.,  Janson M.,  Hippler S.,
  Henning T.,  2009, \mn@doi [Astronomy and Astrophysics, v.498, 567-574
  (2009)] {10.1051/0004-6361/200810988}, 498

\bibitem[\protect\citeauthoryear{{Deeg} et~al.,}{{Deeg}
  et~al.}{2023}]{Deeg2023}
{Deeg} H.~J.,  et~al., 2023, \mn@doi [\aap] {10.1051/0004-6361/202346370},
  \href {https://ui.adsabs.harvard.edu/abs/2023A&A...677A..12D} {677, A12}

\bibitem[\protect\citeauthoryear{{Delgado Mena} et~al.,}{{Delgado Mena}
  et~al.}{2014}]{Delgado-14}
{Delgado Mena} E.,  et~al., 2014, \mn@doi [\aap] {10.1051/0004-6361/201321493},
  \href {http://adsabs.harvard.edu/abs/2014A%26A...562A..92D} {562, A92}

\bibitem[\protect\citeauthoryear{{Delgado Mena}, {Tsantaki}, {Adibekyan},
  {Sousa}, {Santos}, {Gonz{\'a}lez Hern{\'a}ndez}  \& {Israelian}}{{Delgado
  Mena} et~al.}{2017}]{Delgado-17}
{Delgado Mena} E.,  {Tsantaki} M.,  {Adibekyan} V.~Z.,  {Sousa} S.~G.,
  {Santos} N.~C.,  {Gonz{\'a}lez Hern{\'a}ndez} J.~I.,   {Israelian} G.,  2017,
  \mn@doi [\aap] {10.1051/0004-6361/201730535}, \href
  {http://adsabs.harvard.edu/abs/2017A%26A...606A..94D} {606, A94}

\bibitem[\protect\citeauthoryear{{Delgado Mena} et~al.,}{{Delgado Mena}
  et~al.}{2019}]{Delgado-19}
{Delgado Mena} E.,  et~al., 2019, \mn@doi [\aap] {10.1051/0004-6361/201834783},
  \href {https://ui.adsabs.harvard.edu/abs/2019A&A...624A..78D} {624, A78}

\bibitem[\protect\citeauthoryear{{Delgado Mena}, {Adibekyan}, {Santos},
  {Tsantaki}, {Gonz{\'a}lez Hern{\'a}ndez}, {Sousa}  \& {Bertr{\'a}n de
  Lis}}{{Delgado Mena} et~al.}{2021}]{Delgado-21}
{Delgado Mena} E.,  {Adibekyan} V.,  {Santos} N.~C.,  {Tsantaki} M.,
  {Gonz{\'a}lez Hern{\'a}ndez} J.~I.,  {Sousa} S.~G.,   {Bertr{\'a}n de Lis}
  S.,  2021, \mn@doi [\aap] {10.1051/0004-6361/202141588}, \href
  {https://ui.adsabs.harvard.edu/abs/2021A&A...655A..99D} {655, A99}

\bibitem[\protect\citeauthoryear{Dong, Xie, Zhou, Zheng  \& Luo}{Dong
  et~al.}{2018}]{dong_2018}
Dong S.,  Xie J.-W.,  Zhou J.-L.,  Zheng Z.,   Luo A.,  2018, \mn@doi
  [Proceedings of the National Academy of Sciences] {10.1073/pnas.1711406115},
  115, 266

\bibitem[\protect\citeauthoryear{Dormand \& Prince}{Dormand \&
  Prince}{1980}]{Dormand80:rk45}
Dormand J.,  Prince P.,  1980, \mn@doi [Journal of Computational and Applied
  Mathematics] {https://doi.org/10.1016/0771-050X(80)90013-3}, 6, 19

\bibitem[\protect\citeauthoryear{{Doyle}, {Davies}, {Smalley}, {Chaplin}  \&
  {Elsworth}}{{Doyle} et~al.}{2014}]{Doyle2014}
{Doyle} A.~P.,  {Davies} G.~R.,  {Smalley} B.,  {Chaplin} W.~J.,   {Elsworth}
  Y.,  2014, \mn@doi [\mnras] {10.1093/mnras/stu1692}, \href
  {https://ui.adsabs.harvard.edu/abs/2014MNRAS.444.3592D} {444, 3592}

\bibitem[\protect\citeauthoryear{F\H{u}r\'esz}{F\H{u}r\'esz}{2008}]{gaborthesis}
F\H{u}r\'esz G.,  2008, PhD thesis, University of Szeged, Hungary

\bibitem[\protect\citeauthoryear{{Fern{\'a}ndez Fern{\'a}ndez}, {Wheatley}  \&
  {King}}{{Fern{\'a}ndez Fern{\'a}ndez} et~al.}{2023}]{Fernandez23:evap}
{Fern{\'a}ndez Fern{\'a}ndez} J.,  {Wheatley} P.~J.,   {King} G.~W.,  2023,
  \mn@doi [\mnras] {10.1093/mnras/stad1257}, \href
  {https://ui.adsabs.harvard.edu/abs/2023MNRAS.522.4251F} {522, 4251}

\bibitem[\protect\citeauthoryear{Foreman-Mackey, Czekala, Luger, Agol,
  Barentsen  \& Barclay}{Foreman-Mackey et~al.}{2020}]{exoplanet:exoplanet}
Foreman-Mackey D.,  Czekala I.,  Luger R.,  Agol E.,  Barentsen G.,   Barclay
  T.,  2020, exoplanet-dev/exoplanet v0.2.6, \mn@doi{10.5281/zenodo.1998447},
  \url {https://doi.org/10.5281/zenodo.1998447}

\bibitem[\protect\citeauthoryear{{Freedman}, {Lustig-Yaeger}, {Fortney},
  {Lupu}, {Marley}  \& {Lodders}}{{Freedman} et~al.}{2014}]{Freedman2014}
{Freedman} R.~S.,  {Lustig-Yaeger} J.,  {Fortney} J.~J.,  {Lupu} R.~E.,
  {Marley} M.~S.,   {Lodders} K.,  2014, \mn@doi [\apjs]
  {10.1088/0067-0049/214/2/25}, \href
  {https://ui.adsabs.harvard.edu/abs/2014ApJS..214...25F} {214, 25}

\bibitem[\protect\citeauthoryear{{Gaia Collaboration} et~al.,}{{Gaia
  Collaboration} et~al.}{2023}]{GaiaDR3}
{Gaia Collaboration} et~al., 2023, \mn@doi [\aap]
  {10.1051/0004-6361/202243940}, \href
  {https://ui.adsabs.harvard.edu/abs/2023A&A...674A...1G} {674, A1}

\bibitem[\protect\citeauthoryear{Gardner et~al.,}{Gardner
  et~al.}{2006}]{Gardner2006}
Gardner J.~P.,  et~al., 2006, \mn@doi [Space Science Reviews]
  {10.1007/s11214-006-8315-7}, 123, 485–606

\bibitem[\protect\citeauthoryear{{Ginzburg} \& {Chiang}}{{Ginzburg} \&
  {Chiang}}{2020}]{2020MNRAS.498..680G}
{Ginzburg} S.,  {Chiang} E.,  2020, \mn@doi [\mnras] {10.1093/mnras/staa2500},
  \href {https://ui.adsabs.harvard.edu/abs/2020MNRAS.498..680G} {498, 680}

\bibitem[\protect\citeauthoryear{{Guerrero} et~al.,}{{Guerrero}
  et~al.}{2021}]{guerrero2021}
{Guerrero} N.~M.,  et~al., 2021, \mn@doi [\apjs] {10.3847/1538-4365/abefe1},
  \href {https://ui.adsabs.harvard.edu/abs/2021ApJS..254...39G} {254, 39}

\bibitem[\protect\citeauthoryear{{Guillot}}{{Guillot}}{2010}]{2010A&A...520A..27G}
{Guillot} T.,  2010, \mn@doi [\aap] {10.1051/0004-6361/200913396}, \href
  {https://ui.adsabs.harvard.edu/abs/2010A&A...520A..27G} {520, A27}

\bibitem[\protect\citeauthoryear{{Guillot}, {Santos}, {Pont}, {Iro}, {Melo}  \&
  {Ribas}}{{Guillot} et~al.}{2006}]{2006A&A...453L..21G}
{Guillot} T.,  {Santos} N.~C.,  {Pont} F.,  {Iro} N.,  {Melo} C.,   {Ribas} I.,
   2006, \mn@doi [\aap] {10.1051/0004-6361:20065476}, \href
  {https://ui.adsabs.harvard.edu/abs/2006A&A...453L..21G} {453, L21}

\bibitem[\protect\citeauthoryear{Hartman et~al.,}{Hartman
  et~al.}{2019}]{Hartman2019}
Hartman J.~D.,  et~al., 2019, \mn@doi [The Astronomical Journal]
  {10.3847/1538-3881/aaf8b6}, 157, 55

\bibitem[\protect\citeauthoryear{{Hasegawa}, {Bryden}, {Ikoma}, {Vasisht}  \&
  {Swain}}{{Hasegawa} et~al.}{2018}]{2018ApJ...865...32H}
{Hasegawa} Y.,  {Bryden} G.,  {Ikoma} M.,  {Vasisht} G.,   {Swain} M.,  2018,
  \mn@doi [\apj] {10.3847/1538-4357/aad912}, \href
  {https://ui.adsabs.harvard.edu/abs/2018ApJ...865...32H} {865, 32}

\bibitem[\protect\citeauthoryear{{Helled}}{{Helled}}{2023}]{2023A&A...675L...8H}
{Helled} R.,  2023, \mn@doi [\aap] {10.1051/0004-6361/202346850}, \href
  {https://ui.adsabs.harvard.edu/abs/2023A&A...675L...8H} {675, L8}

\bibitem[\protect\citeauthoryear{{Helled} et~al.,}{{Helled}
  et~al.}{2014}]{2014prpl.conf..643H}
{Helled} R.,  et~al., 2014, in {Beuther} H.,  {Klessen} R.~S.,  {Dullemond}
  C.~P.,   {Henning} T.,  eds, Protostars and Planets VI. pp 643--665
  (\mn@eprint {arXiv} {1311.1142}),
  \mn@doi{10.2458/azu_uapress_9780816531240-ch028}

\bibitem[\protect\citeauthoryear{{Helled} et~al.,}{{Helled}
  et~al.}{2022a}]{2022ExA....53..323H}
{Helled} R.,  et~al., 2022a, \mn@doi [Experimental Astronomy]
  {10.1007/s10686-021-09739-3}, \href
  {https://ui.adsabs.harvard.edu/abs/2022ExA....53..323H} {53, 323}

\bibitem[\protect\citeauthoryear{{Helled} et~al.,}{{Helled}
  et~al.}{2022b}]{2022Icar..37814937H}
{Helled} R.,  et~al., 2022b, \mn@doi [\icarus] {10.1016/j.icarus.2022.114937},
  \href {https://ui.adsabs.harvard.edu/abs/2022Icar..37814937H} {378, 114937}

\bibitem[\protect\citeauthoryear{{Howard} \& {Guillot}}{{Howard} \&
  {Guillot}}{2023}]{2023A&A...672L...1H}
{Howard} S.,  {Guillot} T.,  2023, \mn@doi [\aap]
  {10.1051/0004-6361/202244851}, \href
  {https://ui.adsabs.harvard.edu/abs/2023A&A...672L...1H} {672, L1}

\bibitem[\protect\citeauthoryear{{Hoyer} et~al.,}{{Hoyer}
  et~al.}{2023}]{Hoyer23:LTT-9779-metal-rich}
{Hoyer} S.,  et~al., 2023, \mn@doi [\aap] {10.1051/0004-6361/202346117}, \href
  {https://ui.adsabs.harvard.edu/abs/2023A&A...675A..81H} {675, A81}

\bibitem[\protect\citeauthoryear{{Huang} et~al.,}{{Huang}
  et~al.}{2020}]{Huang2020}
{Huang} C.~X.,  et~al., 2020, \mn@doi [Research Notes of the American
  Astronomical Society] {10.3847/2515-5172/abca2e}, \href
  {https://ui.adsabs.harvard.edu/abs/2020RNAAS...4..204H} {4, 204}

\bibitem[\protect\citeauthoryear{{Husser, T.-O.}, {Wende-von Berg, S.},
  {Dreizler, S.}, {Homeier, D.}, {Reiners, A.}, {Barman, T.}  \& {Hauschildt,
  P. H.}}{{Husser, T.-O.} et~al.}{2013}]{husser2013}
{Husser, T.-O.} {Wende-von Berg, S.} {Dreizler, S.} {Homeier, D.} {Reiners, A.}
  {Barman, T.}  {Hauschildt, P. H.} 2013, \mn@doi [A\&A]
  {10.1051/0004-6361/201219058}, 553, A6

\bibitem[\protect\citeauthoryear{{Jenkins}}{{Jenkins}}{2002}]{Jenkins2002}
{Jenkins} J.~M.,  2002, \mn@doi [\apj] {10.1086/341136}, \href
  {http://adsabs.harvard.edu/abs/2002ApJ...575..493J} {575, 493}

\bibitem[\protect\citeauthoryear{{Jenkins} et~al.,}{{Jenkins}
  et~al.}{2010}]{jenkins2010}
{Jenkins} J.~M.,  et~al., 2010, in {Radziwill} N.~M.,  {Bridger} A.,  eds,
  Society of Photo-Optical Instrumentation Engineers (SPIE) Conference Series
  Vol. 7740, Software and Cyberinfrastructure for Astronomy. p. 77400D,
  \mn@doi{10.1117/12.856764}

\bibitem[\protect\citeauthoryear{{Jenkins}, {Tenenbaum}, {Seader}, {Burke},
  {McCauliff}, {Smith}, {Twicken}  \& {Chandrasekaran}}{{Jenkins}
  et~al.}{2020}]{jenkins_2020}
{Jenkins} J.~M.,  {Tenenbaum} P.,  {Seader} S.,  {Burke} C.~J.,  {McCauliff}
  S.~D.,  {Smith} J.~C.,  {Twicken} J.~D.,   {Chandrasekaran} H.,  2020,
  {Kepler Data Processing Handbook: Transiting Planet Search}, Kepler Science
  Document KSCI-19081-003

\bibitem[\protect\citeauthoryear{{Jensen}}{{Jensen}}{2013}]{Jensen:2013}
{Jensen} E.,  2013, {Tapir: A web interface for transit/eclipse observability},
  Astrophysics Source Code Library (\mn@eprint {ascl} {1306.007})

\bibitem[\protect\citeauthoryear{{Jermyn} et~al.,}{{Jermyn}
  et~al.}{2023}]{2023ApJS..265...15J}
{Jermyn} A.~S.,  et~al., 2023, \mn@doi [\apjs] {10.3847/1538-4365/acae8d},
  \href {https://ui.adsabs.harvard.edu/abs/2023ApJS..265...15J} {265, 15}

\bibitem[\protect\citeauthoryear{{Johansen} \& {Lambrechts}}{{Johansen} \&
  {Lambrechts}}{2017}]{2017AREPS..45..359J}
{Johansen} A.,  {Lambrechts} M.,  2017, \mn@doi [Annual Review of Earth and
  Planetary Sciences] {10.1146/annurev-earth-063016-020226}, \href
  {https://ui.adsabs.harvard.edu/abs/2017AREPS..45..359J} {45, 359}

\bibitem[\protect\citeauthoryear{{Johnstone}, {Bartel}  \&
  {G{\"u}del}}{{Johnstone} et~al.}{2021}]{Johnstone21:rotation-model}
{Johnstone} C.~P.,  {Bartel} M.,   {G{\"u}del} M.,  2021, \mn@doi [\aap]
  {10.1051/0004-6361/202038407}, \href
  {https://ui.adsabs.harvard.edu/abs/2021A&A...649A..96J} {649, A96}

\bibitem[\protect\citeauthoryear{Kempton et~al.,}{Kempton
  et~al.}{2018}]{Kempton_2018}
Kempton E. M.-R.,  et~al., 2018, \mn@doi [Publications of the Astronomical
  Society of the Pacific] {10.1088/1538-3873/aadf6f}, 130, 114401

\bibitem[\protect\citeauthoryear{{King} \& {Wheatley}}{{King} \&
  {Wheatley}}{2021}]{King21:euv-gyr}
{King} G.~W.,  {Wheatley} P.~J.,  2021, \mn@doi [\mnras]
  {10.1093/mnrasl/slaa186}, \href
  {https://ui.adsabs.harvard.edu/abs/2021MNRAS.501L..28K} {501, L28}

\bibitem[\protect\citeauthoryear{{Kipping}}{{Kipping}}{2013}]{exoplanet:kipping13}
{Kipping} D.~M.,  2013, \mn@doi [mnras] {10.1093/mnras/stt1435}, \href
  {http://adsabs.harvard.edu/abs/2013MNRAS.435.2152K} {435, 2152}

\bibitem[\protect\citeauthoryear{{Kubyshkina} et~al.,}{{Kubyshkina}
  et~al.}{2018}]{Kubyshkina18:mass-loss-model}
{Kubyshkina} D.,  et~al., 2018, \mn@doi [\apjl] {10.3847/2041-8213/aae586},
  \href {https://ui.adsabs.harvard.edu/abs/2018ApJ...866L..18K} {866, L18}

\bibitem[\protect\citeauthoryear{Kurucz}{Kurucz}{1992}]{kurucz1992}
Kurucz R.~L.,  1992, in Barbuy B.,  Renzini A.,  eds, The Stellar Populations
  of Galaxies. Springer Netherlands, Dordrecht, pp 225--232

\bibitem[\protect\citeauthoryear{{Kurucz}}{{Kurucz}}{1993}]{Kurucz-93}
{Kurucz} R.~L.,  1993, {SYNTHE spectrum synthesis programs and line data}

\bibitem[\protect\citeauthoryear{{Li} et~al.,}{{Li}
  et~al.}{2018}]{2018NatCo...9.3709L}
{Li} L.,  et~al., 2018, \mn@doi [Nature Communications]
  {10.1038/s41467-018-06107-2}, \href
  {https://ui.adsabs.harvard.edu/abs/2018NatCo...9.3709L} {9, 3709}

\bibitem[\protect\citeauthoryear{{Li}, {Tenenbaum}, {Twicken}, {Burke},
  {Jenkins}, {Quintana}, {Rowe}  \& {Seader}}{{Li} et~al.}{2019}]{Li2019}
{Li} J.,  {Tenenbaum} P.,  {Twicken} J.~D.,  {Burke} C.~J.,  {Jenkins} J.~M.,
  {Quintana} E.~V.,  {Rowe} J.~F.,   {Seader} S.~E.,  2019, \mn@doi [\pasp]
  {10.1088/1538-3873/aaf44d}, \href
  {https://ui.adsabs.harvard.edu/abs/2019PASP..131b4506L} {131, 024506}

\bibitem[\protect\citeauthoryear{{Lightkurve Collaboration}
  et~al.,}{{Lightkurve Collaboration} et~al.}{2018}]{lightkurve}
{Lightkurve Collaboration} et~al., 2018, {Lightkurve: Kepler and TESS time
  series analysis in Python}, Astrophysics Source Code Library, record
  ascl:1812.013 (\mn@eprint {ascl} {1812.013})

\bibitem[\protect\citeauthoryear{Lillo-Box, Barrado  \& Bouy}{Lillo-Box
  et~al.}{2012}]{LilloBox2012}
Lillo-Box J.,  Barrado D.,   Bouy H.,  2012, \mn@doi [Astronomy &amp;
  Astrophysics] {10.1051/0004-6361/201219631}, 546, A10

\bibitem[\protect\citeauthoryear{{Luger}, {Agol}, {Foreman-Mackey}, {Fleming},
  {Lustig-Yaeger}  \& {Deitrick}}{{Luger} et~al.}{2019}]{exoplanet:luger18}
{Luger} R.,  {Agol} E.,  {Foreman-Mackey} D.,  {Fleming} D.~P.,
  {Lustig-Yaeger} J.,   {Deitrick} R.,  2019, \mn@doi [aj]
  {10.3847/1538-3881/aae8e5}, \href
  {http://adsabs.harvard.edu/abs/2019AJ....157...64L} {157, 64}

\bibitem[\protect\citeauthoryear{{Mayor} \& {Queloz}}{{Mayor} \&
  {Queloz}}{1995}]{mayorqueloz}
{Mayor} M.,  {Queloz} D.,  1995, \mn@doi [\nat] {10.1038/378355a0}, \href
  {https://ui.adsabs.harvard.edu/abs/1995Natur.378..355M} {378, 355}

\bibitem[\protect\citeauthoryear{Mayor et~al.,}{Mayor et~al.}{2003a}]{harps}
Mayor M.,  et~al., 2003a, The Messenger, 114, 20

\bibitem[\protect\citeauthoryear{{Mayor} et~al.,}{{Mayor}
  et~al.}{2003b}]{mayor2003}
{Mayor} M.,  et~al., 2003b, The Messenger, \href
  {https://ui.adsabs.harvard.edu/abs/2003Msngr.114...20M} {114, 20}

\bibitem[\protect\citeauthoryear{{Mazeh}, {Holczer}  \& {Faigler}}{{Mazeh}
  et~al.}{2016}]{Mazeh2016}
{Mazeh} T.,  {Holczer} T.,   {Faigler} S.,  2016, \mn@doi [\aap]
  {10.1051/0004-6361/201528065}, \href
  {https://ui.adsabs.harvard.edu/abs/2016A&A...589A..75M} {589, A75}

\bibitem[\protect\citeauthoryear{{McCully}, {Volgenau}, {Harbeck}, {Lister},
  {Saunders}, {Turner}, {Siiverd}  \& {Bowman}}{{McCully}
  et~al.}{2018}]{McCully:2018}
{McCully} C.,  {Volgenau} N.~H.,  {Harbeck} D.-R.,  {Lister} T.~A.,  {Saunders}
  E.~S.,  {Turner} M.~L.,  {Siiverd} R.~J.,   {Bowman} M.,  2018, in \procspie.
  p. 107070K (\mn@eprint {arXiv} {1811.04163}), \mn@doi{10.1117/12.2314340}

\bibitem[\protect\citeauthoryear{{Miller} \& {Fortney}}{{Miller} \&
  {Fortney}}{2011}]{2011ApJ...736L..29M}
{Miller} N.,  {Fortney} J.~J.,  2011, \mn@doi [\apjl]
  {10.1088/2041-8205/736/2/L29}, \href
  {https://ui.adsabs.harvard.edu/abs/2011ApJ...736L..29M} {736, L29}

\bibitem[\protect\citeauthoryear{{Mousis}, {Marboeuf}, {Lunine}, {Alibert},
  {Fletcher}, {Orton}, {Pauzat}  \& {Ellinger}}{{Mousis}
  et~al.}{2009}]{2009ApJ...696.1348M}
{Mousis} O.,  {Marboeuf} U.,  {Lunine} J.~I.,  {Alibert} Y.,  {Fletcher} L.~N.,
   {Orton} G.~S.,  {Pauzat} F.,   {Ellinger} Y.,  2009, \mn@doi [\apj]
  {10.1088/0004-637X/696/2/1348}, \href
  {https://ui.adsabs.harvard.edu/abs/2009ApJ...696.1348M} {696, 1348}

\bibitem[\protect\citeauthoryear{Mulders, Pascucci, Apai, Frasca  \&
  Molenda-Żakowicz}{Mulders et~al.}{2016}]{Mulders_2016}
Mulders G.~D.,  Pascucci I.,  Apai D.,  Frasca A.,   Molenda-Żakowicz J.,
  2016, \mn@doi [The Astronomical Journal] {10.3847/0004-6256/152/6/187}, 152,
  187

\bibitem[\protect\citeauthoryear{{M{\"u}ller} \& {Helled}}{{M{\"u}ller} \&
  {Helled}}{2021}]{2021MNRAS.507.2094M}
{M{\"u}ller} S.,  {Helled} R.,  2021, \mn@doi [\mnras]
  {10.1093/mnras/stab2250}, \href
  {https://ui.adsabs.harvard.edu/abs/2021MNRAS.507.2094M} {507, 2094}

\bibitem[\protect\citeauthoryear{{M{\"u}ller} \& {Helled}}{{M{\"u}ller} \&
  {Helled}}{2023a}]{2023FrASS..1079000M}
{M{\"u}ller} S.,  {Helled} R.,  2023a, \mn@doi [Frontiers in Astronomy and
  Space Sciences] {10.3389/fspas.2023.1179000}, \href
  {https://ui.adsabs.harvard.edu/abs/2023FrASS..1079000M} {10, 1179000}

\bibitem[\protect\citeauthoryear{{M{\"u}ller} \& {Helled}}{{M{\"u}ller} \&
  {Helled}}{2023b}]{2023A&A...669A..24M}
{M{\"u}ller} S.,  {Helled} R.,  2023b, \mn@doi [\aap]
  {10.1051/0004-6361/202244827}, \href
  {https://ui.adsabs.harvard.edu/abs/2023A&A...669A..24M} {669, A24}

\bibitem[\protect\citeauthoryear{{M{\"u}ller}, {Helled}  \&
  {Cumming}}{{M{\"u}ller} et~al.}{2020a}]{2020A&A...638A.121M}
{M{\"u}ller} S.,  {Helled} R.,   {Cumming} A.,  2020a, \mn@doi [\aap]
  {10.1051/0004-6361/201937376}, \href
  {https://ui.adsabs.harvard.edu/abs/2020A&A...638A.121M} {638, A121}

\bibitem[\protect\citeauthoryear{{M{\"u}ller}, {Ben-Yami}  \&
  {Helled}}{{M{\"u}ller} et~al.}{2020b}]{2020ApJ...903..147M}
{M{\"u}ller} S.,  {Ben-Yami} M.,   {Helled} R.,  2020b, \mn@doi [\apj]
  {10.3847/1538-4357/abba19}, \href
  {https://ui.adsabs.harvard.edu/abs/2020ApJ...903..147M} {903, 147}

\bibitem[\protect\citeauthoryear{Müller \& Helled}{Müller \&
  Helled}{2024}]{2024arXiv240316273M}
Müller S.,  Helled R.,  2024, Can Jupiter's atmospheric metallicity be
  different from the deep interior? (\mn@eprint {arXiv} {2403.16273})

\bibitem[\protect\citeauthoryear{{Noyes}, {Weiss}  \& {Vaughan}}{{Noyes}
  et~al.}{1984}]{Noyes1984}
{Noyes} R.~W.,  {Weiss} N.~O.,   {Vaughan} A.~H.,  1984, \mn@doi [\apj]
  {10.1086/162735}, \href
  {https://ui.adsabs.harvard.edu/abs/1984ApJ...287..769N} {287, 769}

\bibitem[\protect\citeauthoryear{{Otegi}, {Bouchy}  \& {Helled}}{{Otegi}
  et~al.}{2020}]{Otegi2020}
{Otegi} J.~F.,  {Bouchy} F.,   {Helled} R.,  2020, \mn@doi [\aap]
  {10.1051/0004-6361/201936482}, \href
  {https://ui.adsabs.harvard.edu/abs/2020A&A...634A..43O} {634, A43}

\bibitem[\protect\citeauthoryear{{Owen} \& {Jackson}}{{Owen} \&
  {Jackson}}{2012}]{OwenJackson12:xray-evap}
{Owen} J.~E.,  {Jackson} A.~P.,  2012, \mn@doi [\mnras]
  {10.1111/j.1365-2966.2012.21481.x}, \href
  {https://ui.adsabs.harvard.edu/abs/2012MNRAS.425.2931O} {425, 2931}

\bibitem[\protect\citeauthoryear{{Owen} \& {Lai}}{{Owen} \&
  {Lai}}{2018}]{OwenLai2018}
{Owen} J.~E.,  {Lai} D.,  2018, \mn@doi [\mnras] {10.1093/mnras/sty1760}, \href
  {https://ui.adsabs.harvard.edu/abs/2018MNRAS.479.5012O} {479, 5012}

\bibitem[\protect\citeauthoryear{{Owen} \& {Murray-Clay}}{{Owen} \&
  {Murray-Clay}}{2018}]{OwenMurrayClay18}
{Owen} J.~E.,  {Murray-Clay} R.,  2018, \mn@doi [\mnras]
  {10.1093/mnras/sty1943}, \href
  {https://ui.adsabs.harvard.edu/abs/2018MNRAS.480.2206O} {480, 2206}

\bibitem[\protect\citeauthoryear{{Owen} \& {Wu}}{{Owen} \&
  {Wu}}{2017}]{OwenWu2017}
{Owen} J.~E.,  {Wu} Y.,  2017, \mn@doi [\apj] {10.3847/1538-4357/aa890a}, \href
  {https://ui.adsabs.harvard.edu/abs/2017ApJ...847...29O} {847, 29}

\bibitem[\protect\citeauthoryear{Parmentier \& Crossfield}{Parmentier \&
  Crossfield}{2018}]{Parmentier2018}
Parmentier V.,  Crossfield I. J.~M.,  2018, Exoplanet Phase Curves:
  Observations and Theory.
Springer International Publishing, Cham, pp 1419--1440,
  \mn@doi{10.1007/978-3-319-55333-7_116}, \url
  {https://doi.org/10.1007/978-3-319-55333-7_116}

\bibitem[\protect\citeauthoryear{{Parmentier} \& {Guillot}}{{Parmentier} \&
  {Guillot}}{2014}]{2014A&A...562A.133P}
{Parmentier} V.,  {Guillot} T.,  2014, \mn@doi [\aap]
  {10.1051/0004-6361/201322342}, \href
  {https://ui.adsabs.harvard.edu/abs/2014A&A...562A.133P} {562, A133}

\bibitem[\protect\citeauthoryear{{Paxton}, {Bildsten}, {Dotter}, {Herwig},
  {Lesaffre}  \& {Timmes}}{{Paxton} et~al.}{2011}]{2011ApJS..192....3P}
{Paxton} B.,  {Bildsten} L.,  {Dotter} A.,  {Herwig} F.,  {Lesaffre} P.,
  {Timmes} F.,  2011, \mn@doi [\apjs] {10.1088/0067-0049/192/1/3}, \href
  {https://ui.adsabs.harvard.edu/abs/2011ApJS..192....3P} {192, 3}

\bibitem[\protect\citeauthoryear{{Paxton} et~al.,}{{Paxton}
  et~al.}{2013}]{2013ApJS..208....4P}
{Paxton} B.,  et~al., 2013, \mn@doi [\apjs] {10.1088/0067-0049/208/1/4}, \href
  {https://ui.adsabs.harvard.edu/abs/2013ApJS..208....4P} {208, 4}

\bibitem[\protect\citeauthoryear{{Paxton} et~al.,}{{Paxton}
  et~al.}{2015}]{2015ApJS..220...15P}
{Paxton} B.,  et~al., 2015, \mn@doi [\apjs] {10.1088/0067-0049/220/1/15}, \href
  {https://ui.adsabs.harvard.edu/abs/2015ApJS..220...15P} {220, 15}

\bibitem[\protect\citeauthoryear{{Paxton} et~al.,}{{Paxton}
  et~al.}{2018}]{2018ApJS..234...34P}
{Paxton} B.,  et~al., 2018, \mn@doi [\apjs] {10.3847/1538-4365/aaa5a8}, \href
  {https://ui.adsabs.harvard.edu/abs/2018ApJS..234...34P} {234, 34}

\bibitem[\protect\citeauthoryear{{Paxton} et~al.,}{{Paxton}
  et~al.}{2019}]{2019ApJS..243...10P}
{Paxton} B.,  et~al., 2019, \mn@doi [\apjs] {10.3847/1538-4365/ab2241}, \href
  {https://ui.adsabs.harvard.edu/abs/2019ApJS..243...10P} {243, 10}

\bibitem[\protect\citeauthoryear{{Pepe} et~al.,}{{Pepe}
  et~al.}{2002}]{Pepe2002}
{Pepe} F.,  et~al., 2002, The Messenger, \href
  {https://ui.adsabs.harvard.edu/abs/2002Msngr.110....9P} {110, 9}

\bibitem[\protect\citeauthoryear{{Pepe} et~al.,}{{Pepe}
  et~al.}{2021}]{Pepe2021}
{Pepe} F.,  et~al., 2021, \mn@doi [\aap] {10.1051/0004-6361/202038306}, \href
  {https://ui.adsabs.harvard.edu/abs/2021A&A...645A..96P} {645, A96}

\bibitem[\protect\citeauthoryear{Petigura et~al.,}{Petigura
  et~al.}{2018}]{Petigura_2018}
Petigura E.~A.,  et~al., 2018, \mn@doi [The Astronomical Journal]
  {10.3847/1538-3881/aaa54c}, 155, 89

\bibitem[\protect\citeauthoryear{{Pizzolato}, {Maggio}, {Micela}, {Sciortino}
  \& {Ventura}}{{Pizzolato} et~al.}{2003}]{Pizzolato03:xray-rotation}
{Pizzolato} N.,  {Maggio} A.,  {Micela} G.,  {Sciortino} S.,   {Ventura} P.,
  2003, \mn@doi [\aap] {10.1051/0004-6361:20021560}, \href
  {https://ui.adsabs.harvard.edu/abs/2003A&A...397..147P} {397, 147}

\bibitem[\protect\citeauthoryear{{Pollacco} et~al.,}{{Pollacco}
  et~al.}{2006}]{Pollacco2006}
{Pollacco} D.~L.,  et~al., 2006, \mn@doi [\pasp] {10.1086/508556}, \href
  {https://ui.adsabs.harvard.edu/abs/2006PASP..118.1407P} {118, 1407}

\bibitem[\protect\citeauthoryear{{Pollack}, {Hubickyj}, {Bodenheimer},
  {Lissauer}, {Podolak}  \& {Greenzweig}}{{Pollack}
  et~al.}{1996}]{1996Icar..124...62P}
{Pollack} J.~B.,  {Hubickyj} O.,  {Bodenheimer} P.,  {Lissauer} J.~J.,
  {Podolak} M.,   {Greenzweig} Y.,  1996, \mn@doi [\icarus]
  {10.1006/icar.1996.0190}, \href
  {https://ui.adsabs.harvard.edu/abs/1996Icar..124...62P} {124, 62}

\bibitem[\protect\citeauthoryear{{Ricker} et~al.,}{{Ricker}
  et~al.}{2015}]{Ricker2015}
{Ricker} G.~R.,  et~al., 2015, \mn@doi [Journal of Astronomical Telescopes,
  Instruments, and Systems] {10.1117/1.JATIS.1.1.014003}, \href
  {https://ui.adsabs.harvard.edu/abs/2015JATIS...1a4003R} {1, 014003}

\bibitem[\protect\citeauthoryear{Salvatier, Wiecki  \& Fonnesbeck}{Salvatier
  et~al.}{2016}]{exoplanet:pymc3}
Salvatier J.,  Wiecki T.~V.,   Fonnesbeck C.,  2016, PeerJ Computer Science, 2,
  e55

\bibitem[\protect\citeauthoryear{{Santos} et~al.,}{{Santos}
  et~al.}{2013}]{Santos-13}
{Santos} N.~C.,  et~al., 2013, \mn@doi [\aap] {10.1051/0004-6361/201321286},
  \href {http://adsabs.harvard.edu/abs/2013A%26A...556A.150S} {556, A150}

\bibitem[\protect\citeauthoryear{{Santos} et~al.,}{{Santos}
  et~al.}{2015}]{Santos2015}
{Santos} N.~C.,  et~al., 2015, \mn@doi [\aap] {10.1051/0004-6361/201526850},
  \href {https://ui.adsabs.harvard.edu/abs/2015A&A...580L..13S} {580, L13}

\bibitem[\protect\citeauthoryear{{Santos} et~al.,}{{Santos}
  et~al.}{2017}]{Santos2017}
{Santos} N.~C.,  et~al., 2017, \mn@doi [\aap] {10.1051/0004-6361/201731359},
  \href {https://ui.adsabs.harvard.edu/abs/2017A&A...608A..94S} {608, A94}

\bibitem[\protect\citeauthoryear{{Schlegel}, {Finkbeiner}  \&
  {Davis}}{{Schlegel} et~al.}{1998}]{Schlegel1998}
{Schlegel} D.~J.,  {Finkbeiner} D.~P.,   {Davis} M.,  1998, \mn@doi [\apj]
  {10.1086/305772}, \href
  {https://ui.adsabs.harvard.edu/abs/1998ApJ...500..525S} {500, 525}

\bibitem[\protect\citeauthoryear{{Sestovic}, {Demory}  \& {Queloz}}{{Sestovic}
  et~al.}{2018}]{2018A&A...616A..76S}
{Sestovic} M.,  {Demory} B.-O.,   {Queloz} D.,  2018, \mn@doi [\aap]
  {10.1051/0004-6361/201731454}, \href
  {https://ui.adsabs.harvard.edu/abs/2018A&A...616A..76S} {616, A76}

\bibitem[\protect\citeauthoryear{{Shibata}, {Helled}  \& {Ikoma}}{{Shibata}
  et~al.}{2020}]{2020A&A...633A..33S}
{Shibata} S.,  {Helled} R.,   {Ikoma} M.,  2020, \mn@doi [\aap]
  {10.1051/0004-6361/201936700}, \href
  {https://ui.adsabs.harvard.edu/abs/2020A&A...633A..33S} {633, A33}

\bibitem[\protect\citeauthoryear{{Skrutskie} et~al.,}{{Skrutskie}
  et~al.}{2006}]{Skrutskie2006}
{Skrutskie} M.~F.,  et~al., 2006, \mn@doi [\aj] {10.1086/498708}, \href
  {https://ui.adsabs.harvard.edu/abs/2006AJ....131.1163S} {131, 1163}

\bibitem[\protect\citeauthoryear{{Smith} et~al.,}{{Smith}
  et~al.}{2012}]{Smith2012}
{Smith} J.~C.,  et~al., 2012, \mn@doi [\pasp] {10.1086/667697}, \href
  {https://ui.adsabs.harvard.edu/abs/2012PASP..124.1000S} {124, 1000}

\bibitem[\protect\citeauthoryear{{Sneden}}{{Sneden}}{1973}]{Sneden-73}
{Sneden} C.~A.,  1973, PhD thesis, THE UNIVERSITY OF TEXAS AT AUSTIN.

\bibitem[\protect\citeauthoryear{Sousa}{Sousa}{2014}]{Sousa-14}
Sousa S.~G.,  2014, ARES + MOOG: A Practical Overview of an Equivalent Width
  (EW) Method to Derive Stellar Parameters.
Springer International Publishing, p. 297–310,
  \mn@doi{10.1007/978-3-319-06956-2_26}, \url
  {http://dx.doi.org/10.1007/978-3-319-06956-2_26}

\bibitem[\protect\citeauthoryear{{Sousa}, {Santos}, {Israelian}, {Mayor}  \&
  {Monteiro}}{{Sousa} et~al.}{2007}]{Sousa-07}
{Sousa} S.~G.,  {Santos} N.~C.,  {Israelian} G.,  {Mayor} M.,   {Monteiro}
  M.~J.~P.~F.~G.,  2007, \mn@doi [A\&A] {10.1051/0004-6361:20077288}, \href
  {http://adsabs.harvard.edu/abs/2007A%26A...469..783S} {469, 783}

\bibitem[\protect\citeauthoryear{{Sousa} et~al.,}{{Sousa}
  et~al.}{2008}]{Sousa-08}
{Sousa} S.~G.,  et~al., 2008, \mn@doi [\aap] {10.1051/0004-6361:200809698},
  \href {https://ui.adsabs.harvard.edu/abs/2008A&A...487..373S} {487, 373}

\bibitem[\protect\citeauthoryear{{Sousa}, {Santos}, {Adibekyan}, {Delgado-Mena}
   \& {Israelian}}{{Sousa} et~al.}{2015}]{Sousa-15}
{Sousa} S.~G.,  {Santos} N.~C.,  {Adibekyan} V.,  {Delgado-Mena} E.,
  {Israelian} G.,  2015, \mn@doi [\aap] {10.1051/0004-6361/201425463}, \href
  {http://adsabs.harvard.edu/abs/2015A%26A...577A..67S} {577, A67}

\bibitem[\protect\citeauthoryear{{Sousa} et~al.,}{{Sousa}
  et~al.}{2021}]{Sousa-21}
{Sousa} S.~G.,  et~al., 2021, arXiv e-prints, \href
  {https://ui.adsabs.harvard.edu/abs/2021arXiv210904781S} {p. arXiv:2109.04781}

\bibitem[\protect\citeauthoryear{{Stassun} \& {Torres}}{{Stassun} \&
  {Torres}}{2016}]{StassunTorres2016}
{Stassun} K.~G.,  {Torres} G.,  2016, \mn@doi [\aj]
  {10.3847/0004-6256/152/6/180}, \href
  {https://ui.adsabs.harvard.edu/abs/2016AJ....152..180S} {152, 180}

\bibitem[\protect\citeauthoryear{{Stassun}, {Collins}  \& {Gaudi}}{{Stassun}
  et~al.}{2017}]{Stassun2017}
{Stassun} K.~G.,  {Collins} K.~A.,   {Gaudi} B.~S.,  2017, \mn@doi [\aj]
  {10.3847/1538-3881/aa5df3}, \href
  {https://ui.adsabs.harvard.edu/abs/2017AJ....153..136S} {153, 136}

\bibitem[\protect\citeauthoryear{{Stassun}, {Corsaro}, {Pepper}  \&
  {Gaudi}}{{Stassun} et~al.}{2018}]{Stassun2018}
{Stassun} K.~G.,  {Corsaro} E.,  {Pepper} J.~A.,   {Gaudi} B.~S.,  2018,
  \mn@doi [\aj] {10.3847/1538-3881/aa998a}, \href
  {https://ui.adsabs.harvard.edu/abs/2018AJ....155...22S} {155, 22}

\bibitem[\protect\citeauthoryear{{Stassun} et~al.,}{{Stassun}
  et~al.}{2019}]{Stassun2019}
{Stassun} K.~G.,  et~al., 2019, \mn@doi [\aj] {10.3847/1538-3881/ab3467}, \href
  {https://ui.adsabs.harvard.edu/abs/2019AJ....158..138S} {158, 138}

\bibitem[\protect\citeauthoryear{{Stumpe} et~al.,}{{Stumpe}
  et~al.}{2012}]{Stumpe2012}
{Stumpe} M.~C.,  et~al., 2012, \mn@doi [\pasp] {10.1086/667698}, \href
  {https://ui.adsabs.harvard.edu/abs/2012PASP..124..985S} {124, 985}

\bibitem[\protect\citeauthoryear{{Stumpe}, {Smith}, {Catanzarite}, {Van Cleve},
  {Jenkins}, {Twicken}  \& {Girouard}}{{Stumpe} et~al.}{2014}]{Stumpe2014}
{Stumpe} M.~C.,  {Smith} J.~C.,  {Catanzarite} J.~H.,  {Van Cleve} J.~E.,
  {Jenkins} J.~M.,  {Twicken} J.~D.,   {Girouard} F.~R.,  2014, \mn@doi [\pasp]
  {10.1086/674989}, \href
  {https://ui.adsabs.harvard.edu/abs/2014PASP..126..100S} {126, 100}

\bibitem[\protect\citeauthoryear{{Szab{\'o}} \& {Kiss}}{{Szab{\'o}} \&
  {Kiss}}{2011}]{SzaboKiss2011}
{Szab{\'o}} G.~M.,  {Kiss} L.~L.,  2011, \mn@doi [\apjl]
  {10.1088/2041-8205/727/2/L44}, \href
  {https://ui.adsabs.harvard.edu/abs/2011ApJ...727L..44S} {727, L44}

\bibitem[\protect\citeauthoryear{{Thorngren} \& {Fortney}}{{Thorngren} \&
  {Fortney}}{2018}]{2018AJ....155..214T}
{Thorngren} D.~P.,  {Fortney} J.~J.,  2018, \mn@doi [\aj]
  {10.3847/1538-3881/aaba13}, \href
  {https://ui.adsabs.harvard.edu/abs/2018AJ....155..214T} {155, 214}

\bibitem[\protect\citeauthoryear{{Thorngren} \& {Fortney}}{{Thorngren} \&
  {Fortney}}{2019}]{2019ApJ...874L..31T}
{Thorngren} D.,  {Fortney} J.~J.,  2019, \mn@doi [\apjl]
  {10.3847/2041-8213/ab1137}, \href
  {https://ui.adsabs.harvard.edu/abs/2019ApJ...874L..31T} {874, L31}

\bibitem[\protect\citeauthoryear{{Thorngren}, {Fortney}, {Murray-Clay}  \&
  {Lopez}}{{Thorngren} et~al.}{2016}]{2016ApJ...831...64T}
{Thorngren} D.~P.,  {Fortney} J.~J.,  {Murray-Clay} R.~A.,   {Lopez} E.~D.,
  2016, \mn@doi [\apj] {10.3847/0004-637X/831/1/64}, \href
  {https://ui.adsabs.harvard.edu/abs/2016ApJ...831...64T} {831, 64}

\bibitem[\protect\citeauthoryear{{Tokovinin}}{{Tokovinin}}{2018}]{tokovinin}
{Tokovinin} A.,  2018, \mn@doi [\pasp] {10.1088/1538-3873/aaa7d9}, \href
  {https://ui.adsabs.harvard.edu/abs/2018PASP..130c5002T} {130, 035002}

\bibitem[\protect\citeauthoryear{{Torres}, {Andersen}  \&
  {Gim{\'e}nez}}{{Torres} et~al.}{2010}]{Torres2010}
{Torres} G.,  {Andersen} J.,   {Gim{\'e}nez} A.,  2010, \mn@doi [\aapr]
  {10.1007/s00159-009-0025-1}, \href
  {https://ui.adsabs.harvard.edu/abs/2010A&ARv..18...67T} {18, 67}

\bibitem[\protect\citeauthoryear{{Tsantaki}, {Sousa}, {Adibekyan}, {Santos},
  {Mortier}  \& {Israelian}}{{Tsantaki} et~al.}{2013}]{Tsantaki-2013}
{Tsantaki} M.,  {Sousa} S.~G.,  {Adibekyan} V.~Z.,  {Santos} N.~C.,  {Mortier}
  A.,   {Israelian} G.,  2013, \mn@doi [\aap] {10.1051/0004-6361/201321103},
  \href {http://adsabs.harvard.edu/abs/2013A%26A...555A.150T} {555, A150}

\bibitem[\protect\citeauthoryear{{Twicken}, {Chandrasekaran}, {Jenkins},
  {Gunter}, {Girouard}  \& {Klaus}}{{Twicken} et~al.}{2010}]{Twicken2010}
{Twicken} J.~D.,  {Chandrasekaran} H.,  {Jenkins} J.~M.,  {Gunter} J.~P.,
  {Girouard} F.,   {Klaus} T.~C.,  2010, in {Radziwill} N.~M.,  {Bridger} A.,
  eds,  Society of Photo-Optical Instrumentation Engineers (SPIE) Conference
  Series Vol. 7740, Software and Cyberinfrastructure for Astronomy. p. 77401U,
  \mn@doi{10.1117/12.856798}

\bibitem[\protect\citeauthoryear{{Twicken} et~al.,}{{Twicken}
  et~al.}{2018}]{Twicken_2018}
{Twicken} J.~D.,  et~al., 2018, \mn@doi [\pasp] {10.1088/1538-3873/aab694},
  \href {https://ui.adsabs.harvard.edu/abs/2018PASP..130f4502T} {130, 064502}

\bibitem[\protect\citeauthoryear{Vehtari, Gelman, Simpson, Carpenter  \&
  Bürkner}{Vehtari et~al.}{2021}]{rhat}
Vehtari A.,  Gelman A.,  Simpson D.,  Carpenter B.,   Bürkner P.-C.,  2021,
  \mn@doi [Bayesian Analysis] {10.1214/20-ba1221}, 16

\bibitem[\protect\citeauthoryear{{Vissapragada} et~al.,}{{Vissapragada}
  et~al.}{2022}]{Vissapragada2022}
{Vissapragada} S.,  et~al., 2022, \mn@doi [\aj] {10.3847/1538-3881/ac92f2},
  \href {https://ui.adsabs.harvard.edu/abs/2022AJ....164..234V} {164, 234}

\bibitem[\protect\citeauthoryear{Wakeford et~al.,}{Wakeford
  et~al.}{2017}]{Wakeford_2018}
Wakeford H.~R.,  et~al., 2017, \mn@doi [The Astronomical Journal]
  {10.3847/1538-3881/aa9e4e}, 155, 29

\bibitem[\protect\citeauthoryear{{Wilson} et~al.,}{{Wilson}
  et~al.}{2022}]{Wilson22:toi-1064}
{Wilson} T.~G.,  et~al., 2022, \mn@doi [\mnras] {10.1093/mnras/stab3799}, \href
  {https://ui.adsabs.harvard.edu/abs/2022MNRAS.511.1043W} {511, 1043}

\bibitem[\protect\citeauthoryear{Winn}{Winn}{2014}]{winn2014transits}
Winn J.~N.,  2014, Transits and Occultations (\mn@eprint {arXiv} {1001.2010})

\bibitem[\protect\citeauthoryear{{Wright}, {Drake}, {Mamajek}  \&
  {Henry}}{{Wright} et~al.}{2011}]{Wright11:rotation-xrays}
{Wright} N.~J.,  {Drake} J.~J.,  {Mamajek} E.~E.,   {Henry} G.~W.,  2011,
  \mn@doi [\apj] {10.1088/0004-637X/743/1/48}, \href
  {https://ui.adsabs.harvard.edu/abs/2011ApJ...743...48W} {743, 48}

\bibitem[\protect\citeauthoryear{{Zechmeister} \& {K{\"u}rster}}{{Zechmeister}
  \& {K{\"u}rster}}{2009}]{zechmeister_kurster}
{Zechmeister} M.,  {K{\"u}rster} M.,  2009, \mn@doi [\aap]
  {10.1051/0004-6361:200811296}, \href
  {https://ui.adsabs.harvard.edu/abs/2009A&A...496..577Z} {496, 577}

\bibitem[\protect\citeauthoryear{{Ziegler}, {Tokovinin}, {Brice{\~n}o}, {Mang},
  {Law}  \& {Mann}}{{Ziegler} et~al.}{2020}]{ziegler}
{Ziegler} C.,  {Tokovinin} A.,  {Brice{\~n}o} C.,  {Mang} J.,  {Law} N.,
  {Mann} A.~W.,  2020, \mn@doi [\aj] {10.3847/1538-3881/ab55e9}, \href
  {https://ui.adsabs.harvard.edu/abs/2020AJ....159...19Z} {159, 19}

\bibitem[\protect\citeauthoryear{{da Silva} et~al.,}{{da Silva}
  et~al.}{2006}]{daSilva-2006}
{da Silva} L.,  et~al., 2006, \mn@doi [\aap] {10.1051/0004-6361:20065105},
  \href {https://ui.adsabs.harvard.edu/abs/2006A&A...458..609D} {458, 609}

\makeatother
\end{thebibliography}

%%%%%%%%%%%%%%%%%%%%%%%%%%%%%%%%%%%%%%%%%%%%%%%%%%

%%%%%%%%%%%%%%%%% APPENDICES %%%%%%%%%%%%%%%%%%%%%

\appendix

\section{Spectroscopy}

The HARPS RV data (described in Section\,\ref{harps}) are presented in tables \ref{tab:harps_3071} (TOI-3071) anb \ref{tab:harps} (TOI-2374). The PFS RV data (described in Section\,\ref{pfs}) are presented in Table \ref{tab:pfs}.

\begin{table}
    \scriptsize
	\caption{TOI-3071 HARPS radial velocities.}
	\label{tab:harps_3071}
	\begin{threeparttable}
	\begin{tabular}{cccccc}
	\toprule
	Time                & RV            & $\sigma_\textrm{RV}$  & FWHM          & Bisector      & Contrast  \\
	(BJD-2457000)               & ($ms^{-1}$)   & ($ms^{-1}$)           & ($kms^{-1}$)   & ($ms^{-1}$)   &           \\
	\midrule
    2657.758 & 9353.8 & 4.5 & 7.945(11) & 45(6) & 43.199052(1) \\ 
    2658.831 & 9375.9 & 4.8 & 7.938(11) & 48(7) & 43.167730(1) \\ 
    2660.804 & 9390.7 & 3.8 & 7.957(11) & 43(6) & 43.186370(1) \\ 
    2664.689 & 9410.5 & 3.1 & 7.958(11) & 45(4) & 43.381653(1) \\ 
    2664.818 & 9415.3 & 4.3 & 7.985(11) & 33(6) & 43.070969(1) \\ 
    2665.622 & 9358.9 & 2.9 & 7.954(11) & 21(4) & 43.494550(1) \\ 
    2665.806 & 9391.0 & 4.3 & 7.960(11) & 23(6) & 43.497998(1) \\ 
    2666.774 & 9347.8 & 3.7 & 7.941(11) & 40(5) & 43.568806(1) \\ 
    2667.597 & 9402.3 & 3.9 & 7.958(11) & 30(5) & 43.480382(1) \\ 
    2667.759 & 9376.5 & 3.9 & 7.938(11) & 38(5) & 43.521020(1) \\ 
    2668.766 & 9408.2 & 5.1 & 7.953(11) & 30(7) & 43.469858(1) \\ 
    2669.802 & 9416.1 & 4.2 & 7.951(11) & 26(6) & 43.498341(1) \\ 
    2670.753 & 9366.7 & 3.8 & 7.950(11) & 20(5) & 43.505071(1) \\
    2671.658 & 9358.0 & 3.3 & 7.961(11) & 24(4) & 43.514566(1) \\ 
	\bottomrule
	\end{tabular}
    \begin{tablenotes}
    \item {The full HARPS data products can be found on the ESO archive}
    \end{tablenotes}
    \end{threeparttable}
\end{table}

\begin{table}
    \scriptsize
	\caption{TOI-2374 HARPS radial velocities.}
	\label{tab:harps}
	\begin{threeparttable}
	\begin{tabular}{cccccc}
	\toprule
	Time                & RV            & $\sigma_\textrm{RV}$  & FWHM          & Bisector      & Contrast  \\
	(BJD-2457000)               & ($ms^{-1}$)   & ($ms^{-1}$)           & ($kms^{-1}$)   & ($ms^{-1}$)   &           \\
	\midrule
    2678.897 & 14983.6 & 3.4 & 6.6264(88) & 35(5) & 51.054114(1) \\ 
    2679.890 & 15008.6 & 3.7 & 6.2425(88) & 39(5) & 51.041247(1) \\ 
    2681.905 & 14961.4 & 3.6 & 6.2461(88) & 34(5) & 51.061773(1) \\ 
    2682.913 & 14963.8 & 2.8 & 6.2338(88) & 35(4) & 50.923438(1) \\ 
    2711.852 & 14006.3 & 1.7 & 9.193(13) & 81(2) & 21.564691(1) \\ 
    2712.822 & 14938 & 28 & 6.2944(88) & -22(39) & 52.935152(1) \\ 
    2714.869 & 15012.4 & 2.8 & 6.2385(88) & 42(4) & 50.732359(1) \\ 
    2715.792 & 14969.3 & 2.7 & 6.2179(88) & 26(4) & 50.174915(1) \\ 
    2716.806 & 14902 & 20 & 6.3119(89) & 42(28) & 51.119746(1) \\ 
    2723.924 & 14994.7 & 3.8 & 6.2257(88) & 27(5) & 51.121218(1) \\ 
    2725.869 & 14958.4 & 2.6 & 6.2410(88) & 25(4) & 51.017728(1) \\ 
    2726.847 & 14997.4 & 2.9 & 6.2614(88) & 23(4) & 51.007037(1) \\ 
    2727.894 & 15006.0 & 2.7 & 6.2535(88) & 30(4) & 51.031929(1) \\ 
    2728.850 & 14961.7 & 5.0 & 6.2662(88) & 17(7) & 51.068559(1) \\ 
    2750.724 & 14974.5 & 7.8 & 6.3386(90) & 59(11) & 50.319108(1) \\ 
    2759.766 & 14958.7 & 3.5 & 6.2457(88) & 24(5) & 51.046751(1) \\ 
    2760.767 & 14983.7 & 7.9 & 6.2861(88) & 25(11) & 51.429800(1) \\ 
    2760.788 & 14978.0 & 4.6 & 6.2310(88) & 34(6) & 51.064180(1) \\ 
    2761.722 & 14992.3 & 6.6 & 6.2332(88) & 39(9) & 51.422302(1) \\ 
    2766.754 & 14997.3 & 5.0 & 6.2391(88) & 18(7) & 51.411205(1) \\ 
    2767.864 & 14959.4 & 3.2 & 6.2342(88) & 23(4) & 51.197217(1) \\ 
	\bottomrule
	\end{tabular}
     \begin{tablenotes}
    \item {The full HARPS data products can be found on the ESO archive}
    \end{tablenotes}
    \end{threeparttable}
\end{table}

\begin{table}
    \scriptsize
	\caption{TOI-2374 PFS radial velocities.}
	\label{tab:pfs}
	\begin{threeparttable}
	\begin{tabular}{ccccc}
	\toprule
	Time                & $\Delta$RV$^{*}$            & $\sigma_\textrm{RV}$          & Bisector  & $\sigma_\textrm{bis}$\\
	(BJD-2457000)               & ($ms^{-1}$)   & ($ms^{-1}$)           & ($ms^{-1}$) & ($ms^{-1}$)      \\
	\midrule
        2471.603 & 4.45 & 2.3 & 46.6 & 31.4 \\ 
        2535.533 & -20.59 & 3.3 & 94.4 & 118.8 \\ 
        2472.547 & 43.89 & 2.06 & 100.5 & 24.9 \\ 
        2504.567 & -2.89 & 5.02 & -179.6 & 173.2 \\ 
        2509.534 & 0 & 3.41 & -69.6 & 59.0 \\ 
        2473.515 & 36.13 & 2.25 & -43.4 & 32.1 \\ 
        2534.528 & -5.46 & 2.67 & -36.9 & 53.6 \\ 
	\bottomrule
	\end{tabular}
    \begin{tablenotes}
    \item $^*$ Relative RVs (see Sect. \ref{pfs})
    \item The full PFS data products can be found on ExoFOP-TESS at \url{https://exofop.ipac.caltech.edu/tess/target.php?id=439366538}
    \end{tablenotes}
    \end{threeparttable}
\end{table}

\section{Stellar abundances and chemical clocks}
The stellar abundances of the elements derived in Sect. \ref{stellar-abundances} are shown in Table \ref{tab:abundances} and the stellar ages derived from chemical clocks as described in Sect. \ref{stellar-physical-parameters} are shown in Table \ref{tab:clocks}.

\begin{table}
    \centering
    \caption{Stellar abundances weighted average and standard deviation, derived with the logg from Gaia DR3. Li is depleted in both targets. The abundances of carbon and oxygen for TOI-2374 cannot be determined for this cool star with our methodology. For TOI-3071, the abundance of oxygen has a high error because the lines are very weak and hard to measure in its spectrum with low S/N. Moreover, one of the lines was contaminated by a telluric line in many of the spectra so it was measured using a small number of spectra.}
    \label{tab:abundances}
    \begin{threeparttable}
    \begin{tabular}{ccc}
	\toprule
        \textbf{Star} & \textbf{TOI-2374} & \textbf{TOI-3071} \\ \hline
        \textbf{[Fe/H]  } & 0.15 $\pm$ 0.04  & 0.35 $\pm$ 0.04   \\ 
        \textbf{[O/H]} & - & 0.332 $\pm$  0.154   \\ 
        \textbf{[C/H]} & - & 0.241 $\pm$ 0.026    \\ 
        \textbf{[Cu/H]} & 0.293 $\pm$ 0.100 & 0.358 $\pm$  0.028    \\ 
        \textbf{[Zn/H]} &  0.088 $\pm$ 0.079 & 0.284 $\pm$  0.058   \\ 
        \textbf{[Sr/H]} & 0.248 $\pm$ 0.167 & 0.272 $\pm$ 0.077     \\ 
        \textbf{[Y/H] } & 0.090 $\pm$ 0.155 & 0.327 $\pm$ 0.023     \\ 
        \textbf{[Zr/H]  } & 0.321 $\pm$ 0.109 & 0.339 $\pm$ 0.025  \\ 
        \textbf{[Ba/H] } & 0.026 $\pm$ 0.087 & 0.280 $\pm$ 0.039 \\ 
        \textbf{[Ce/H]   } & 0.120 $\pm$ 0.117 & 0.219 $\pm$  0.036   \\ 
        \textbf{[Nd/H]} & 0.323 $\pm$ 0.084  &   0.209 $\pm$ 0.034  \\ 
        \textbf{[Mg/H]} & 0.07 $\pm$  0.09 & 0.30 $\pm$  0.06 \\ 
        \textbf{[Al/H]} & 0.17 $\pm$  0.07 & 0.30 $\pm$  0.08 \\ 
        \textbf{[Si/H]} & 0.21 $\pm$  0.09 & 0.33 $\pm$  0.04 \\ 
        \textbf{[Ti/H]} & 0.20 $\pm$  0.12 & 0.33 $\pm$  0.04 \\ 
        \textbf{[Ni/H]} & - & 0.36 $\pm$  0.02  \\ \hline
    \end{tabular}
    \end{threeparttable}
\end{table}

\begin{table}
    \centering
    \caption{Stellar ages derived from chemical clocks (see Sect. \ref{stellar-physical-parameters}).}
    \label{tab:clocks}
    \begin{threeparttable}
    \begin{tabular}{cc}
    \toprule    
    %     \textbf{TOI-2374} & \textbf{1D clocks} \\
    % \midrule
    %     \textbf{YMg} & 2.8414000 $\pm$4.2890812  \\ 
    %     \textbf{MgFe} & 1.4519998 $\pm$2.8934489 \\ 
    %     \textbf{YTi} & 5.7325001 $\pm$5.4396371 \\ 
    %     \textbf{TiFe} & 4.9349999 $\pm$ 4.4475704 \\ 
    %     \textbf{SiFe} &  5.9095994 $\pm$ 4.4259591  \\ 
    %     \textbf{YZn} &5.0669000 $\pm$ 4.6189370 \\ 
    %     \textbf{ZnFe} & 4.0805398 $\pm$ 3.0237977   \\ 
    %     \textbf{Ysi} & 6.5499998 $\pm$ 5.4218431 \\ 
    % \midrule
        \textbf{TOI-2374}: \textbf{2D clocks} & Age (Gyr)  \\
    \midrule
        \textbf{[Y/Zn]} & 3.6911800 $\pm$ 3.7638734  \\ 
        \textbf{[Zn/Fe]} & 2.8571398 $\pm$ 2.1241537  \\ 
        \textbf{[Y/al]} & 4.4640000 $\pm$ 2.7333606   \\ 
        \textbf{[Y/Mg]} & 2.6525000 $\pm$ 3.7252200  \\ 
    \midrule
        \textbf{TOI-3071}: \textbf{3D clocks} & Age (Gyr)  \\
    \midrule
        \textbf{[Y/Zn]} & 1.0152100 $\pm$ 1.0256072  \\ 
        \textbf{[Y/Ti]} &   1.1382002 $\pm$ 0.87969329 \\ 
        \textbf{[Y/Mg]} & 0.94835018 $\pm$ 0.98228770   \\ 
        \textbf{[Sr/Ti]} & 1.7918402 $\pm$ 1.4564712  \\
        \textbf{[Sr/Mg]} &  1.4527202 $\pm$ 1.3351688    \\
        \textbf{[Y/Si]} &  0.93273022 $\pm$ 0.85117129      \\
        \textbf{[Sr/Si]} &1.1112002 $\pm$ 1.3381209  \\
        \textbf{[Y/Al]} &  1.4601001  $\pm$ 0.80915273 \\
    \bottomrule
    \end{tabular}
    \end{threeparttable}
\end{table}

\section{The joint fit}

\begin{figure*}
    \centering
    \includegraphics[width=0.4\textwidth]{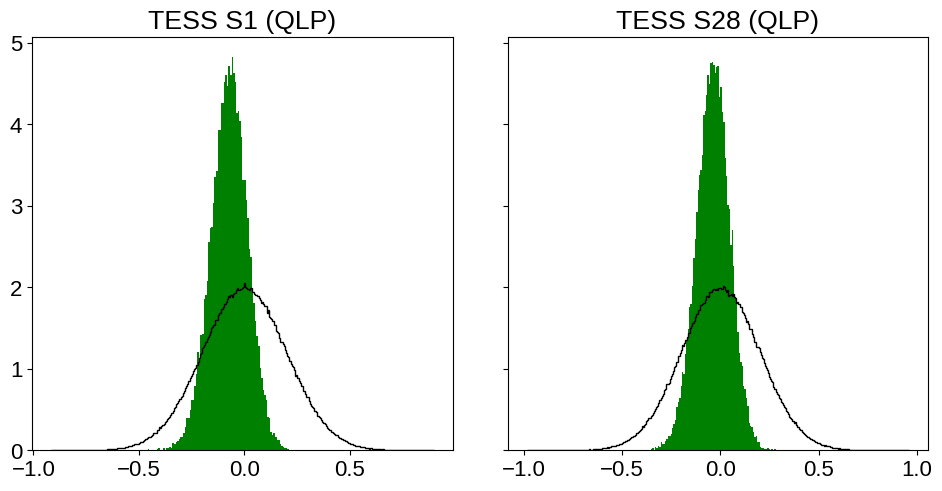}
    \caption{Posterior distributions for the dilution correction factors of the TOI-2374 TESS light curves contrasted with the prior distributions.}
    \label{fig:dilution_hist_2374}
\end{figure*}

\begin{figure*}
    \centering
    \includegraphics[width=0.6\textwidth]{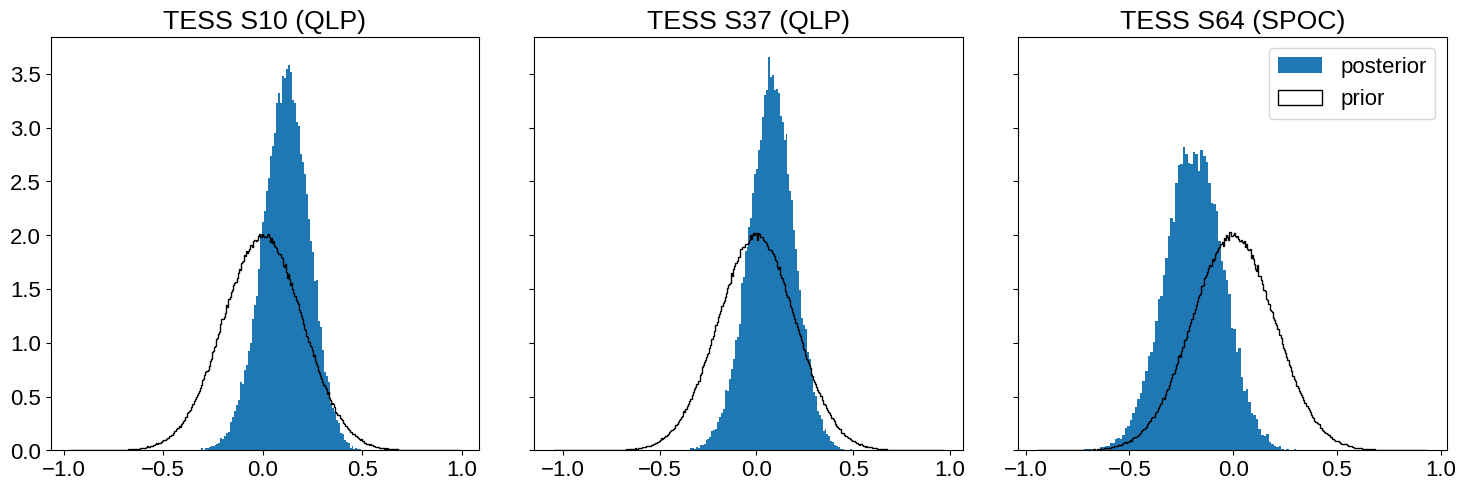}
    \caption{Posterior distributions for the dilution correction factors of the TOI-3071 light curves contrasted with the prior distributions.}
    \label{fig:dilution_hist_3071}
\end{figure*}

The full set of priors and fit values from our joint fit model described in Section\,\ref{jointfit} are presented in tables \ref{tab:jointfit2374} and \ref{tab:jointfit3071}. The posterior distributions for the dilution correction factors are shown in figures \ref{fig:dilution_hist_2374} and \ref{fig:dilution_hist_3071}.

\begin{table*}
    \renewcommand{\arraystretch}{1.1}
    \small
    \caption{Prior distributions used in our joint fit model for the TOI-2374 system, fully described in Section\,\ref{jointfit}, and the fit values resulting from the model. The priors are created using distributions in \textsc{PyMC3} with the relevant inputs to each distribution described in the table footer. The fit values are given as the median values of the samples, and the uncertainties as the 16th and 84th percentiles. Further (derived) system parameters can be found in Table \ref{tab:params}.}
	\label{tab:jointfit2374}
	\begin{threeparttable}
	\begin{tabular}{l p{1.3cm} cc}
	\toprule
	\textbf{Parameter} & \textbf{(unit)} & \textbf{Prior Distribution} & \textbf{Fit value}  \\
	\midrule
	\multicolumn{4}{l}{\textbf{Planet}} \\
	$\log{(P)}$                     & (days)               & $\mathcal{N}(1.461777, 1.0)$       & $1.461776\pm0.000002 $\\
	Epoch $t_0$                 & (BJD-2457000)        & $\mathcal{N}(1326.555387, 1.0)$   & $1326.563 \pm 0.002$ \\
	$\log{(R_p)}$                   & (\mbox{R$_{\odot}$}) & $\mathcal{N}(-2.6998^*, 1.0)$          & $-2.40\pm0.03$\\
	$\sqrt{e}\sin{\omega}$                &                      & $\mathcal{U}(-1,1)$                             & $0.20^{+0.06}_{-0.09}$ \\
	$\sqrt{e}\cos{\omega}$                &         &  $\mathcal{U}(-1,1)$                             & $0.10^{+0.13}_{-0.17}$ \\
	\midrule
 
	\multicolumn{4}{l}{\textbf{Star}} \\
	Mass $M_\star$                    & (\mbox{M$_{\odot}$}) & $\mathcal{N_B}(0.75, 0.01, 0.0, 3.0)$  & $0.75 \pm 0.01$ \\
	Radius $R_\star$                  & (\mbox{R$_{\odot}$}) & $\mathcal{N_B}(0.69, 0.02, 0.0, 3.0)$  & $0.69\pm0.02$ \\
	\midrule
  	\multicolumn{4}{l}{\textbf{Limb darkening coefficients}} \\
	TESS filter $u_1$                    &  & \citet{exoplanet:kipping13}   & $0.7 ^{+0.5}_{-0.4}$ \\
    TESS filter $u_2$                    &  & \citet{exoplanet:kipping13}  & $0.0 ^{+0.5}_{-0.4}$ \\
    B filter $u_1$                    &  & \citet{exoplanet:kipping13}  & $0.9 ^{+0.5}_{-0.6}$ \\
    B filter $u_2$                    &  & \citet{exoplanet:kipping13}  & $-0.1 \pm 0.5$ \\
    ip filter $u_1$                    &  & \citet{exoplanet:kipping13}  & $0.5 \pm 0.3$ \\
    ip filter $u_2$                    &  & \citet{exoplanet:kipping13}  & $0.2 \pm 0.4$ \\
    gp filter $u_1$                    &  & \citet{exoplanet:kipping13}  & $0.9 ^{+0.3}_{-0.4}$ \\
    gp filter $u_2$                    &  & \citet{exoplanet:kipping13}  & $-0.2 ^{+0.4}_{-0.3}$ \\
    \midrule
	\multicolumn{4}{l}{\textbf{Photometry}} \\
	{\it TESS}$_{S1}$ transit normalisation factor $f_0$         &                      & $\mathcal{N}(1.0, 0.1)$                & $1.00023 \pm 0.00008$ \\
	{\it TESS}$_{S28}$ transit normalisation factor $f_0$          &                      & $\mathcal{N}(1.0, 0.1)$                & $1.00011 \pm 0.00008$\\
	{\it Brierfield} transit normalisation factor $f_0$         &                      & $\mathcal{N}(1.0, 0.1)$                & $1.0000 \pm 0.0005$ \\
	{\it LCO-SAAO} transit normalisation factor $f_0$          &                      & $\mathcal{N}(1.0, 0.1)$                & $1.0001 \pm 0.0001$\\
	{\it LCO-SSO (ip filter)} transit normalisation factor $f_0$          &                      & $\mathcal{N}(1.0, 0.1)$                & $1.0000 \pm 0.0002$\\
	{\it LCO-SSO (gp filter)} transit normalisation factor $f_0$          &                      & $\mathcal{N}(1.0, 0.1)$                & $1.0000 \pm 0.0002$\\ 

	{\it TESS}$_{S1}$ dilution correction factor $d$     &       & $\mathcal{N}(0.0,0.2)$   & $-0.07 ^{+0.08}_{-0.09}$ \\
	{\it TESS}$_{S28}$ dilution correction factor $d$   &       & $\mathcal{N}(0.0,0.2)$  &   $-0.04 ^{+0.06}_{-0.08}$\\	
	{\it Brierfield} dilution correction factor $d$   &       & $0.0$ (fixed)  & -  \\
	{\it LCO-SAAO} dilution correction factor $d$    &       & $0.0$ (fixed)  & - \\	
	{\it LCO-SSO (ip filter)} dilution correction factor $d$   &   & $0.0$ (fixed) & - \\
	{\it LCO-SSO (gp filter)} dilution correction factor $d$     &    & $0.0$ (fixed) & - \\	

	{\it TESS}$_{S1}$ log (Jitter)         &                      & $\mathcal{N}(-6.3000^{\dagger}, 0.01)$                & $-6.32 \pm 0.01$ \\
	{\it TESS}$_{S28}$ log (Jitter)          &                      & $\mathcal{N}(-5.8934^{\dagger}, 0.01)$                & $-5.94 \pm 0.01$\\
	{\it Brierfield} log (Jitter)         &                      & $\mathcal{N}(-5.8482^{\dagger}, 0.01)$                & $-5.85 \pm 0.01$ \\
	{\it LCO-SAAO} log (Jitter)          &                      & $\mathcal{N}(-6.7355^{\dagger}, 0.01)$                & $-6.74 \pm 0.01$\\
	{\it LCO-SSO (ip filter)} log (Jitter)         &                      & $\mathcal{N}(-6.8753^{\dagger}, 0.01)$                & $-6.88 \pm 0.01$ \\
	{\it LCO-SSO (gp filter)} log (Jitter)          &                      & $\mathcal{N}(-6.7380^{\dagger}, 0.01)$                & $-6.74 \pm 0.01$\\
	\midrule
	\multicolumn{4}{l}{\textbf{RVs}} \\
	$\log{(K)}$                     & (m\,s$^{-1}$)        & $\mathcal{U}(0.0, 5.0)$               & $3.30 ^{+0.06}_{-0.08}$ \\
	{\it HARPS} offset                          & (m\,s$^{-1}$)        & $\mathcal{N}(0.0, 100.0)$          & $7.2 \pm 4.4$ \\
 	{\it PFS} offset                          & (m\,s$^{-1}$)        & $\mathcal{N}(0.0, 100.0)$          & $-14.8 ^{+11.5}_{-10.8}$ \\
 	Linear trend                          & (m\,s$^{-1}$\,d$^{-1}$)        & $\mathcal{N}(0.0, 100.0)$          & $-0.12 \pm 0.06$ \\
  	Quadratic trend                          & (m\,s$^{-1}$\,d$^{-2}$)        & $\mathcal{N}(0.0, 10.0)$          & $0.0004 \pm 0.0003$ \\	
    {\it HARPS} $\log{(\rm{Jitter)}}$           & (m\,s$^{-1}$)        & $\mathcal{U}(-5.0, 3.0)$ & $1.0^{+0.6}_{-3.1}$\\
 	{\it PFS} $\log{(\rm{Jitter)}}$           & (m\,s$^{-1}$)        & $\mathcal{U}(-5.0, 3.0)$ & $2.0\pm0.4$\\
	\bottomrule
	\end{tabular}
	\begin{tablenotes}
	\item \textbf{Distributions:}
	\item $\mathcal{N}(\mu, \sigma)$: a normal distribution with a mean $\mu$ and a standard deviation $\sigma$;
	\item $\mathcal{N_B}(\mu, \sigma, a, b)$: a bounded normal distribution with a mean $\mu$, a standard deviation $\sigma$, a lower bound $a$, and an upper bound $b$ (bounds optional);
    \item $\mathcal{U}(a,b)$: a uniform distribution between $a$ and $b$;
    \item Distributions for limb darkening coefficients $u_1$ and $u_2$ are built into the \texttt{exoplanet} package and based on \citet{exoplanet:kipping13}. 
	\item \textbf{Prior values:}
	\item $^*$ equivalent to $0.5(\log{(D)}) + \log{(R_\star)}$ where $D$ is the transit depth (ppm multiplied by $10^{-6}$) and $R_\star$ is the mean of the prior on the stellar radius (\mbox{R$_{\odot}$});
 	\item $^{\dagger}$ equivalent to the log of the error median on the data of the correspondent instrument.
	\end{tablenotes}
	\end{threeparttable}
\end{table*}

\begin{table*}
    \renewcommand{\arraystretch}{1.4}
    \small
    \caption{Same as table \ref{tab:jointfit2374} but for the TOI-3071 system. Further (derived) system parameters can be found in Table \ref{tab:params}.}
	\label{tab:jointfit3071}
	\begin{threeparttable}
	\begin{tabular}{l p{1.3cm} cc}
	\toprule
	\textbf{Parameter} & \textbf{(unit)} & \textbf{Prior Distribution} & \textbf{Fit value}  \\
	\midrule
	\multicolumn{4}{l}{\textbf{Planet}} \\
	$\log{(P)}$                     & (days)               & $\mathcal{N}(0.2367297, 1.0)$       & $0.236603\pm0.000002$ \\
	Epoch $t_0$                 & (BJD-2457000)        & $\mathcal{N}(1594.611536, 1.0)$   & $1594.608 \pm 0.002$\\
	$\log{(R_p)}$                   & (\mbox{R$_{\odot}$}) & $\mathcal{N}(-2.709956^*, 1.0)$          & $-2.99\pm0.07$\\
	$\sqrt{e}\sin{\omega}$                &                      & $\mathcal{U}(-1,1)$                             & $-0.11^{+0.10}_{-0.07}$  \\
	$\sqrt{e}\cos{\omega}$                &         &  $\mathcal{U}(-1,1)$                             & $0.07^{+0.16}_{-0.14}$\\
	\\
	\midrule
 
	\multicolumn{4}{l}{\textbf{Star}} \\
	Mass $M_\star$                    & (\mbox{M$_{\odot}$}) & $\mathcal{N_B}(1.29, 0.02, 0.0, 3.0)$  & $1.29 \pm 0.02$ \\
	Radius $R_\star$                  & (\mbox{R$_{\odot}$}) & $\mathcal{N_B}(1.31, 0.04, 0.0, 3.0)$  & $1.31\pm0.04$ \\
    \midrule
 	\multicolumn{4}{l}{\textbf{Limb darkening coefficients}} \\
	TESS filter $u_1$                    &  & \citet{exoplanet:kipping13}  & $0.28^{+0.31}_{-0.20}$ \\
    TESS filter $u_2$                    &  & \citet{exoplanet:kipping13}  & $0.31^{+0.34}_{-0.37}$  \\
	\midrule	
	\multicolumn{4}{l}{\textbf{Photometry}} \\
	{\it TESS}$_{S10}$ transit normalisation factor $f_0$         &                      & $\mathcal{N}(1.0, 0.1)$                & $1.00013 \pm 0.00004$ \\
	{\it TESS}$_{S37}$ transit normalisation factor $f_0$          &                      & $\mathcal{N}(1.0, 0.1)$                & $1.00010 \pm 0.00004$\\
	{\it TESS}$_{S64}$ transit normalisation factor $f_0$          &                      & $\mathcal{N}(1.0, 0.1)$                & $1.00018 \pm 0.00003$\\ 
	{\it TESS}$_{S10}$ dilution correction factor $d$         &                      & $\mathcal{N}(0.0,0.2)$                & $0.1 \pm 0.1$ \\
	{\it TESS}$_{S37}$ dilution correction factor $d$          &                      & $\mathcal{N}(0.0,0.2)$                & $0.1 \pm 0.1 $\\
	{\it TESS}$_{S64}$ dilution correction factor $d$          &                      & $\mathcal{N}(0.0,0.2)$                & $-0.2 \pm 0.1$\\	
	{\it TESS}$_{S10}$ log (Jitter)         &                      & $\mathcal{N}(-6.67^{\dagger}, 1.0)$                & $-8.7 ^{+0.4}_{-0.5}$ \\
	{\it TESS}$_{S37}$ log (Jitter)          &                      & $\mathcal{N}(-6.02^{\dagger}, 1.0)$                & $-8.3^{+0.4}_{-0.5}$\\
	{\it TESS}$_{S64}$ log (Jitter)         &                      & $\mathcal{N}(-5.52^{\dagger}, 1.0)$                & $-8.7 ^{+0.3}_{-0.4}$ \\
    \midrule
	\multicolumn{4}{l}{\textbf{HARPS RVs}} \\
	$\log{(K)}$                     & (m\,s$^{-1}$)        & $\mathcal{U}(0.0, 5.0)$               & $3.52 \pm 0.05$ \\
	Offset                          & (m\,s$^{-1}$)        & $\mathcal{N}(0.0, 100.0)$          & $0.7 \pm 1.6$ \\
 	Linear trend                          & (m\,s$^{-1}$\,d$^{-1}$)         & $\mathcal{N}(0.0, 10.0)$          & $0.45 \pm 0.28$ \\
  	Quadratic trend                          & (m\,s$^{-1}$\,d$^{-2}$)         & $\mathcal{N}(0.0, 1.0)$          & $-0.09 \pm 0.07$ \\
	$\log{(\rm{Jitter)}}$           & (m\,s$^{-1}$)        & $\mathcal{U}(-5.0, 3.0)$ & $-2.1 \pm 2.0$\\
	\bottomrule
    \end{tabular}
    \begin{tablenotes}
	\item \textbf{Distributions:}
	\item $\mathcal{N}(\mu, \sigma)$: a normal distribution with a mean $\mu$ and a standard deviation $\sigma$;
	\item $\mathcal{N_B}(\mu, \sigma, a, b)$: a bounded normal distribution with a mean $\mu$, a standard deviation $\sigma$, a lower bound $a$, and an upper bound $b$ (bounds optional);
    \item $\mathcal{U}(a,b)$: a uniform distribution between $a$ and $b$;
    \item Distributions for limb darkening coefficients $u_1$ and $u_2$ are built into the \texttt{exoplanet} package and based on \citet{exoplanet:kipping13}.    
	\item \textbf{Prior values:}
	\item $^*$ equivalent to $0.5(\log{(D)}) + \log{(R_\star)}$ where $D$ is the transit depth (ppm multiplied by $10^{-6}$) and $R_\star$ is the mean of the prior on the stellar radius (\mbox{R$_{\odot}$});
 	\item $^{\dagger}$ equivalent to the log of the error median on the data of the correspondent instrument.
	\end{tablenotes}
	\end{threeparttable}
\end{table*}

\section{HEAVY ELEMENT MASS}
\begin{figure}
    \centering
    \includegraphics[width=0.85\columnwidth]{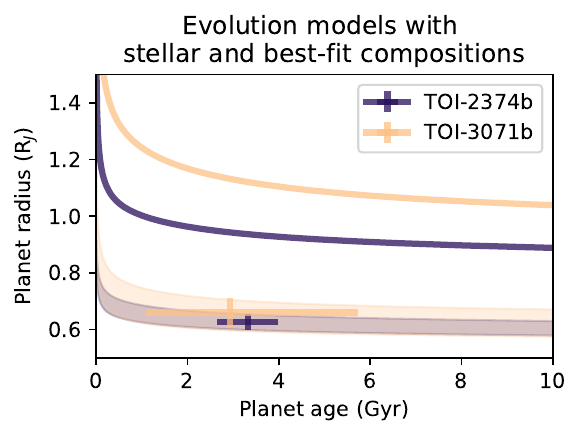}
    \caption{Cooling curves for TOI-2374\,b (dark blue) and TOI-3071\,b (yellow). The solid lines assume they have the same composition as their stars. The shaded regions are cooling curves using the best-fit inferred compositions within $1 \sigma$. The error bars show the observational data. Both planets are much smaller than the predictions from the evolution models, suggesting super-stellar bulk metallicities.}
    \label{fig:age_radius_stellar}
\end{figure}

\begin{figure}
    \centering
    \includegraphics[width=0.75\columnwidth]{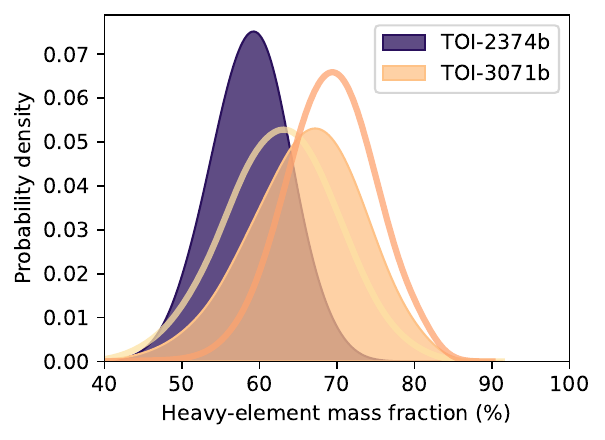}
    \caption{Inferred heavy-element mass fraction posteriors for the two planets. For TOI-3071b, the two curves show the results for the two different stellar age estimates (chemical clocks and gyrochronology), and the filled area is their combined posterior.}
    \label{fig:heavy_element_mass_fraction_posteriors}
\end{figure}

The cooling curves for TOI-2374\,b and TOI-3071\,b described in section \ref{sec:heavymasses} are shown in Fig. \ref{fig:age_radius_stellar}, together with the planet radii and stellar ages from observations. The inferred heavy-element mass fraction posteriors described in the same section are shown in Figure \ref{fig:heavy_element_mass_fraction_posteriors}

\section{Author affiliations}\label{sec:affiliations}

$^{1}$International Center for Advanced Studies (ICAS) and ICIFI (CONICET), ECyT-UNSAM, Campus Miguelete, 25 de Mayo y Francia, (1650) Buenos Aires, Argentina\\
$^{2}$Department of Physics, University of Warwick, Gibbet Hill Road, Coventry CV4 7AL, UK\\
$^{3}$Centre for Exoplanets and Habitability, University of Warwick, Gibbet Hill Road, Coventry CV4 7AL, UK\\
$^{4}$Department of Astrophysics, University of Zürich, Winterthurerstrasse 190, 8057 Zürich, Switzerland\\
$^{5}$Instituto de Astrof\'isica e Ci\^encias do Espa\c{c}o, Universidade do Porto, CAUP, Rua das Estrelas, 4150-762 Porto, Portugal\\
$^{6}$Departamento de F\'isica e Astronomia, Faculdade de Ci\^encias, Universidade do Porto, Rua do Campo Alegre, 4169-007 Porto, Portugal\\
$^{7}$Department of Physics and Astronomy, Vanderbilt University, Nashville, TN 37235, USA\\
$^{8}$Center for Astrophysics \textbar \ Harvard \& Smithsonian, 60 Garden Street, Cambridge, MA 02138, USA\\
$^{9}$Department of Astrophysical Sciences, Princeton University, 4 Ivy Lane, Princeton, NJ 08544, USA\\
$^{10}$Observatoire de Genève, Université de Genève, Chemin Pegasi 51, 1290 Versoix, Switzerland\\
$^{11}$Earth and Planets Laboratory, Carnegie Institution for Science, 5241 Broad Branch Road, NW, Washington, DC 20015, USA\\
$^{12}$Observatories of the Carnegie Institution for Science, 813 Santa Barbara Street, Pasadena, CA 91101, USA\\
$^{13}$NASA Ames Research Center, Moffett Field, CA 94035, USA\\
$^{14}$Space Telescope Science Institute, 3700 San Martin Drive, Baltimore, MD 21218, USA\\
$^{15}$Centro de Astrobiolog\'ia (CAB, CSIC-INTA), Depto. de Astrof\'isica, ESAC campus, 28692, Villanueva de la Ca\~nada (Madrid), Spain\\
$^{16}$NASA Exoplanet Science Institute, IPAC, California Institute of Technology, Pasadena, CA 91125 USA\\
$^{17}$University Observatory Munich, Ludwig-Maximilians-Universit\"{a}t, Scheinerstr. 1, 81679 Munich, Germany\\
$^{18}$European Southern Observatory, Karl-Schwarzschild-Str. 2, 85748 Garching, Germany\\
$^{19}$Department of Physics and Astronomy, McMaster University, 1280 Main Street West, Hamilton, Ontario, L8S 4L8, Canada\\
$^{20}$Université Grenoble Alpes, CNRS, IPAG, 38000 Grenoble, France\\
$^{21}$Las Campanas Observatory, Carnegie Institution of Washington, Colina el Pino, Casilla 601 La Serena, Chile\\
$^{22}$Brierfield Observatory Bowral, NSW, 2576, Australia
$^{23}$Department of Physics and Kavli Institute for Astrophysics and Space Research, Massachusetts Institute of Technology, Cambridge, MA 02139, USA\\
$^{24}$Department of Earth, Atmospheric and Planetary Sciences, Massachusetts Institute of Technology, Cambridge, MA 02139, USA\\
$^{25}$Department of Aeronautics and Astronautics, MIT, 77 Massachusetts Avenue, Cambridge, MA 02139, USA\\
$^{26}$Kotizarovci Observatory, Sarsoni 90, 51216 Viskovo, Croatia\\
$^{27}$Perth Exoplanet Survey Telescope, Perth, 6152 Western Australia\\
$^{28}$Planetary Discoveries, 28935 Via Adelena, Santa Clarita, CA 91354, USA\\
$^{29}$SETI Institute, Mountain View, CA 94043 USA\\
$^{30}$ Department of Physics, Engineering and Astronomy, Stephen F. Austin State University, 1936 North St, Nacogdoches, TX 75962, USA\\

\bsp	
\label{lastpage}
\end{document}